# THE NATIONAL UNDERGROUND SCIENCE AND ENGINEERING LABORATORY AT HOMESTAKE: SCIENCE BOOK: PHYSICS CHAPTER

July, 2003

by

The Homestake Collaboration

# The Homestake Collaboration

**Executive Committee**

Baha Balantekin, University of Wisconsin
Thomas Bowles, Los Alamos National Laboratory
Janet Conrad, Columbia University
Sherry Farwell, South Dakota School of Mines and Technology
Wick Haxton, University of Washington
Ken Lande, University of Pennsylvania
Kevin Lesko, Lawrence Berkeley Laboratory
Bill Marciano, Brookhaven National Laboratory
Marvin Marshak, University of Minnesota
Tullis Onstott, Princeton University
Michael Shaevitz, Columbia University
John Wilkerson, University of Washington

**Scientific Advisor to the Executive Committee**

John Bahcall, Institute for Advanced Study

**Earth Science Steering Committee**

Brian McPherson, New Mexico Tech
Tullis Onstott, Princeton
Tommy Phelps, ORNL
Bill Roggenthen, SDSM&T
Herb Wang, Wisconsin
Joe Wang, LBNL

**Collaboration Members**

Craig Aalseth, Pacific Northwest National Laboratory
Daniel S. Akerib, Case Western Reserve University
Steven Anderson, Black Hills State University
Elena Aprile, Columbia University
Frank T. Avignone III, University of South Carolina
Tom Barket, South Dakota Science Teachers Association
Laura Baudis, Stanford University
John F. Beacom, Fermilab
Mark Boulay, Los Alamos National Laboratory
George Brimhall, UC Berkeley
Len Bugel, Fermilab
Thomas Campbell, SDSM&T
Art Champagne, University of North Carolina
Juan I. Collar, University of Chicago
F.S. Colwell, Idaho National Engineering and Environmental Laboratory
Peter Doe, University of Washington
Michael Dragowsky, Case Western Reserve University
Ed Duke, SDSM&T
Dan Durben, Black Hills State University
Tom Durkin, SDSM&T
Hiro Ejiri, RCNP, Osaka University
Steve Elliott, University of Washington
Royce Engstrom, University of South Dakota
Joseph Formaggio, University of Washington

Jim Fredrickson, Pacific Northwest National Laboratory
George M. Fuller, University of California, San Diego
Richard Gaitskell, Brown University
Maury Goodman, Argonne National Laboratory
Uwe Greife, Colorado School of Mines
Alec Habig, University of Minnesota Duluth
Andre Hamer (deceased), Los Alamos National Laboratory
Frank Hartmann, Max Planck Institute, Heidelberg
Karsten M. Heeger, Lawrence Berkeley National Laboratory
Andrew Hime, Los Alamos National Laboratory
Zbignew (Ziggy) J. Hladysz, Mining Engineering Program, SDSM&T
Chang Kee Jung, The State University of New York at Stony Brook
Jon Kellar, SDSM&T
Thomas L. Kieft, New Mexico Institute of Mining and Technology
Sally Koutsoliotas, Bucknell University
Robert Lanou, Brown University
Barbara Sherwood Lollar, University of Toronto
Clark McGrew, SUNY at Stony Brook
Harry Miley, Pacific Northwest National Laboratory
Jeffrey S. Nico, National Institute of Standards and Technology
Bob Noiva, University of South Dakota
Peter Parker, Yale University
Tommy J. Phelps, Oak Ridge National Laboratory
Andreas Piepke, University of Alabama
Alan Poon, Lawrence Berkeley National Laboratory
Lisa M. Pratt, Indiana University
Jan Puszynski, SDSM&T
Bill Roggenthen, SDSM&T
Bernard Sadoulet, University of California, Berkeley
Ben Sayler, Black Hills State University
Richard Schnee, Case Western Reserve University
Kate Scholberg, MIT
Tom Shutt, Princeton University
Panagiotis Spentzouris, Fermilab
Robert Svoboda, LSU
Joseph S. Y. Wang, Lawrence Berkeley National Laboratory
Peter J. Wierenga, University of Arizona
Raymond Wildung, Pacific Northwest National Laboratory
Paul Wildenhain, University of Pennsylvania
Patrick R Zimmerman, Institute of Atmospheric Sciences, SDSM&T

**Collaboration Engineers**

Jerry Aberle, Lead
John Marks, Lead
Gary Kuhl, Skyline Engineering
Jamie Stampe, Skyline Engineering

# I. Science Book: Physics

**Neutrino Physics:** Several Nobel Prizes in the last 15 years (Lederman, Schwartz, and Steinberger, 1988; Reines, 1995; Davis and Koshiba, 2002) celebrate the fundamental contributions neutrino physics made in two areas: the development of the electroweak "standard model" of particle physics and the inauguration of a new field, neutrino astrophysics. What is more remarkable, however, is that during this period, a series of new discoveries have been made in neutrino physics and astrophysics – discoveries that may help resolve some of the deepest questions in physics. The demonstration that neutrinos are massive provided the first proof that the standard model is incomplete. In fact, most theorists believe that the pattern of neutrino masses is providing our first window on new physics residing at the Grand Unified Scale, an energy 100 billion times that of our most power accelerators. We now know that neutrinos are a component of particle dark matter, at least as important as the visible stars in their contribution to the universe's mass-energy budget, but not the dominant component.

But like a detective who has found valuable clues but cannot yet resolve the mystery, current discoveries have posed important new questions. We do not yet know the absolute scale of neutrino mass, as the solar and atmospheric results probe $m^2$ differences, not absolute masses. The absolute mass scale is central to understanding what neutrinos imply about beyond-the-standard-model physics, and to the role of neutrinos in cosmology. Crucial in understanding the mechanism responsible for neutrino mass is the nature of this particle under particle-antiparticle conjugation: we do not yet know whether the antineutrino is identical to (Majorana) or distinct from (Dirac) the neutrino. While we know the first and second neutrino generations mix, as do the second and third, we have not yet measured the mixing of the first and third generations. This mixing is important in the astrophysics of supernova explosions and will determine whether future long-baseline oscillation experiments will succeed in measuring neutrino properties important to the very early universe. The long-standing problem of why we are here – why our primordial universe was not symmetric in its matter and antimatter composition, with these two components later annihilating – may very likely have to do with CP violation among neutrinos. The "seesaw" mass pattern suggested by experiments fits nicely into models where the baryon number asymmetry was created through the process of leptogenesis.

The current status is often summarized in this way. We know, within experimental accuracies that must be greatly improved, $m_2^2 - m_1^2$ and the magnitude of $m_3^2 - m_2^2$, but not the overall scale $(m_1 + m_2 + m_3)/3$. We do not know if these three masses are (dominantly) Majorana. The mixing matrix describing the relationship between flavor and mass eigenstates is

$$\begin{pmatrix} \nu_e \\ \nu_\mu \\ \nu_\tau \end{pmatrix} = \begin{pmatrix} c_{12}c_{13} & s_{12}c_{13} & s_{13}e^{-i\delta} \\ -s_{12}c_{23} - c_{12}s_{23}s_{13}e^{i\delta} & c_{12}c_{23} - s_{12}s_{23}s_{13}e^{i\delta} & s_{23}c_{13} \\ s_{12}s_{23} - c_{12}c_{23}s_{13}e^{i\delta} & -c_{12}s_{23} - s_{12}c_{23}s_{13}e^{i\delta} & c_{23}c_{13} \end{pmatrix} \begin{pmatrix} \nu_1 \\ e^{i\phi_2}\nu_2 \\ e^{i\phi_3}\nu_3 \end{pmatrix}$$

$$= \begin{pmatrix} 1 & & \\ & c_{23} & s_{23} \\ & -s_{23} & c_{23} \end{pmatrix} \begin{pmatrix} c_{13} & & s_{13}e^{-i\delta} \\ & 1 & \\ -s_{13}e^{i\delta} & & c_{13} \end{pmatrix} \begin{pmatrix} c_{12} & s_{12} & \\ -s_{12} & c_{12} & \\ & & 1 \end{pmatrix} \begin{pmatrix} \nu_1 \\ e^{i\phi_2}\nu_2 \\ e^{i\phi_3}\nu_3 \end{pmatrix}$$

where $c_{12} = \cos\theta_{12}$, etc., and where the second line represents the successive solar, 13, and atmospheric mixings. We know $\theta_{12} \sim 30°$ and $\theta_{23} \sim 45°$, but have only the limit $\theta_{13} < 10°$. Furthermore, we have no information on the CP violating Dirac phase $\delta$ and the Majorana phases

ϕ*₂* and ϕ*₃*.

Most significant to neutrino sleuths is that nature has been kind: the wonderful discoveries that have come from neutrino astrophysics imply that new experiments can be done on earth, with terrestrial neutrino beams, as well as with astrophysical neutrino sources. This means that two communities, underground scientists who build ultraclean detectors and accelerator physicists who create new neutrino beams, can work together to resolve many of the open neutrino questions. The required experiments are difficult, requiring a new generation of cleaner, deeper, and larger underground detectors for neutrino physics, as well as unprecedented neutrino superbeams. But with the right combination of experiments, the questions that need to be answered can be answered: most of the unknown parameters can be determined.

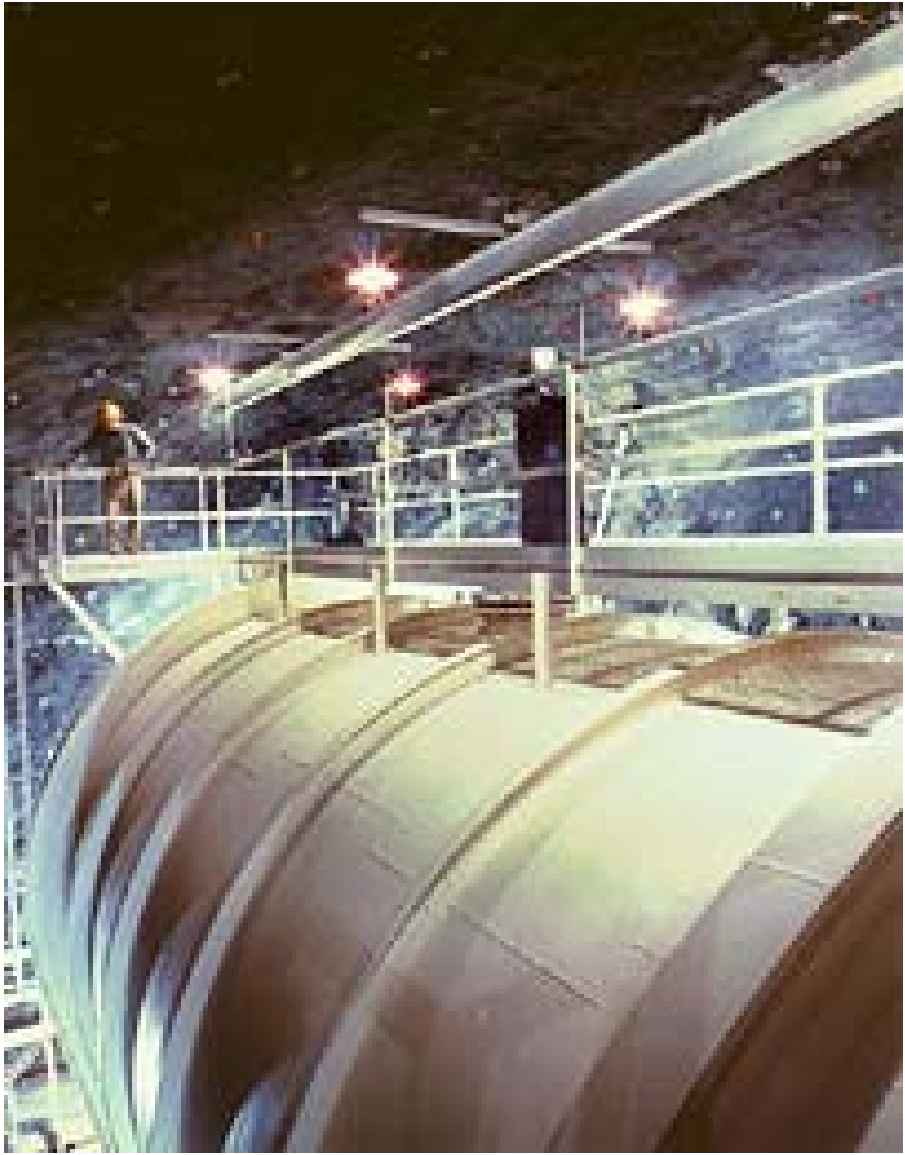

Figure C.1: Ray Davis's Homestake chlorine experiment.

***A.1 Double Beta Decay: The Importance of the Science:*** Double beta decay arises from a special feature of the nuclear force: nuclei with even numbers of protons and neutrons are more tightly bound than other nuclei. This frequently produces a mass pattern where an even-even nucleus with N protons and Z neutrons, N+Z=A, denoted (A,Z), as well as the nucleus (A,Z+2) are well bound, but the intermediate odd-odd nucleus (A,Z+1) is less so, and thus has a larger mass. Consequently, the ordinary first-order β decay process (A,Z) → (A,Z+1) + $\nu_e$ + $e^-$ is energetically forbidden, while a second-order-weak process in which (A,Z) decays spontaneously to (A,Z+2), with the emission of two electrons, is allowed. Thus nature isolates a very rare phenomenon – there is only one other example of a measurable sensitive to second-order weak effects.

The rates for such "double beta decay" are exceedingly tiny, corresponding to nuclear life times many orders of magnitude greater than the age of our universe. Interest in this process comes from its connections to neutrino physics. Nature appears to obey a symmetry, CPT, than assigns to each particle an antiparticle, a particle with the same mass and spin, but with opposite charge (or more correctly, with all of the particle's additive quantum numbers reversed). If we consider standard-model fermions – quarks, charged leptons, neutrinos – it is clear that the quarks and charged leptons have distinct antiparticles: under particle-antiparticle conjugation, the electron becomes a positron, clearly distinct because its charge is positive, opposite that of the electron. However, this question is more difficult for the neutrino.

The neutrino carries no charge (and has no other obvious additive quantum numbers). Thus its behavior under particle-antiparticle conjugation is unclear – one could imagine the neutrino having a distinct antiparticle, one that behaves in reactions in a way that distinguishes the antineutrino from the neutrino. (In this case one would invent an additively conserved quantum number, called lepton number, to distinguish the two particles, perhaps with the neutrino and electron assigned $l = +1$ and the antineutrino and positron $l = -1$.) Or the neutrino could be identical to the antineutrino. The first case corresponds to a Dirac neutrino, the second to a Majorana neutrino.

If neutrinos are Dirac particles, similar to other standard model fermions, then one would expect these particles would share a common mechanism for generating masses. Efforts to generalize the standard model then lead to a puzzle. For example, many grand unified theories (GUTs) enlarge the first-generation doublets of the standard model, (e,$\nu_e$), to include the other first-generation particles. Then one would expect that these first-generation Dirac particles would couple to the mass-generating fields similarly (up to group theory factors), and thus have comparable masses. However, while the u and d quarks and electron all have masses ~ 1 MeV, tritium beta decay studies tell us that the electron neutrino mass is less than 2.2 eV. An attractive resolution to this puzzled was offered by Gell-Mann, Ramond, and Slansky and by Yanagida some years ago. If neutrinos also have Majorana mass terms, then it becomes natural to generate a light neutrino mass of size

$$m_\nu = m_D \left( \frac{m_D}{m_R} \right)$$

where $m_D$ is the Dirac mass and $m_R$ the right-handed Majorana mass. Thus, if the right-handed Majorana mass scale is large, perhaps at the GUT scale, then *($m_D/m_R$)* is the needed small parameter explaining why the neutrino is so light. This seesaw mechanism – the higher the scale $m_R$, the smaller the neutrino mass $m_\nu$ -- then relates small neutrino masses to new physics residing beyond the standard model. The Majorana mass terms explicitly violate lepton number conservation.

Double beta decay probes this physics. If neutrinos are Dirac particles, then only the standard-model two-neutrino process contributes to double beta decay: two correlated single beta decays occur, with virtual excitation of the intermediate nuclear state, producing two electrons and two electron antineutrinos in the final state. Thus the final-state total lepton number $l = 0$ and lepton number is conserved. If lepton number is violated – as it would if Majorana mass terms are included – then neutrinoless ββ decay can also occur, producing a final state with $l = 2$. That is, the antineutrino produced in the first β decay is (at least partially) identical to the neutrino, and thus can be reabsorbed on a second nucleon, with the emission of a second electron. The final state contains just two electrons, which carry off the total nuclear decay energy. This is a very distinctive experimental signal: plotting the summed energy spectrum, neutrinoless double beta decay produces a line at the endpoint energy, while two-neutrino ββ decay produces a continuous spectrum peaked at an intermediate energy, with the phase space going to zero at the endpoint.

The neutrinoless process measures the Majorana neutrino mass. If parity is violated maximally, as in the standard model, the antineutrino produced in β decay is righthanded, while the neutrino that induces the ($v_e$,$e^-$) reaction on the second nucleon must be lefthanded. Thus the exact V-A character of the weak interaction would, for massless neutrinos, forbid neutrinoless ββ decay, regardless of lepton number conservation. But if the neutrino is massive, the neutrinoless process is not forbidden, but merely suppressed by the factor $m_v/E_v$, where $E_v$ is the typical energy of the exchanged neutrino. Because phase space favors the neutrinoless process, searches for neutrinoless ββ decay are very sensitive probes of small Majorana masses, despite this suppression.

Experimental progress in this field has been rapid. Thirty years of effort were required before the standard model two-neutrino process was finally observed in 1987. Now accurate lifetimes and decay spectra are known for about a dozen nuclei. The standard-model process is crucial to theory, providing important benchmarks for the nuclear physics matrix element calculations that are done to relate neutrinoless double beta decay limits to the underlying neutrino mass.

Progress in neutrinoless ββ decay has been even more impressive. Extraordinary efforts to reduce backgrounds through ultrapure isotopically enriched materials, improved energy resolution (important in distinguishing 0ν from 2ν decay), and siting experiments in deep underground locations (to avoid cosmogenic backgrounds) has produced a "Moore's Law" for double beta decay, a factor-of-two improvement in lifetime limits every two years. The bound from the nucleus $^{76}$Ge, $\sim 2 \times 10^{25}$ years, corresponds to a Majorana mass limit of (300-1300) meV (milli-eV), depending on the nuclear matrix elements employed. The goal of the next generation of experiments is very ambitious, pushing lifetimes by $\sim 10^3$ and probing masses as low as 10-30 meV.

These ambitious new efforts are driven by the recent discoveries of Super-Kamiokande, SNO, KamLAND, and K2K that establish neutrinos are massive. The questions remaining:

- **Are light neutrinos Majorana particles?** There are strong theoretical prejudices that this is the case, but the observation of neutrinoless double beta decay is the proof of Majorana masses and lepton number violation.
- **What is the absolute scale of neutrino mass?** The measured oscillations determine mass differences, but not the overall scale of neutrino mass. Neutrinoless double beta decay is our most promising tool for determining that scale. As a virtual process it probes all neutrinos that couple to the electron. The neutrino mass scale is a crucial parameter in cosmology and astrophysics: without this parameter, the interpretation of current (e.g., WMAP) and future cosmic microwave background and large-scale structure surveys will be less certain. (The initial WMAP analysis quoted a mass sensitivity of 0.7 eV, though later analyses have claimed significantly less sensitivity.) The Majorana mass tested in ββ decay is the sum over $2n$ mass eigenstates, weighted by the coupling probability to the electron and by the relative

CP eigenvalue λ (assuming CP conservation), where *n* is the number of generations:

$$<m_\nu^{Maj}> = \sum_{i=1}^{2n} \lambda_i U_{ei}^2 m_i$$

- **What is the neutrino mass hierarchy?** Assuming three light neutrinos, the oscillation results allow three possible mass patterns: a standard seesaw hierarchy; a degenerate scheme where the various mass splittings are small compared to the overall scale; and an inverted hierarchy, where the (nearly degenerate) neutrino pair responsible for solar neutrino oscillations is heavier than the third neutrino. Assuming Majorana neutrinos, both of the later would produce neutrinoless ββ decay within the reach of next generation experiments. In fact, the next experiment could settle this issue, with a large Majorana mass indicating a degenerate scheme, and a null result indicating a standard seesaw hierarchy (or a Dirac neutrino).
- **CP violation in ββ decay.** Many theorists believe that the lepton number violation probed in double beta decay, together with CP violation, are the key to understanding baryogenesis. The mass pattern emerging from recent oscillation experiments, which suggests a seesaw Majorana mass scale ~ $0.3 \times 10^{15}$ GeV ~ $M_{GUT}$, is quite consistent with leptogenesis models. While leptogenesis occurred at very high energies early in the big bang, the low-energy neutrino phases can be related to leptogenesis through models. The three-neutrino mass matrix contains three phases, one of which can be probed in long-baseline experiments, and two of which can only be probed in double beta decay. Extraction of these phases is a great challenge, made difficult by the nuclear structure uncertainties.
- **Nuclear physics.** Aside from its significance in neutrino physics, double beta decay is a fundamental nuclear decay mode for about 50 otherwise stable nuclei. The nuclear decay amplitude, a two-nucleon matrix element involving a sum over virtual intermediate nuclear states, is a challenging many-body problem. The quality of neutrino mass limits, or neutrino mass values, depends on the accuracy with which matrix element can be calculated. Double beta decay has stimulated important advances in nuclear structure theory, including Monte Carlo and Lanczos shell model techniques and quasiparticle RPA.

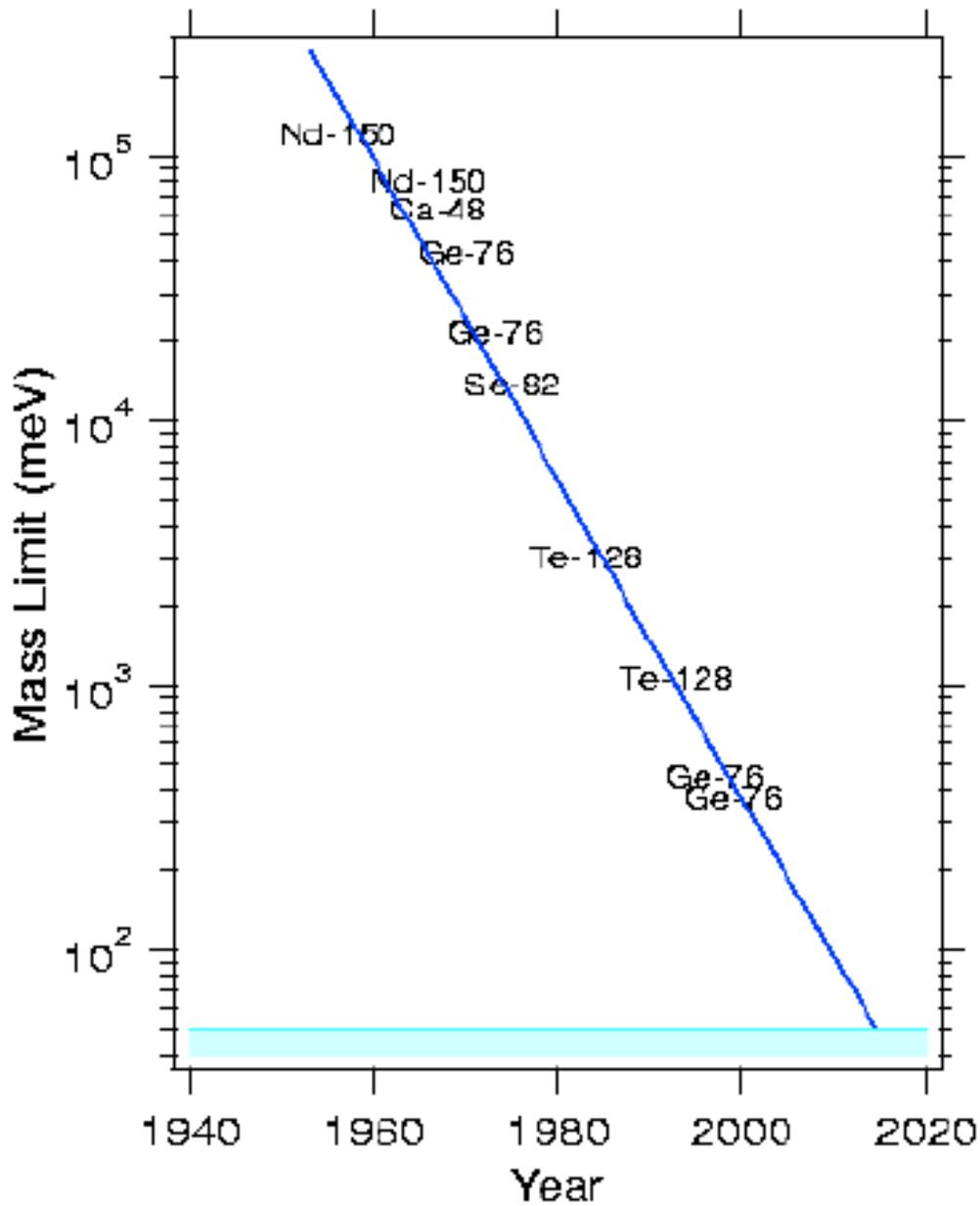

Figure C.2: Improvements in the Majorana mass limits from neutrinoless double beta decay experiments. As the half life is proportional to the square of this quantity, the trend represents a improvement in experimental sensitivity of five orders of magnitude over four decades. Next-generation experiments employing ton quantities of enriched isotopes have the potential to push sensitivities two or three orders of magnitude further.

| Double Beta Decay Experiments |||||||||
|---|---|---|---|---|---|---|---|---|
| Experiment | Isotope | Technique | Isotope Mass (kg) | Enriched | $Q_{\beta\beta}$ (MeV) | $<m_{ee}>$ (eV) 90%CL | Overhead (mwe) | Location |
| Heidelberg-Moscow | $^{76}$Ge | 5 Ge crystals | 9.9 | 86% | 2.04 | < 0.40 | 2700 | Gran Sasso, Italy |
| IGEX | $^{76}$Ge | 6 Ge crystals | ~9 | 86% | 2.04 | < 0.44 | 2450 | Canfranc, Spain |
| UCI | $^{82}$Se | TPC with foils | 0.014 | 97% | 2.99 | < 7.7 | 290 | Hoover Dam, US |
| ELEGANT | $^{100}$Mo | drift chamber-scintillators | 0.20 | 94.5% | 3.03 | < 2.7 | 1800 | Oto, Japan |
| Kiev | $^{116}$Cd | CdWO$_4$ crystals | 0.09 | 83% | 2.8 | < 3.3 | 1000 | Slotvinia, Ukraine |
| Missouri | $^{128}$Te | Geochemical | Te Ore | No | .87 | < 1.5 | N/A | N/A |
| Milano | $^{130}$Te | Cryogenic 20 TeO$_2$ crystals | 2.3 | No | 2.53 | < 2.6 | 2700 | Gran Sasso, Italy |
| Cal-UN-PSI | $^{136}$Xe | High Pres. TPC | 2.1 | 62.5% | 2.47 | < 3.5 | 3000 | Switzerland |
| UCI | $^{150}$Nd | TPC foils | 0.015 | 91% | 3.37 | < 7.1 | 290 | Hoover Dam, US |
| NEMO3 | $^{82}$Se, $^{100}$Mo, $^{116}$Cd, $^{150}$Nd | drift chamber-scintillator | 1, 10, 1, 1 | Yes | 3.0, 3.0, 2.8, 3.4 | ~0.1 | 4800 | Frejus France |
| CUORICINO | $^{130}$Te | Cryogenic 56 TeO$_2$ crystals | 11.5 | No | 2.6 | ~0.1 | 2700 | Gran Sasso, Italy |
| GENIUS | $^{76}$Ge | 400 Ge crystals | 1000 | Yes | 2.04 | 0.01 | | Gran Sasso, Italy |
| MAJORANA | $^{76}$Ge | 210 Ge crystals | 500 | Yes | 2.04 | 0.02 | ≥ 4000 | |
| CAMEO | $^{82}$Se, $^{100}$Mo, $^{116}$Cd | Borexino CTF | ~1, 1, 1 | Yes | 2.99, 3.0, 2.8 | ~1 | | Gran Sasso, Italy |
| MOON | $^{100}$Mo | Scint+Foils | 3400 | No | 3.03 | 0.03 | ≥ 2500 | |
| CUORE | $^{130}$Te | Cryogenic 1020 TeO$_2$ crystals | 210 | No | 2.53 | 0.02 | | Gran Sasso, Italy |
| EXO | $^{136}$Xe | High Pres. TPC | 10000 | Yes | 2.47 | 0.01 | ≥ 2000 | |
| DBCA-II(2) | $^{150}$Nd | Drift chamber | 18 | Yes | 3.37 | ~0.05 | | Oto, Japan |

Figure C.3: A table of operating (blue), developing (red), and proposed (black) double beta decay experiments. The most noticeable change is the large masses proposed for future experiments, motivated by the mass scales deduced from atmospheric neutrino oscillation experiments.

***A.2 Double Beta Decay: The Readiness of Next-Generation Experiments.*** Over the past 30 years a variety of experimental techniques for double beta decay searches have been developed and applied to favorable cases. Early work relied heavily on geochemical measurements; there are a couple of cases amenable to radiochemical techniques, as well. A great deal of progress has been made in recent years in experiments where the ββ decay source is also the detector. In particular, the most stringent limits have come from Ge detectors enriched in the double beta decay isotope of interest, $^{76}$Ge (natural abundance 7.58%). The two most recent experiments, by the IGEX and Heidelberg-Moscow collaborations, have produced the most restrictive lifetime limits, and thus the best constraints on $<m_\nu^{Maj}>$, less than 300-1300 meV, with the range reflecting nuclear matrix element uncertainties.

The next-generation detectors now under consideration have as their goal mass sensitivities of 10-50 meV. All such experiments must have the following characteristics:
- Sufficient mass, ~ 1 ton, to acquire reasonable statistics.
- Good energy resolution, in order to separate the 0ν line (in the summed energy of the two electrons) from the tail of the 2ν ββ decay spectrum.
- Extremely low backgrounds: the detector must be located at sufficient depth to reduce cosmogenic backgrounds to an acceptable level and be made of sufficiently pure materials to render natural radioactivity backgrounds tolerable.

Several other considerations are important:
- The cost of next-generation experiments is such that the proposed technology must be thoroughly demonstrated before a full-scale experiment is undertaken.
- Although the use of an abundant natural isotope would be ideal, there are very few candidates. Experiments at the one-ton scale are practical with enriched isotopes, reducing the detector volume.
- A small detector volume generally minimizes internal backgrounds, which scale as the detector volume (provided the enrichment process does not concentrate some troublesome activity). It also minimizes external backgrounds by reducing the shielding volume necessary for the detector. By designing an apparatus where the detector is also the source, the detector size can be further reduced.
- A large Q value increases the phase space for ββ decay, and thus generally the rate. A large Q value also places the two-electron energy above many potential low-energy backgrounds.
- For a fixed 0ν ββ decay rate, nuclei with smaller 2ν ββ decay rates are favored, as this reduces possible confusion from 2ν events near the endpoint.
- Identifying the daughter nucleus in coincidence with the produced electrons is a very effective background suppression technique, eliminating virtually all backgrounds except 2ν ββ decay.
- Event reconstruction, providing kinematic data such as opening angles and the single-electron energy spectrum, can aid in the elimination of backgrounds.
- Good spatial resolution (detector granularity) and timing information are also effective in rejecting backgrounds.
- Certain nuclei – lighter isotopes or nuclei near closed shells – may be more tractable for nuclear structure studies. In such cases experimental limits (or an observation of double beta decay) can be more reliably translated into neutrino mass bounds (or results).

No experiment, past or proposed, has succeeded in optimizing all these characteristics simultaneously. Currently there are approximate 15 proposals for experiments hoping to achieve next-generation sensitivity. Of these proposals, there are four where the level of effort could, in the near term, result in a proven technique capable of reaching the 10-50 meV mass sensitivity goal. Each of these efforts has significant US involvement, and three of the four have indicated their desire for an underground location in the United States:

***Majorana ($^{76}$Ge):*** Majorana is a next-generation enriched $^{76}$Ge experiment that, in the view of many experts, is ***now ready for construction***. The experimenters propose to reach the 0.5-1.0 ton detector mass goal by staging, building additional modules as the enriched Ge becomes available. The readiness is based on the demonstrated success of very similar detectors at the 10 kg scale: the technical extrapolations required in scaling up to a multi-crystal array of the necessary mass are modest. The staging is important, allowing the experimenters to verify their techniques as detector mass is added. The results from the first modules, however, will quickly surpass current Moscow-Heidelberg and IGEX limits.

Worldwide there are two schemes being studied. The Majorana collaboration has proposed a 0.5-ton detector of segmented, 86% enriched $^{76}$Ge diodes in multi-crystal arrays. A group at the Max Planck Institute, Heidelberg, is investigating a scheme for suspending the Ge crystals in a liquid nitrogen shield (GENIUS). The Heidelberg group has a small test facility operating at Gran Sasso, while the Majorana group is testing a prototype enriched segmented detector and is constructing an 18-crystal array of natural Ge diodes.

Several hundred enriched $^{76}$Ge crystals, all segmented and instrumented for pulse-shape analysis, will reside in Majorana's modules. The modular design allows for easy access to the crystals while still minimizing the detector footprint. Although alternative methods are being explored, the low-risk baseline cooling method is conventional (unlike the GENIUS design). While the detector segmentation and improved pulse discrimination are advances over past designs, the vast bulk of the technology has been taken from earlier Ge experiments. The IGEX members of the Majorana collaboration have 25 years of experience in collaborating with industry on custom-designed Ge detectors and cryostats. The collaboration has previously fabricated electroformed copper parts for the cryostats and the shield that meet the activity limits for Majorana. The proposed Majorana electronics scheme has been used previously in Ge detectors, and the pulse-shape analysis algorithms have been tested in numerous detector installations. The ECP in Karsnoyarsk, which previously produced the enriched Ge used in IGEX, is the isotope supplier for Majorana. Thus the bulk of the technology necessary for Majorana has been tested and, in most cases, used in previous ββ decay experiments.

The Majorana collaboration – approximately 50 scientists from 12 institutions – intends to submit its proposal in spring, 2003. Within 1.5 years of initial funding the group will have at least one 50 kg module commissioned underground and operating. One year of data will produce a half-life limit one order of magnitude beyond current experiments. The cost to complete the full array is estimated to be $75-100M. The experiment requires a deep underground site, with an overburden of at least 4500 mwe, as discussed below.

***Cuore ($^{130}$Te):*** The success of the MIBETA experiment, an array of TeO$_2$ cryogenic detectors with a total mass of 6.8 kg, led to the CUORE proposal (Cryogenic Underground Observatory for Rare Events). In CUORE 1000 750-g TeO$_2$ crystals will be operated as a collection of bolometers. The detectors will be grouped into 25 separate towers of 40 crystals, with each tower arranged in 10 planes of four crystals each. One such plane with a mass of 40 kg has now been successfully tested and a single tower prototype, called CUORICINO, began operations at the Gran Sasso laboratory in January 2003. CUORICINO will provide an important test of backgrounds and other experimental parameters.

The energy resolution at the 0ν ββ decay peak (2.529 MeV) is expected to be about 5 keV FWHM (~ 0.2%). The background measured in the first plane is ~ 0.5 counts/(keV kg y). However a major contributor to this background was surface contamination arising from the use of a cerium oxide polishing compound high in thorium. With this problem eliminated, the experimenters anticipate that backgrounds will be reduced below ~ 0.1 counts/(keV kg y).

A major advantage of CUORE is that the isotope of interest, $^{130}$Te, has a natural abundance of 34%. Thus no enrichment is needed, resulting in lower costs. The MIBETA results indicate that cosmogenic activities within the $TeO_2$ crystals are not a serious concern. On the other hand, the crystal mounts and cryostat form a significant amount of material close to the bolometers. Much of the cryostat is shielded with low-activity lead from Roman times, but quantities of copper and Teflon remain close to the crystals.

CUORE is likely **close to being ready for construction**. If CUORICINO is not background limited and reaches it design sensitivity goals, the group will then be ready to proceed with CUORE construction. This evaluation should be possible after about a year of CUORICINO operations, which places the CUORE decision point early in 2004. The CUORE collaboration is led by Italian groups, but there are members from the US (LBNL and Univ. of South Carolina), Spain, and the Netherlands. The collaboration estimates construction costs of $8-10M. The proposed site for CUORE is Gran Sasso.

*EXO ($^{136}$Xe):* The Enriched Xenon Observatory (EXO) is a proposal to use up to 10 tons of 60-80% enriched $^{136}$Xe in a scintillation detector. The unique aspect of the experiment is the proposed detection of the $^{136}$Ba daughter ion in coincidence with the decay event. If this technique were perfected, it would eliminate virtually all background except that associated with 2ν ββ decay. The real-time optical detection of the daughter Ba ion might be possible if the ion can be localized and probed with lasers. This spectroscopy has been used for $Ba^+$ ions in atom traps. However the additional technology to detect single Ba ions in a condensed medium or to extract single Ba ions from a condensed medium and trap them has yet to be demonstrated. The optical detection of the $Ba^+$ ion is accomplished by pumping the ion from the $6^2S_{1/2}$ ground state to the $6^2P_{1/2}$ excited state via a 493-nm laser. As the excited state has a 30% branching ratio to the $5^4D_{3/2}$ metastable state, the ion can be detected by re-exciting this metastable state to the 6P state via a 650-nm laser, then observing the decay back to the ground state. The procedure can be repeated millions of times per second on a single ion to produce a significant signal.

EXO will use liquid Xe scintillator, as a gaseous TPC would require a larger detector to accommodate the same Xe mass. However, at liquid densities the laser light scattering is too great to permit optical detection of the Ba ions *in situ*. Thus the proposed technique, once the candidate ββ decay event is localized via its scintillation and ionization, is extraction of the ion via a cold-finger electrode coated in frozen Xe. The ion is electrostatically attracted to the cold finger, which later can be heated to evaporate the Xe, releasing the Ba ion into a radiofrequency quadrupole trap. At this point the $Ba^{++}$ ion is neutalized to $Ba^+$, laser cooled, and optically detected. The focus of current research is to demonstrate each of the steps in this procedure.

Achieving sufficient energy resolution to avoid interference from 2ν ββ decay is important to EXO. The collaboration recently demonstrated, by measuring scintillation light and ionization simultaneously, that the necessary resolution could be achieved. The EXO collaboration has received DOE HEP funding to proceed with a 200-kg enriched Xe detector without Ba tagging. This initial prototype, which will be sited in the Waste Isolation Pilot Plant (WIPP) in New Mexico, will help the collaboration evaluate backgrounds and other aspects of the experiment.

While the EXO experiment still has **technical issues remaining to be resolved**, the possibility of nearly background-free observation of ββ decay in a very large (10-ton) detector is exciting. As in the case of the experiment discussed below, this new technology may ultimately prove to have greater reach that any currently existing, and thus could provide the foundation for next-to-next-generation efforts, too.

*MOON ($^{100}$Mo):* The MOON (Mo Observatory of Neutrinos) collaboration will use $^{100}$Mo as a 0ν ββ decay source and as a target for solar neutrinos. These two science goals (as well as the sensitivity to low-energy supernova electron neutrinos) make the detector an enticing idea. $^{100}$Mo has a high Q-value (3.034 MeV), enhancing the phase space for ββ decay and placing the summed electron energy peak well above most radioactivity backgrounds. It appears to have favorable matrix elements for 0ν and 2ν ββ decay, which will enhance the neutrino mass sensitivity, provided sufficient resolution is achieved to reduce 2ν backgrounds. Energy and angular correlation measurements in the detector will be used to identify 0ν ββ decay events and reject backgrounds. The initial MOON concept was a supermodule of scintillator and Mo ensembles. One option is a module of plastic fiber scintillators with thin (0.03 g/cm$^2$) layers of cladded Mo, arranged to achieve position resolution comparable to the fiber diameter (2-3 mm). A total of 34 tons of natural Mo would provide ~ 3.3 tons of $^{100}$Mo (abundance 9.63%).

As a solar neutrino detector $^{100}$Mo has a low threshold (168 keV), a known and favorable cross section, and a charged-current event rate of ~ 160/($^{100}$Mo ton y) (without neutrino oscillations). The subsequent delayed decay (15.8 s) of the daughter nucleus $^{100}$Tc to $^{100}$Ru provides a coincidence. The primary background to this coincidence is thought to be 2ν ββ decay.

Radiopurity levels of better than 1 mBq/ton must be achieved in the Mo and scintillator. Plastics of this cleanliness can be produced, while the Mo requirements can be met with carbonyl chemistry. However, the total surface area of the Mo-scintillator modules, ~ 26000 m$^2$, poses difficulties. Dust, being electrostatically charged, tends to garner Rn daughters and become radioactive. Keeping these surfaces free of dust during production and assembly will be a challenge. Another challenge is the resolution. Simulations indicate that energy resolution for the 0ν ββ decay peak will be ~ 7%, which is at the upper end of the range of feasibility for sub-50-meV sensitivity to neutrino mass.

A more attractive option may be a bolometer, yielding improved energy resolution and a smaller surface area. This would likely require an insulating compound substituting for the pure Mo. MOON is in an R&D phase with **significant technical challenges not yet resolved**. The Japanese-US collaboration that is developing MOON has not yet developed a firm cost estimate, but the experiment will require funding in excess of $50M. It will require a very deep site, preferably in a US underground laboratory.

*A.3 Double Beta Decay: Facility Requirements.* This section summarizes the facility requirements for double beta decay: what must an underground laboratory provide to optimize the prospects for next-generation experiments?

*Cosmogenic backgrounds and depth requirements:* Current double beta decay experiments generally employ active shields to reduce cosmic ray backgrounds. MOON and Majorana background estimates include the use of such shields in combination with rather stringent data cuts. This will reduce backgrounds to an acceptable but not negligible level: Majorana anticipates seven background events in the region of the double beta decay peak, assuming a 10-year exposure. For this reason there is concern about backgrounds not yet identified and/or that evade the veto. Of particular worry are cosmic-ray-associated high-energy neutrons, which can induce spallation, capture γs, recoils, and inelastic reactions.

An important benchmark for Majorana comes from the current IGEX $^{76}$Ge experiment, which is located in the Canfranc tunnel at a depth of 2450 mwe. Approximately one-third of the total count rate in the endpoint region of interest (2038 keV) is correlated with the cosmic ray muon veto. From this rate the Majorana collaboration has estimated that similar events occurring near (but outside) the veto shield could, via neutron secondaries, double the anticipated Majorana background rate. Increasing the depth is significantly easier than other possible mitigating steps, such as direct detection of high-energy neutrons. A site at 4000 mwe will reduce cosmic-ray muon levels to $2\times10^3$/m$^2$/y, about a factor of 12 below Canfranc levels. Today this would require siting Majorana outside the US.

Proposed double beta decay experiments involve a variety of heavy nuclei including Ca, Ge, Se,

Mo, Cd, Te, and Xe. A detailed catalog of potential radioisotope production from such targets is not presently available. This has placed an additional burden on experimentalists, who individually must estimate cross sections and identify isotopes of concern. Ultimately crucial reactions must be tested with accelerator experiments. The associated uncertainties make deep sites attractive.

Spallation reactions on carbon in the scintillator used in MOON may be the most serious depth-related background in this detector, affecting both double beta decay and pp neutrino detection. A worrisome activity is $^{10}$C produced by (n,3n). $^{10}$C produces a positron and a 718 keV γ after 0.71 ns. (Recall that the pp signal is a prompt low-energy electron followed by a β decay delayed by ~ 15 s.) The effectiveness of possible cuts will depend on the position resolution achieved in MOON; estimates of the depth required to reduce this background to a negligible level currently range from 5000-6000 mwe.

EXO may have less stringent depth requirements. The published (hep-ex/0002003) background study of EXO considered a penetrating muon flux of $0.01/m^2/s$, which corresponds to an overburden of about 2400 mwe. Based on the Gothard group's experience with TPC muon rejection, it is estimated that the laser system would be needed less than once per hour for muons entering the detector but not otherwise rejected. Muons interacting outside the detector can produce neutrons that would enter EXO undetected. The discussion in hep-ex/002003/ noted that many of the resulting spallation reactions would be rejected by the associated high-ionization short-track events. We are not aware of detailed discussions of (n,2n) and similar neutron reactions that could evade such detection.

The backgrounds that have received the most attention from the EXO collaboration are not depth related. These include γ rays associated with radon or other radioactive gases leaking into the detector volume and 2ν ββ decay. For an assumed energy resolution of σ ~ 44 keV, considered to be conservative, 23 2ν events would occur within 2σ of the endpoint per ton-year. The ultimate sensitivity of the EXO technology may depend on improvements in the resolution that come from the additional event kinematics cuts possible in a TPC.

In summary, Majorana background estimates suggest that 4500 mwe would be an adequate depth, though the collaboration favors a deeper site, if available, to provide some margin of safety. MOON depth requirements are still quite uncertain, with the possibility that very great overburdens of 6000 mwe will be necessary. Provided the laser tagging is perfected and the prototype experiment reveals no unanticipated cosmogenic backgrounds, EXO could function at depths as shallow as 2000 mwe.

*Double beta decay space requirements:* The space requirements for the experiments, given as length × width × height, are:
- Majorana: $5 \times 4 \times 3$ m$^3$ apparatus and $4 \times 4 \times 3$ m$^3$ control systems.
- EXO: $5 \times 5 \times 5$ m$^3$ apparatus, $5 \times 4 \times 3$ m$^3$ control systems, and $4 \times 4 \times 3$ m$^3$ cryogenic purification systems.
- MOON: $5 \times 8 \times 5$ m$^3$ apparatus and $8 \times 11 \times 6$ m$^3$ lab area and control systems.

*Basic facilities needs:* Most of the general site requirements of double beta decay experiments are modest. They include:
- Power requirements of 10-25 kW.
- Stable temperatures, usually 20°C or less, with air conditioning.
- Scrubbed air, with residual radon levels below 1Bq/m$^3$.
- A clean room for preparations, assembly and cleaning.
- DI water system.
- Cranes for assembly and manipulation of detector and shielding elements.
- Radon-free materials storage area.

- Machine shop, both general and for ultra-low background work.
- Low-level counting capabilities for materials screening.
- A source of radon-free liquid nitrogen or nitrogen gas. (Most experiments have a critical volume needing to be purged of radon. Pure nitrogen purge gas is commonly used, e.g., boil-off $N_2$ gas from liquid nitrogen.)

*Special facilities needs (experiment specific):* Majorana and EXO have identified special facilities needs and special safety concerns.
- An underground copper electroforming facility. The acids and plating baths used in the electroforming require special safety procedures. (Majorana)
- Possibly, underground Ge crystal growth and detector preparation facilities. (Majorana)
- Large-volume liquid Xe containment. Large quantities of liquid cryogens present an oxygen displacement hazard. (EXO)
- A system for continuous cryogenic purification. (EXO)

***A.4 Double Beta Decay: Summary.*** The importance of the science and the readiness of key next-generation experiments like Majorana argue that NUSEL must be able to host such experiments as early as possible. In addition there are exciting new technologies under development that could exceed existing techniques in reach. Thus NUSEL should be prepared to host at least two such experiments. There are many shared resources that would benefit all 0ν ββ decay experiments and R&D efforts, as well as other activities within NUSEL. These should be part of basic NUSEL infrastructure. There are also some technical needs and safety concerns that are experiment specific.

***B.1 Solar Neutrinos: The Importance of the Science.*** The field of neutrino astrophysics began in 1965 with the efforts of Ray Davis Jr. and his colleagues to measure solar neutrinos, a byproduct of thermonuclear energy generation in the solar core. The subsequent history of this subject demonstrates the rapidly accelerating pace of technical innovation in underground science, the increasing breadth of the scientific issues, and the deepening connections with both conventional astrophysics and accelerator experiments.

Solar neutrinos offer unique opportunities for testing both electroweak physics and the nuclear reactions occurring in the interior of our best known star. The neutrino flux predictions come from the standard solar model (SSM), an application of the theory of main-sequence stellar evolution to our nearest star. The SSM traces the evolution of the sun over the past 4.6 billion years, thereby predicting the present-day temperature and composition profiles of the solar core that govern neutrino production. The SSM is based of four assumptions:

- The sun evolves in hydrostatic equilibrium, maintaining a local balance between the gravitational force and the pressure gradient. To implement this condition in a calculation, one must specify the equation of state as a function of temperature, density, and composition.
- Energy is transported by radiation and convection. While the solar envelope is convective, radiative transport dominates in the core region where thermonuclear reactions take place. The opacity depends sensitively on the solar composition, particularly the abundances of heavier elements.
- Thermonuclear reaction chains generate solar energy. The SSM predicts that the energy is produced from hydrogen burning, $4p \rightarrow {}^4He + 2e^+ + 2\nu_e$, through the pp chain (98% of the time) and CNO cycle. The core temperature, $\sim 1.5 \times 10^7$ K, results in typical center-of-mass energies for reacting particles of $\sim 10$ keV, much less than most of the Coulomb barriers inhibiting charged particle reactions. Thus cross sections are small, and generally must be estimated from laboratory data taken at higher energies.

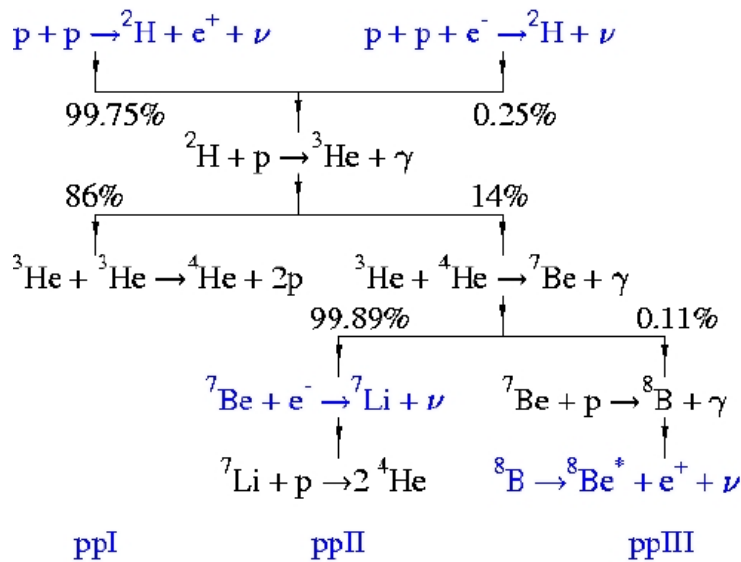

Figure C.4: The pp chain, showing the three cycles and the neutrino-producing reactions that probe the competition between the cycles.

- The SSM is constrained to produce today's solar radius, mass, and luminosity. The sun is believed to have been highly convective when it first entered the main sequence. Thus, with the assumption that the surface abundances of metals (A>5) were undisturbed by the subsequent evolution, the SSM equates the initial core metallicity to the observed surface abundances. The remaining parameter, the initial abundance ratio $^4$He/H, is adjusted until the model reproduces the present solar luminosity after 4.6 b.y. of evolution.

Experimental tests of the standard model, including measurements of the sound speed throughout most of the solar interior, are in good agreement with the SSM. Thus we can use this model with some confidence in predicting the flux of $\nu_e$s from the sun. The principal neutrino sources come from p+p β decay (maximum neutrino energy of 420 keV), from p+p+e$^-$ (producing a line source of neutrinos at 1.44 MeV), from $^8$B β decay (maximum energy ~ 15 MeV), and from electron capture on $^7$Be (producing line sources at 860 and 380 keV). There are also significant fluxes from CNO cycle reactions.

Davis used 610 tons of perchloroethylene ($C_2Cl_4$) in a radiochemical experiment to look for the solar-neutrino- induced reaction $^{37}$Cl($\nu_e$,e$^-$)$^{37}$Ar. The experiment, which was sensitive primarily to the high-energy $^8$B and $^7$Be neutrino fluxes, was mounted at the 4850 ft level of Homestake. The results led to the recognition of the "solar neutrino problem" – the observation of an electron neutrino flux substantially (~ factor of three) below the predictions of the SSM. The Davis experiment was followed, in the 1980s, by the Kamiokande experiment in Japan, where solar neutrinos were observed in a 4.5-kiloton water Cerenkov detector. Solar neutrinos elastically scatter (ES) off of electrons in the water, with the recoiling electron then detected because of the Cerenkov radiation it produces. The direction of the recoil electron – away from the sun – is an important aid in separating events from background. This event-by-event direct detection experiment was sensitive to the high-energy portion of the $^8$B solar neutrino flux. In the 1990s two new radiochemical experiments, SAGE and GALLEX, used $^{71}$Ga as the target to make integral measurements primarily sensitive to the low energy pp and $^7$Be neutrino fluxes. SAGE was mounted in the Baksan Laboratory in Russia, and used gallium metal, which is a liquid slightly above room temperature. GALLEX was performed in Gran Sasso and used a $GaCl_3$ solution. In each of these experiments substantially lower neutrino fluxes were found than predicted by the SSM. The pattern of discrepancies was difficult to explain by making plausible adjustments to the SSM, such as changes that might lower the core temperature.

An elegant particle-physics solution to these discrepancies is provided by neutrino oscillations – a phenomenon requiring massive neutrinos and mixing (that is, mass eigenstates not coincident with the flavor eigenstates). The oscillation of solar electron neutrinos into ones of another flavor would result in reduced detection rates in all of the detectors discussed above. As massive neutrinos are not allowed in the standard electroweak model – a consequence of the lack of a right-handed neutrino field and the requirement of renormalizability – neutrino oscillations would imply new physics.

The sun is an extraordinary neutrino source for investigating neutrino oscillations. The long baseline (~ $10^8$ km) between the source (the solar core) and the earth, combined with the low energies of solar neutrinos, implies sensitivity to the differences in the squares of neutrino masses of $\delta m^2$ ~ $10^{-12}$ eV$^2$. Such a value is many orders of magnitude beyond the reach of terrestrial experiments. Moreover, as the neutrinos pass through ~ $10^{10}$ g/cm$^2$ of matter before exiting the sun, their interactions can lead to great enhancements in oscillation probabilities. This phenomenon, the Mikheyev-Smirnov-Wolfenstein (MSW) mechanism, arises because the effective masses of neutrinos are altered by the presence of matter. In particular, the electron neutrino mass increases relative to those of the other neutrino flavors, producing a level crossing (nearly degenerate neutrino masses) at some critical solar density.

The most decisive solar neutrino experiments were mounted recently. Super-Kamiokande is the successor to the Kamioka experiment, with a mass of 50 kilotons of ultrapure water and a solar neutrino ES event rate of many thousands per year. This detector is primarily sensitive to electron neutrinos as the heavy-flavor cross section is only 0.15 that for electron neutrinos. This experiment provided a very precise constraint on the flux, on the spectrum of recoil electrons, and on day-night differences arising from MSW effects in the earth. The Sudbury Neutrino Observatory is using 1 kiloton of heavy water to distinguish electron and heavy-flavor solar neutrino reactions. The heavy water is contained within a central acrylic vessel, which is surrounded by seven kilotons of ordinary water. SNO detects $\nu_e$s via the charged-current (CC) reaction $\nu_e+d \rightarrow p+p+e^-$ and neutrinos of any flavor by the neutral-current (NC) breakup reaction $\nu_x+d \rightarrow p+n+\nu_x$. The first reaction provides excellent spectral information, as the electron carries off most of the energy of the incident neutrino, while the NC measurement is an integral one (only the neutron is measured). By comparing the CC and NC rates, one can separately determine electron and muon/tauon neutrino components of the solar neutrino flux (measured at Earth). The CC results show a reduced flux and a spectral shape consistent with matter-enhanced neutrino oscillations. SNO's NC measurements yielded a total $^8$B neutrino flux (summed over flavors) in good agreement with the predictions of the SSM. The comparison of the CC and NC results shows that approximately 2/3rds of the solar neutrinos arrive on earth as heavy-flavor neutrinos.

These new results are quite spectacular. With very high confidence the solar neutrino problem is due to flavor oscillations with parameters in the so-called large-mixing-angle (LMA) region of the $\delta m^2$-$\sin^2 2\theta$ plane. The SNO NC agreement with the SSM is gratifying because this flux depends critically on the solar core temperature, varying approximately as $T_c^{22}$, where $T_c$ is the solar core temperature. The LMA solution has recently been verified by a terrestrial experiment, KamLAND, which measures the electron antineutrino flux produced by Japanese power reactors at various distances from the detector. The KamLAND results further restrict the values of $\delta m^2$ allowed within the LMA region.

Because of the importance of the solution uncovered – massive neutrinos and neutrino oscillations are the first phenomena we have found that require physics beyond the standard electroweak model – there is great interest in extending such measurements to lower solar neutrino energies. The primary goal of the field is to make precise measurements of the fluxes and flavors of the lower energy solar neutrinos, particularly the pp neutrinos, the dominant solar neutrino flux (90% of the total) produced in the initial step of hydrogen burning. Arguably the pp neutrino flux, at its solar source, is known more precisely than any other, terrestrial or astrophysical. This opens possibilities for probing fundamental questions in both astrophysics and physics:

*Physics*
*For the foreseeable future, solar neutrinos will provide the only intense source of electron neutrinos.* These neutrinos are essential in efforts to understand the properties of neutrinos and in probing new phenomena like CPT violation.

*A low-energy solar neutrino measurement provides an important test of the LMA solution.* One feature of this solution is that the survival probability for low energy neutrinos is substantially higher than that for the $^8$B neutrinos measured in SNO. Verifying this prediction is important. A low-energy measurement, when combined with SNO data, will also help to narrow the uncertainty on $\delta m_{12}^2$.

*Future high-statistics pp solar neutrino experiments provide the best opportunity for improving our knowledge of $\theta_{12}$.* Assuming that there are only three active neutrino flavors, solar neutrino oscillations are dominated by the mixing of two mass eigenstates, $\nu_1$ and $\nu_2$. Solar neutrino experiments are the primary source of information on $\delta m_{12}^2$ and $\theta_{12}$. While KamLAND has

helped to narrow the LMA solution region in $\delta m_{12}^2$, neither it nor Borexino will appreciably improve our knowledge of the mixing angle. A pp solar neutrino measurement with an uncertainty of 3% would constrain $\theta_{12}$ with an accuracy comparable to that possible with the entire existing set of solar neutrino data. Thus a goal of next-generation pp solar neutrino experiments is to achieve 1% uncertainty, thus substantially tightening the constraints on $\theta_{12}$.

*Precision flux measurements of the low-energy (pp and $^7$Be) solar neutrinos will substantial improve limits on sterile neutrinos coupling to the electron neutrino.* Sterile neutrinos with even small couplings to active species can have profound cosmological effects. The coupling of $\nu_e$s to sterile states can be limited by CC and NC measurements with an accurately known neutrino source. KamLAND (together with the SNO CC and NC data) should ultimately limit the sterile component of $^8$B neutrinos to ~ 13%. This bound can be improved by measuring the NC and CC interactions of pp and $^7$Be neutrinos to accuracies of a few percent. In general one expects the sterile component of solar neutrino fluxes to be energy dependent. Thus low-energy solar neutrino experiments are an important part of such searches.

*Future pp solar neutrino experiments will be able to probe CPT violation with an order of magnitude more sensitivity than has been achieved in the neutral kaon system.* As neutrinos are chargeless, they provide an important testing ground for CPT violation. If CPT is violated, the neutrino mass scale and mass splittings will differ between neutrinos and antineutrinos. A high precision measurement of the pp $\nu_e$ survival probability is an important component of CPT violation searches. In some models this will test CPT violation at a scale $< 10^{-20}$ GeV, which can be compared to the present bound from the neutral kaon system, $< 4.4 \times 10^{-19}$ GeV.

*Preliminary estimates indicate that limits on the neutrino magnetic moment could be improved by an order of magnitude in future pp neutrino ES experiments.* A neutrino magnetic moment will generate an electromagnetic contribution to neutrino ES, distorting the spectrum of scattered electrons. The effect, relative to the usual weak amplitude, is larger at lower energies, making a high-precision pp solar neutrino experiment an attractive testing ground for the magnetic moment. The strongest existing laboratory limit, $\mu(\nu_e) < 1.5 \times 10^{-10}$ Bohr magnetons, could be improved by an order of magnitude in a 1% experiment.

*Solar neutrino detectors are superb supernova detectors.* Many of the most interesting features in the supernova neutrino "light curve" are flavor specific and occur at late times, 10 or more seconds after core collapse. Because of their low thresholds, flavor specificity, large masses, and low backgrounds, solar neutrino detectors are ideal for following the neutrino emission out to late times. Solar neutrino detectors will likely be the only detectors capable of isolating the supernova $\nu_e$ flux during the next supernova. The 3 ms deleptonization burst, important in kinematic tests of neutrino mass, is mostly of this flavor. Electron-flavor neutrinos also control the isospin of the nucleon gas – the so called "hot bubble" – that expands off the neutron star. The hot bubble is the likely site of the r-process. Currently we lack a robust model of supernova explosions. The open questions include the nature of the explosion mechanism, the possibility of kaon or other phase transitions in the high density protoneutron star matter, the effects of mixed phases on neutrino opacities and cooling, possible signatures of such phenomena in the neutrino light curve, and signals for black hole formation. Supernovae are ideal laboratories for neutrino oscillation studies. One expects an MSW crossing governed by $\theta_{13}$, opening opportunities to probe this unknown mixing angle. The MSW potential of a supernova is different from any we have explored thus far because of neutrino-neutrino scattering contributions. Oscillation effects can be unraveled because the different neutrino flavors have somewhat different average temperatures. All of this makes studies of supernova neutrino arrival times, energy and time spectra, and flavor composition critically important. Finally, the detection of the neutrino burst from a galactic supernova will provide an early warning to optical astronomers: the shock wave takes from hours to a day to reach the star's surface.

*Some proposed solar neutrino detectors serve as double beta decay and particle dark matter detectors.* Examples are MOON (double beta decay) and CLEAN and Majorana (WIMP dark matter).

**Astrophysics**
*Solar neutrino experiments provide crucial tests of the SSM.* The delicate competition between the ppI, ppII, and ppIII cycles comprising the pp chain is sensitive to many details of solar physics, including the core temperature, the sun's radial temperature profile, the opacity, and the metalicity. This competition can be probed experimentally by measuring the $^7$Be flux (which tags the ppII cycle), the $^8$B flux (which tags the ppIII cycle), and the pp flux (which effectively tags the sum of ppI, ppII, and ppIII). The pp neutrino flux is a crucial test of the SSM due to the accuracy with which it is predicted (~ 1%). Present experiments have left substantial uncertainties in flux determinations. While the situation will improve with new data from SNO and Borexino, future high-precision pp flux measurements (~ 1%) are crucial, as the table indicates.

|  | pp | $^7$Be | CNO | $^8$B |
|---|---|---|---|---|
| Present experiments | 18% | 35% | 100% | 13% |
| Near-term experiment | 12% | 8% | 100% | 8% |
| Future experiments | 1-3% | 2-5% | 5-10% | 2-5% |

Table IB.1  Present and anticipated precision of solar neutrino experiments.

*The CNO neutrino flux is an important test of stellar evolution.* Approximately 1.5% of the sun's energy is produced through CNO-cycle hydrogen burning. The CNO cycle is important in the early evolution of the sun as out-of-equilibrium burning of C, N, and O powers an initial convective solar stage, thought to last about $10^8$ years. Furthermore, one of the key SSM assumptions equates the initial core metalicity to today's surface abundances. A measurement of CNO cycle neutrinos would quantitatively test this assumption.

***B.2 Solar Neutrinos: The Readiness of Next-Generation Experiments.*** The difficulty of proposed new experiments to characterize the low-energy solar neutrinos is high. The requirement is to develop new-generation detectors with high statistics, thresholds in the 100 keV range, and abilities to reject backgrounds that rise exponentially as thresholds are lowered. There are excellent R&D programs, some of which are quite mature, focused on meeting these requirements. But no project has resolved all of the questions of background suppression, systematics, and cost-effective detector design, and thus none is ready for construction.

The proposed detectors include both ES and CC schemes in which the end goal is sensitivity to low-energy neutrinos with 1-3% uncertainty. Both types of detectors are needed to determine the flavor content of the solar flux. Both types of experiments generally require specialized underground sites, with the depth requirements for ES detectors being particularly stringent (~ 6000 mwe, according to the recent NRC Neutrino Facilities report).

The ES pp neutrino experiments discussed below are HERON, TPC, and CLEAN, the leading US R&D efforts. There is also an important Japanese effort, XMASS. The XMASS detector will record scintillation light produced by ES in a few tons of isotopically enriched liquid Xe. XMASS will also be used in double beta decay and dark matter searches, though with different Xe isotopes as the source/target. An important advantage of ES experiments is the certainty with which the ES cross section is known. This makes it possible to contemplate a 1% measurement of the pp flux, which would substantially narrow the uncertainty on $\theta_{12}$, without calibration by an artificial neutrino source.

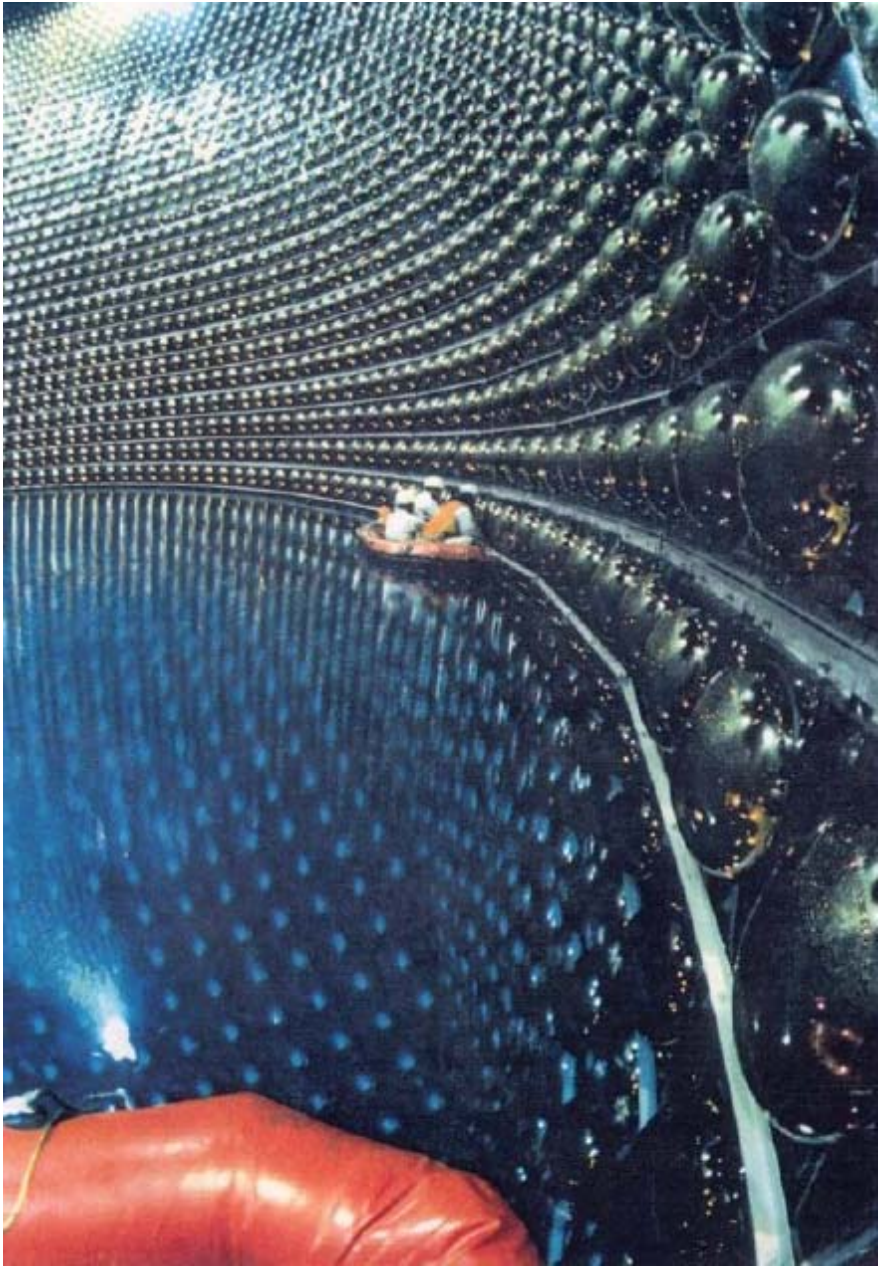

Figure C.5: The Super-Kamiokande detector, as it was being filled for the first time. This is an example of a large (50 kton) underground detector requiring a specialized excavation. Next-generation experiments now under consideration are an order of magnitude larger and must be sited at a depth of at least 4000 mwe, so that the cosmic ray muon background is an order of magnitude smaller than in Super-Kamiokande.

LENS and MOON are the two CC R&D efforts discussed below. These use inverse β decay in combination with a delayed decay of the daughter nucleus, with this coincidence helping to distinguish events from background. Both detectors employ highly segmented designs in order to reduce backgrounds from accidental coincidences.

***CLEAN (ES in Liquid Neon):*** The goal of CLEAN is a 1% measurement of the ES of pp and $^7$Be solar neutrinos in liquid neon. CLEAN will also search for particle dark matter by observing the recoil of neon nuclei after the scattering of weakly interacting massive particles (WIMPs). The recoil electrons produced by neutrino ES produce about 15000 extreme ultraviolet photons/MeV, split between two components: two-thirds are in the short (5 ns) singlet state and one-third is in the long (3.9 μs) triplet state. This yields an estimated 1.4 photoelectrons detected per keV, implying 17% energy resolution for a 100 keV β. The expected threshold for CLEAN is ~ 20 keV. The scintillation light is isotropic and thus provides no directional information. The position of an event can be reconstructed with a resolution of 20-30 cm, thus allowing one to use fiducial volume cuts effectively.

The proposed detector consists of a cylinder, 6m in diameter and 6m high, containing liquid neon at 27K. A spherical array of PMTs is suspended within the cylinder. The neon-filled cylinder is housed in a liquid nitrogen shield and surrounded by an ultrapure water shield to reduce the external gamma ray flux (e.g., from the rock walls in NUSEL). The detector and shield together fill a cylindrical volume, 12m in diameter and 12m high. The fiducial volume of 10 tons is defined to be a 2.5m diameter sphere at the center of the detector.

The advantages of liquid neon include the absence of long-lived radioisotopes, high density (1.2 g/cm$^3$), high scintillation yield, low cost of the target material ($40K/ton), and the ease with which neon can be purified by distillation and cold traps. The dominant background in CLEAN is expected to be γs from the PMTs. The wavelength shifter employed in the detector can be coated on a low-radioactivity substrate well removed from the PMTs, allowing a very low total count rate. The fiducial volume cut (the restriction to the 2.5m diameter sphere at the center) yields an estimated background/signal ratio of a few percent. It has been estimated that a depth of 4500 mwe is sufficient to reduce cosmic backgrounds to less than 1% of the signal. (However, uncertainties in this estimate are discussed in the next section.)

CLEAN also has potential as a supernova neutrino and particle dark matter detector. Supernova neutrinos could be recorded by ES and by nuclear recoil, following coherent scattering of the neutrino off a neon nucleus. A comparison of the two rates will provide important information on the energy spectrum of heavy flavor supernova neutrinos. WIMP scattering on neon also produces observable recoil energies. If CLEAN reaches its background goals, it will be fully competitive with the next generation of ~ 1 ton dedicated dark matter detectors.

While **significant R&D issues remain to be resolved** in the CLEAN project, progress has been very good. The collaboration has purchased and successfully tested low-background PMTs at liquid-neon temperatures. Measurements of the photon yield and attenuation lengths in liquid neon are underway. A proposal for a 1-ton prototype will be submitted in 2003. The collaboration will study light attenuation and detector cryogenics with the prototype, and may convert the prototype into a particle dark matter detector by refilling it with liquid xenon. The collaboration is expanding as the R&D is carried out. With good progress the group expects to submit a construction proposal a few years from now. While a careful cost estimate cannot be made at this time, expectations are in the $40-60M range.

***HERON (ES in Helium):*** HERON is a proposal to observe the ES of pp and $^7$Be solar neutrinos in superfluid liquid helium. The recoil electron produces both ionization and rotons that propagate throughout the detector. Recombination of the ionization cloud produces in excess of 30,000 photons/MeV of extreme ultraviolet (16 eV) photons. The photons are detected in a set of bolometers mounted above the surface of the liquid helium. The rotons that reach the superfluid surface eject a burst of energetic $^4$He atoms, which then condense onto the silicon bolometers mounted above the surface, producing a temperature rise in the bolometers. The energy resolution is extremely good, ~ 6 eV for a 6 keV x-ray, and the threshold is expected to be no more than 50 keV. Position information is provided by a system of coded apertures mounted in front of the array of Si bolometer wafers.

HERON will use 20 tons of superfluid liquid helium, a volume 5m in diameter and 5m high. This volume is surrounded by a liquid nitrogen cryostat, which also serves as passive shielding, and by an ultrapure water shield to reduce the gamma ray flux from the underground site's rock walls. As foreign materials immediately precipitate out of the superfluid, the liquid helium is free of any internal radioactive contaminants. The detector requires an overburden of approximately 4500 mwe to reduce cosmic ray backgrounds to an acceptable level.

The dominant background in HERON are the $\gamma$s external to the superfluid, particularly from the cryostat. Using the position information available from the coded aperture system, a fiducial volume cut can reduce this background. In addition, many of the $\gamma$s Compton-scatter multiple times in the detector, enabling one to reject the events. The estimate background rejection efficiency is a promising $3 \times 10^{-3}$. A final cryostat design and an assessment of materials radiopurity is needed before one can assess the adequacy of this rejection. The necessary engineering studies are underway.

The counting rate for a 10-ton fiducial volume will be ~ 5000 pp events/y. Angular information is not available, but the rate is sufficient to allow measurement of the 7% annual flux variation due to the earth's orbital eccentricity, thereby establishing the solar origin of the events.

**Significant R&D issues remain to be resolved** before HERON is constructed. The HERON collaboration is carrying out experiments on scintillation and drifted charges, to determine whether single electrons that drift to the surface under an applied electric field can be extracted in He "bubbles" and detected using the Si bolometers. The collaboration is also developing larger Si wafers that are sensitive to single photons and optimizing the coded apertures for improved position information and background pattern recognition. Studies of systematics are being made to determine whether systematic uncertainties can be made commensurate with the SSM pp flux uncertainty. The collaboration plans to construct a prototype within the next two years, in order to prepare the way for a full construction proposal. The estimated cost of HERON is uncertain, but is in excess of $40M.

*Time Projection Chamber (ES in He):* TPC is a proposal to observe pp and $^7$Be solar neutrinos in a high-pressure time projection chamber. The recoil electrons produce ionization tracks that are drifted to a set of anode wires, where the signals are amplified and read out. The scattering angle can be reconstructed from the TPC's three-dimensional tracking data. The recoil electrons are concentrated in a 15-20° forward cone with respect to the direction of the sun. The incident neutrino energy can be determined from the reconstructed angle and the energy deposited along the track. The difference in the ES NC and CC recoil spectra can be used to separate the flux into its electron and heavy-flavor components.

The TPC target is a He/CH$_4$ mixture at 10 atmospheres. The detector's central region is 14m in diameter and 20m in length. The recoil tracks are drifted distances up to 10m to anode planes at the ends of the detector. Extensive Monte Carlo simulations have shown that TPC can reconstruct 100 keV electron tracks.

Two sources of background must be addressed. External $\gamma$s enter the detector, producing

Compton recoil electrons. As the detector is not self-shielding, fiducial cuts are not effective in reducing this background. Instead, the detector components must be fabricated from radiopure materials. Previous experience (most notably the MUNU experiment) suggests the structural elements can be made sufficiently pure. External shielding is also required to reduce γs from the rock walls of the chamber. The second background is due to internal contaminants, with tritium, $^{14}$C, $^{40}$Ar, $^{85}$Kr, and Rn being among the most serious activities. These activities can be reduced by using gases produced from very old petrochemicals from deep underground, by purifying the gases used, and by carefully sealing the detector from the atmosphere. It is known that methane from certain deep wells is depleted in $^{14}$C. Background rates of 100-200/day (20-40/day in the signal region) are expected, while the pp rate is about 5/day after cuts. The experimenters plan to subtract the background using tracks in the hemisphere opposite the forward event cone, resulting in a subtraction error of 6%/√yr.

The detector response will be calibrated using delta rays generated by cosmic ray muons and double Compton scattering. These events are kinematically constrained, providing redundant information that can be used in the calibration, and have rates exceeding the signal rate by two to three orders of magnitude, if the detector is sited between 1500 and 3000 mwe. (In this range the cosmic ray background is sufficient for calibration, but not overwhelming.) By using the two classes of events, the experimenters can measure the detector's energy scale and threshold, its fiducial volume, and its tracking resolution to 0.1-0.2%.

With a target mass of 7 tons, TPC will record about 1500 pp events/y after cuts. As the experiment would then be statistics limited, a much larger detector (70 tons) has been discussed. The increased counting rate would provide a better determination of $\sin^2 2\theta_{12}$, provided systematic uncertainties are small, and of the $^7$Be and CNO neutrino rates (with 1% and 4% accuracy, respectively).

**Significant R&D issues must be resolved** before TPC is ready for construction. Current research includes developing the readout systems and electronics, determining attenuation lengths for drifting the ionization tracks, testing gas purification, and mechanical design. An R&D proposal has been submitted. With continued good progress, a full proposal might be ready in a few years. It is believed that costs will be less than $150M, though no formal cost estimate has been made.

*LENS (CC in $^{115}$In):* The LENS (Low Energy Neutrino Spectroscopy) collaboration is an international group that includes members from Borexino, SAGE, GALLEX, Chooz, Bougey, and SNO. While several targets have been considered, current efforts are focused on $^{115}$In. As the $^{115}$In CC threshold is 114 keV, LENS samples most of the pp spectrum. The signal is a prompt electron in spatial and time-delayed coincidence (τ = 4.7 μs) with a 116 keV event. A triple coincidence with a 497 keV cascade γ in an adjacent cell completes the CC signature. The event rate is ~ 99 pp events/y/ton of In. The primary background in LENS arises from the natural beta decay of $^{115}$In (endpoint energy of 495 keV), which produces a singles rate of ~ 250 kHz/ton. Suppression of the resulting accidental coincidence backgrounds requires good energy and position resolution. An overburden of at least 3000 mwe is necessary to reduce cosmic ray backgrounds to an acceptable level.

Two major technical advances have helped LENS: 1) the development of indium-loaded liquid scintillators based on indium-carboxylates with a light output 3-4 times that of previous scintillators, and 2) a detector architecture with indium-loaded liquid scintillator modules interspersed with indium-free modules. The latter detect the 497 keV tag γ with sufficient energy resolution that only moderate granularity is required to suppress accidental coincidences from indium decays.

The basic architecture of LENS is a closed-packed array of linear modules of indium-loaded liquid scintillator (10% indium by weight), 10cm × 10cm × 300cm, with active indium-free liquid scintillator buffers, 100cm in length, on the two ends. Each module is viewed from the ends by a pair of PMTs. The indium-free modules have a larger cross section, 20cm × 20cm, and are arrayed to envelop each indium-loaded module. With time-of-flight event location, the design produces an overall granularity of ~ 2000 cells/ton In. Monte Carlo simulations predict detection efficiencies of 25-30% for pp neutrinos and ~ 80% for the higher energy $^7$Be and CNO neutrinos.

The LENS collaboration has installed a low background counting facility in Gran Sasso and is now assembling a prototype modular array within that facility. The experimenters will measure the *in situ* detector response for the prototype indium-loaded modules and the level of intrinsic and external backgrounds.

To achieve a 1.5% statistical uncertainty for pp neutrinos after five years of operations, 40 tons of indium is required. The corresponding 200 ton-y uncertainties for the $^7$Be and CNO fluxes are 5% and 10%, respectively. The dominant systematic uncertainty in LENS is the $^{115}$In($\nu_e$,e$^-$) cross section. This could be calibrated with 2% accuracy if a very intense (8 MCi) $^{51}$Cr neutrino source can be produced.

**Significant R&D issues must be resolved** before LENS construction commences. The full detector will have ~ 25000 modules containing ~ 400 tons of indium-loaded scintillator, two kilotons of indium-free scintillator, and 100,000 PMTs. The approximate detector dimensions are 9m × 32m × 32m. The total estimated cost is in the $150M range. The LENS goal is to submit a construction proposal in two years.

*MOON (CC in $^{100}$Mo):* MOON's solar neutrino response is the CC reaction $^{100}$Mo($\nu_e$,e$^-$)$^{100}$Tc. The threshold for this reaction is 168 keV. The signal is the prompt electron followed by the delayed β decay (15.8s half life) of $^{100}$Tc. The MOON detector will contain 34 tons of natural molybdenum, yielding three tons of $^{100}$Mo and a solar neutrino event rate ~ 360/y. The most troublesome background is $^{100}$Mo 2ν ββ decay. High spatial resolution, good energy resolution, and the ability to distinguish two β tracks from a single track will help in suppressing this background. The established technical requirements for MOON include U and Th activities of less than mBq/ton of molybdenum, timing resolution of 2ns, spatial resolution of ±3mm, energy resolution of ~ 7% at 3 MeV, and a dynamic range of 0.1-40 MeV.

The MOON collaboration is investigating three detector systems to achieve the design goals. The first uses 6m × 6m × 50 mg/cm$^2$ thick Mo foils sandwiched between 2.5mm scintillator planes, with signals read out through wavelength shifting fibers connected to a total of 13,600 16-anode segmented PMTs. The primary concern in this design is due to scattering and losses in the Mo foils. The second uses Mo-loaded liquid scintillator with 0.3-0.7% Mo by weight viewed by avalanche photodiodes, connected to a wavelength-shifting fiber optics readout. Additional R&D must be carried out to increase (by a factor of a few) the photon yield of the scintillator. The third option is the construction of a cryogenic calorimeter loaded with Mo. While this design offers the prospect of very good energy resolution, significant R&D will be needed before technical feasibility can be assessed.

All three approaches would be simplified if MOON were to use Mo enriched in $^{100}$Mo. It appears possible to produce several tons of enriched Mo in Russian facilities, but the associated costs have not been assessed.

The primary backgrounds are ββ decay, impurities within the Mo, and cosmic ray backgrounds. The estimated signal/background from ββ decay is ~ 2/1 for the Mo foils design. R&D on Mo purification are ongoing, though it appears the purity requirements can be met. Cosmic ray backgrounds can be reduced to a negligible level by siting the experiment at depths of 5000-6000 mwe.

**Significant R&D must be done** in order to assess competing design options and the ultimate sensitivity of MOON. The MOON collaboration anticipates that this will take a few years. At that point a full detector could be designed and a proposal for funding submitted. Because the design is not fixed, it is difficult to offer even a crude cost estimate at this point.

*B.3 Solar Neutrinos: Facility Requirements.* This section summarizes the facility requirements for future solar neutrino experiments: what must an underground laboratory provide to optimize the prospects for next-generation experiments?

*Cosmogenic backgrounds and depth requirements:* The depth requirements for the ES detectors HERON and CLEAN are estimated to be ~ 4500 mwe, though there are significant uncertainties because the spallation yields from neon, in particular, are poorly known. CLEAN may be susceptible to delayed activities from isotopes of Ne, F, O, and N, some of which have lifetimes of several tens of seconds. A conservative design would site these experiments at ~ 5500 mwe, providing an order-of-magnitude safety margin. In the $\beta\beta$ decay discussion of MOON it was noted that a conservative design would place this detector at great depths, 6000 mwe.

The HELLAZ stated depth requirement is ~ 1900 mwe. However our estimates indicates a 1% deadtime in this experiment would not be achieved without an overburden in excess of 2200 mwe. If cosmic ray calibration were not an issue, HELLAZ would be sited at 4400 mwe in Homestake, to make use of the ground support available on the 4850 ft level. However, many other levels in the Yates formation at lower overburdens are readily accessible from the Yates shaft. HELLAZ requires quite a large room, so an excavation in the Yates formation, with its excellent rock strength, would minimize construction costs.

LENS has relatively modest depth requirements, 3000 mwe, due to the triple coincidence of the signal. The natural site at Homestake would be on the 4850 ft level (4400 mwe).

Finally, we note that Raghavan has recently suggested a 1 kton liquid scintillator detector (similar to Borexino or KamLAND) with directional sensitivity to measured the terrestrial antineutrino flux coming from the earth's core. Such a detector would also be used for a variety of solar neutrino measurements. In the energy region corresponding to the *pep* line and CNO neutrinos, a difficult background is the cosmic ray spallation product $^{11}$C, produced at ~ 50 times the rate of another spallation product, $^{7}$Be, that has been of concern to Borexino experimentalists. Half of the $^{11}$C positrons fall in the *pep* window. A depth of ~ 7000 mwe reduces this background to ~ 5% of the SSM *pep* signal. Thus if a detector for geophysical neutrinos is constructed, a site on the deeper of the two proposed NUSEL main levels (7400 ft, or 6500 mwe) might increase the reach of the experiment.

*Solar neutrino space requirements:* The two cryogenic experiments, HERON and CLEAN, require customized cylindrical cavities ~ few $10^3$ m$^3$ in volume. The remaining experiments can be accommodated nicely in standard "breadloaf" halls. The volumes needed are in the range (0.5-27) × $10^3$ m$^3$, easily within Homestake specifications for both the 4850 ft and 7400 ft levels.

These cavity dimensions take into account the significant amounts of water shielding, for most of the detectors, needed to provide the local shielding from $\gamma$s and neutrons emanating from the rock walls. The possible need for auxiliary space (usually adjacent to the detector) is also included. In general this space has few aspect ratio constraints on its dimensioning as the intended uses include staging during assembly, locations for auxiliary equipment such as refrigerators, pumps, water handling, electronics, storage, etc.

*Basic facilities needs:*
- Power requirements of a few hundred kW.

- Stable temperatures, usually 20°C or less, with air conditioning.
- Scrubbed air, with residual radon levels below 1-10 mBq/m$^3$ (depending on the experiment). This is not a concern for cryogenic detectors because of the lack of diffusion into them.
- A clean room for preparations, assembly and cleaning, and various levels of clean-room conditions in the detector cavity during construction and operations.
- DI water system.
- Cranes for assembly and manipulation of detector and shielding elements. Significant hoist capacity (both weight and volume) during construction. (It is recognized that there will be some compromise between NUSEL desires for experimentalists to "modularize" their designs and collaboration desires to move very large and possibly awkward loads underground.)
- Twenty-four-hour personnel access to detectors.
- A storage area for radon-free materials and to "cool" cosmogenics.
- Machine shop, both general and for ultra-low background work.
- Low-level counting capabilities for materials screening.
- Space for R&D work (surface and underground).

*Special facilities needs (experiment specific):*
- Special materials preparation areas (e.g., electroforming).
- Containment, external dumping, oxygen deficiency equipment, and emergency evacuation procedures to enhance the safety of experiments employing large quantities of flammable gases or scintillation materials, high-pressure gases, and cryogenic liquids such as neon, helium, and nitrogen.
- The necessary permits to allow transportation and importation of intense neutrino calibration sources. (These sources are easily shielded and present no personnel hazard from external radiation.)

***B.4 Solar Neutrinos: Summary.*** There is a great deal of ongoing R&D activity, both in the US and overseas, focused on next-generation high-statistics solar neutrino detectors capable of characterizing the flux, spectrum, and flavor of low-energy solar neutrinos. The most advanced experiments appear to be 2-3 years away from construction. Thus NUSEL should be prepared to accommodate such experiments very early in its lifetime. The facilities requirements of such experiments are significant, with some requiring specialized excavations, and many requiring specialized procedures for flammables, cryogens, and compressed gases. NUSEL should be prepared to host R&D efforts immediately on opening. Many of the proposed experiments are now engaged in, or plan to mount in the near future, underground prototype experiments.

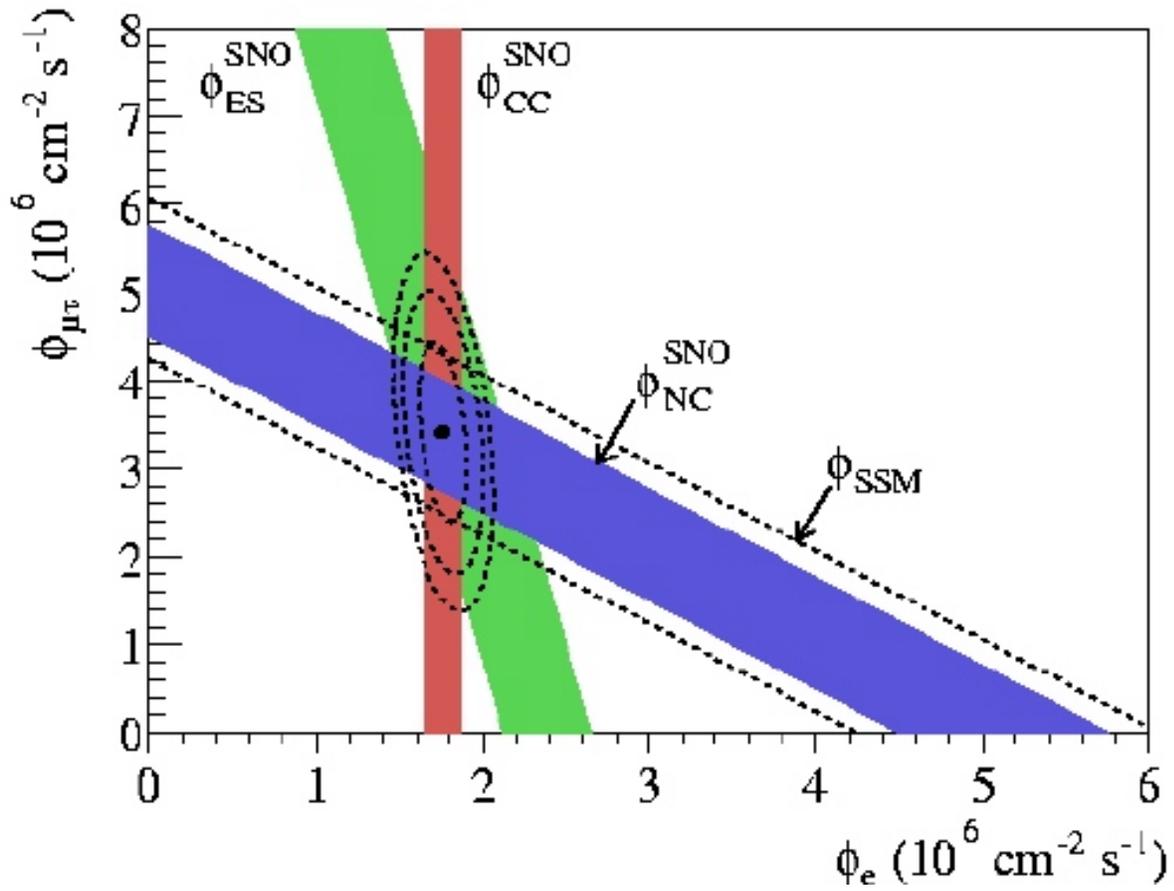

Figure C.6: The Sudbury Neutrino Observatory results, charged-current, neutrino-current, and elastic scattering rates, show that heavy flavor neutrinos account for two-thirds of the solar neutrino flux. The total flux is in remarkable agreement with the predictions of the standard solar model (the area between the dashed lines). Once these data were added to other solar neutrino constraints, the large-mixing-angle (LMA) solution emerged. SNO measures the high-energy portion of the solar neutrino flux, 0.01% of the total.

***C.1 Long-baseline Neutrino Oscillations: Importance of the Science.*** As noted earlier, one of the fortunate aspects of recent solar and atmospheric neutrino oscillation discoveries is the neutrino mass scale. The neutrino $m^2$ differences for atmospheric and solar neutrino oscillations are within the reach of terrestrial experiments using accelerator and reactor neutrino sources. This opens up marvelous opportunities for more precise measurements of poorly known parameters, such as ($\delta m_{12}^2$, $\theta_{12}$) and ($|\delta m_{23}^2|$, $\theta_{23}$), for measuring new neutrino parameters, such as the sign of $\delta m_{23}^2$ and $\theta_{13}$, and for probing new phenomena, such as CP violation, CPT violation, and matter effects on oscillations. The neutrino mass-matrix parameters are currently our only experimental indication of the physics that lies beyond the standard electroweak model. New phenomena such as CP violation could lead to profound discoveries, such as the mechanism responsible for the creation of matter in the early universe.

Terrestrial neutrino experiments offer many advantages. The neutrino propagation distances we can probe are arbitrary (up to the earth's radius), limited only by our ingenuity in creating powerful neutrino beams and sensitive detectors. By using combinations of near and far detectors, beams and detectors can be carefully tested. Systematic effects can be controlled and evaluated, resulting in known uncertainties.

Terrestrial experiments have already had significant impact. The KamLAND reactor neutrino (baseline ~ 180 km) experiment has verified the LMA solution to the solar neutrino problem, while narrowing the allowed range for $\delta m_{12}^2$. The K2K experiment – muon neutrinos of energy ~ 1.5 GeV from the KEK accelerator are observed in the Kamiokande detector, baseline ~ 250 km – has found a reduced flux consistent with the atmospheric neutrino oscillation parameters. Two additional accelerator experiments with baselines ~ 750 km, MINOS (FermiLab to Soudan) and CNGS (CERN to Gran Sasso), will run within the next five years with the goal of narrowing uncertainties on the atmospheric oscillation parameters. The science goals of future experiments include:

*Precise measurements of $\delta m_{12}^2$ and $\theta_{12}$:* It appears unlikely that long-baseline accelerator neutrino experiments will provide precise measurements $\theta_{12}$. KamLAND's results on reactor electron antineutrino disappearance have reduced the LMA allowed range for $\delta m_{12}^2$, however. KamLAND will produce more data, and it is conceivable that future long-baseline reactor experiments could improve on KamLAND's results. Because the SSM pp flux prediction is so precise, high statistics (~ 1%) pp solar neutrino measurements are a promising strategy for further constraining $\theta_{12}$.

*Precise measurements of $\delta m_{23}^2$ and $\theta_{23}$:* In the case of the "atmospheric" parameters the open challenges include definitive observation of oscillatory behavior and a precise measurement of $\delta m_{23}^2$ and $\theta_{23}$, with one question being the extent to which $\theta_{23}$ differs from 45°. Certain models have been offered where the $\theta_{23}$ is precisely 45° as a result of symmetry, so that deviations from this value are a measure of the symmetry breaking.

*Detection of the appearance signal $\nu_\mu \rightarrow \nu_e$:* The appearance of $\nu_e$s in a $\nu_\mu$ beam at an atmospheric length scale would provide a measure of a subdominant oscillation governed by $\theta_{13}$. This mixing angle is currently unknown but limited by the Chooz reactor experiment to < 10°. As the Jarlskog invariant that governs CP violation in neutrino oscillations is proportional to sin $2\theta_{13}$ cos $\theta_{13}$, the size of this mixing angle is crucial to plans to measure leptonic CP violation (and the hope of relating this observation to theories of leptogenesis). Even a very small $\theta_{13}$ ($10^{-3}$ or greater) is important in supernova physics because of the MSW mechanism: this angle governs the high-density crossing between electron and heavy-flavor neutrinos, thereby producing a hotter electron neutrino flux. Note that the sign of $\delta m_{23}$ determines whether electron neutrinos or antineutrinos experience an MSW crossing. The charge-current interactions of electron neutrinos in a supernova influence the proton/neutron ratio important to the r-process and other nucleosynthesis.

*Detection of the matter effects in the appearance signal $\nu_\mu \to \nu_e$:* The matter effects, as noted above, depend on the sign of $\delta m_{23}^2$. Thus a long-baseline experiment in which the beam penetrates a significant distance through the earth will resolve the hierarchy question: is the neutrino pair that dominates solar neutrino oscillations lighter (normal hierarchy) or heavier (inverted hierarchy) than the third neutrino?

*Detection of CP violation in neutrino oscillations:* In the introduction to the neutrino physics discussion of this section we noted that the "Dirac" CP phase – the phase observable in neutrino oscillations – appears in the combination $\sin\theta_{13}\, e^{i\delta}$. This produces a difference between the oscillation probabilities when neutrinos and antineutrinos are interchanged,

$$P(\nu_\mu \to \nu_e) \Leftrightarrow P(\bar\nu_\mu \to \bar\nu_e)$$

As the CP-violating signal grows with the length of the baseline (though of course the flux is decreasing as $1/L^2$), very long baselines are helpful in such experiments. The expected effects are small and, of course, depend crucially on $\sin\theta_{13}$. It is also difficult to unravel many of the parameter ambiguities associated with matter effects and with uncertainties in the CP-conserving neutrino parameters. Thus some of the proposals to measure $\delta$ require measurements with multiple baselines, while others depend on the distinctive oscillation "fingerprint" that is implanted on a well-characterized broad-band neutrino beam.

*The existence of light sterile neutrinos:* The short-baseline neutrino oscillation experiment LSND conducted the appearance experiment using neutrinos from stopped pion decay

$$\bar\nu_\mu \to \bar\nu_e$$

The experimenters found a nonzero oscillation probability of ~ 0.3% and a region of allowed oscillation parameters centered on $\delta m^2 \sim 1$ eV$^2$. As this $\delta m^2$ is distinct from the solar and atmospheric values, LSND requires a fourth light neutrino in a mass range that would influence large-scale structure. No evidence has emerged for such a "sterile" neutrino from other experiments, but neither does any experiment contradict the claim (though the KARMEN and Bugey results limit the allowed range of the LSND oscillation parameters). The FermiLab experiment MiniBoone, now in the commissioning stage, will check the LSND result. If MiniBoone confirms LSND, the pattern of light neutrinos is considerably more complicated than that of three active species. Even if the LSND parameters are ruled out, sterile neutrinos with very small couplings to active species can be important in the early universe. Thus tightening the limits on sterile neutrinos is important, regardless of the validity of LSND claims.

*Tests of CPT violation:* A possibility for accommodating the solar, atmospheric, and LSND results that does not require a fourth neutrino is CPT violation. CPT violation allows different masses for neutrinos and antineutrinos. A CPT-violating oscillation observable is an asymmetry in the probabilities

$$P(\nu_\alpha \to \nu_\beta) \Leftrightarrow P(\bar\nu_\beta \to \bar\nu_\alpha)$$

### C.2 Long-baseline Neutrino Oscillations: The Readiness of Next-Generation Experiments.

The long-baseline oscillation program is quiet complex because it connects to so many other issues. Arguably **the discovery of leptonic CP violation will remain the single most important goal of neutrino physics** for the near future. To reach this goal a number of intermediate steps must be taken, some to clarify the difficulty of the search for CP violation.

The observation of CP violation will require two major advances, the creation of powerful new neutrino beams and the construction of an underground detector of unprecedented size, approaching a megaton. This megadetector is likely to be **the most ambitious and most important project NUSEL undertakes**. Its justification rests not only on long-baseline physics, but also proton decay, supernova neutrino physics, atmospheric neutrinos, and other science. Certain detector parameters, particularly depth, may be governed by these other uses. For this reason, we will delay our discussion of the megadetector to later in the Science Book, after the chapters on proton decay, supernova physics, etc. As the cost of the megadetector is likely ~ $0.5B, a decision to proceed will require very careful consideration by the physics community, independent of NUSEL. The megadetector is not part of the present proposal. However, in our view one of the strongest arguments for Homestake is its obvious suitability as a site for the megadetector, should the community decide to proceed. The Facilities Development Plan shows that the most logical and cost-effective plan for NUSEL-Homestake naturally **preserves an ideal, deep site** for the megadetector – one in the thoroughly studied, exceptionally competent rock of the Yates formation. It also allows NUSEL to provide the megadetector collaboration with a **dedicated hoist** capable of mining (and disposing on site) one megaton of rock/year. That hoist, the largest at Homestake, could later be dedicated to and optimized for construction.

Some of possible steps in a coherent program of long-baseline neutrino physics include:

*Completion of current experiments:* Additional data from KamLAND (and also new data from SNO in the salt and $^3$He NC detection modes) will further narrow uncertainties on the solar neutrino mixing parameters. MINOS and CNGS will start to provide data ~ 2005 and K2K will complete its data taking, narrowing the uncertainty on the atmospheric $\delta m^2$ to ~ 10%. MiniBoone will produce results, and if LSND is verified, many new directions for new neutrino experiments will need to be explored.

*Efforts to measure $\theta_{13}$:* There are at least three suggestions for either measuring $\theta_{13}$, or showing that it is much smaller than the Chooz limit of 10°. One is to exploit the NuMI beam (FermiLab to Soudan, expected to be available in 2005) to measure the appearance of $\nu_e$s in a $\nu_\mu$ beam. The kinematics of the reactions used to produce the neutrino beam provide a relative clean $\nu_\mu$ beam with a well-defined energy at a specific angle "off-axis" from the beam's center. Thus, at the cost of some loss in flux, one gains knowledge of the neutrino energy. This is important in reducing various backgrounds that might be mistaken as signals for $\nu_e$ appearance (e.g., CC $\nu_\mu$ events where the muon is misidentified as an electron, or CC $\nu_\tau$ events where the tauon decays into an electron). The experiment does not require an underground detector. The off-axis angle is chosen to select energies of 1.5-2.0 GeV, so that the detector sits at one oscillation length for the atmospheric mass difference, given the Soudan to FermiLab distance of 735 km. A ~ 50-kiloton detector at this location will be used to identify $\nu_e$ events, and a ~ 500-ton near detector will be used for calibration. The detector options include fine-grained calorimeters, water Cerenkov detectors, and liquid argon detectors. The goal is sensitivity to $\sin^2 2\theta_{13}$ approaching 0.01 – though, because of correlations with $\delta m_{13}^2$, the sensitivity deteriorates if the true $\delta m_{13}^2$ differs from the atmospheric best value. Data could be taken as early as 2007. There is a very similar off-axis proposal involving a neutrino beam from the Japanese Hadron Facility and Super-Kamiokande as the far detector.

A second proposal is a reactor neutrino experiment with either two or three detectors. To circumvent the neutrino spectrum uncertainties, the ratio of event rates is studied as a function of energy. With detectors arranged in a near-far configuration – distances of 0.5 and 3.0 km will encompass the atmospheric mass difference – electron antineutrino disappearance imprints on the ratio an oscillatory pattern. The magnitude of the oscillation depends on the size of $\sin^2 2\theta_{13}$. A third detector further helps with systematics. It appears that the reach of such an experiment in quite competitive with that of off-axis proposals and likely could be mounted in a relatively short time, given the success of KamLAND. The optimal detector sizes depend on the power of the reactor, but 10-100 tons is typical. Most important, the combination of off-axis and reactor antineutrino experiments appears to be considerably more sensitive than either individually, due to different dependencies on uncertainties in quantities like the atmospheric m$^2$ difference.

A third possibility is a very-long-baseline neutrino oscillation experiment with a broad-band neutrino superbeam. This proposal aims at determining parameters like $\theta_{13}$, removing parameter degeneracies, seeing matter effects, and measuring or limiting CP violation (if other parameters are favorable) in one experiment. This idea is discussed below.

*Very long baseline searches for CP violation:* The program to observe CP violation includes the determination of $\theta_{13}$ and the use of matter effects to determine the mass hierarchy (normal vs. inverted). The basic strategy requires intense neutrino beams (superbeams) and very long baselines (greater than 1200 km, so that one can exploit significant matter effects) for comparing

$$P(\nu_\mu \to \nu_e) \Leftrightarrow P(\bar{\nu}_\mu \to \bar{\nu}_e)$$

But to disentangle all of the mixing parameters, multiple measurements are required. Two approaches have been discussed seriously. One is a program of measurements with multiple baselines. Multiple beamlines at different distances to the same detector is an attractive possibility because the detector (~ 0.5 Mtons) may be the most costly part of the program. For Homestake the natural choices are the complementary baselines to FermiLab (1290 km) and Brookhaven (2530). Longer distances (or higher energy beams) allow one to probe oscillations beyond the first oscillation maximum, which provides information simultaneously on the CP phase and the sign of $\delta m_{23}^2$.

Another possibility is the observation of $\nu_\mu$ oscillations, in appearance and disappearance modes, using a single baseline, but employing a wideband neutrino beam. The wideband beam allows one to observe the effects of multiple oscillation lengths as an energy dependent modulation of the beam spectrum. This strategy has been explored in detail by a Brookhaven group, though the approach is general and could be applied elsewhere.

There are two components to this program, the superbeam and the detector. Superbeams have been under study at Brookhaven and FermiLab, as well as overseas. A superbeam requires an intense high-energy proton beam, a high-power target for pion production, and a pion decay channel to produce the neutrinos. The Brookhaven proposal is an upgrade of the AGS to produce a 1 MW proton beam, with a possible future upgrade to 4 MW. FermiLab is exploring two options, one based on a new 8 GeV proton synchrotron, and one based on a new 8 GeV superconducting proton linac, feeding the Main Injector. The detector most often discussed is a multi-100 kiloton water Cerenkov detector like UNO (or an array of small water detectors), but smaller scintillation and liquid argon detectors are also being explored.

Of course, if MiniBoone confirms LSND, there will be (at least) three distinct $\delta m^2$ values, and a more complicated set of long- and short-baseline experiments will be required to understand the associated phenomena.

*The Neutrino Factory Future:* A Neutrino Factory is designed to create very intense, pure beams of electron and muon neutrinos and antineutrinos at relatively high energies. The primary purpose of the facility would be to enable very accurate measurements of the parameters of the neutrino mixing matrix. In particular, if $\theta_{13}$ were very small, this would be the next step in the effort to find leptonic CP violation. The neutrino factory would allow measurement of $\sin^2 2\theta_{13}$ as small as ~ 0.0001, and to see maximal CP violation at that limiting $\theta_{13}$.

***C.3 Long-Baseline Neutrino Oscillations: Facility Requirements.*** The issue for NUSEL is the requirements for constructing the megadetector, which depend not only on long-baseline uses of the detector, but also the demands of proton decay, supernova neutrino detection, etc. We address this issue in our later megadetector discussion.

***C.4 Long-Baseline Neutrino Oscillations: Summary.*** The above program is exceedingly rich. It promises to answer many of the unanswered questions we have about the neutrino mass matrix, including the crucial question of CP violation. It makes excellent use of the facilities at FermiLab and Brookhaven, and takes advantage of the very long baselines to Homestake. And as will be apparent in later discussions, the megadetector needed for this program will advance studies of proton decay, supernova physics, and other underground science fields.

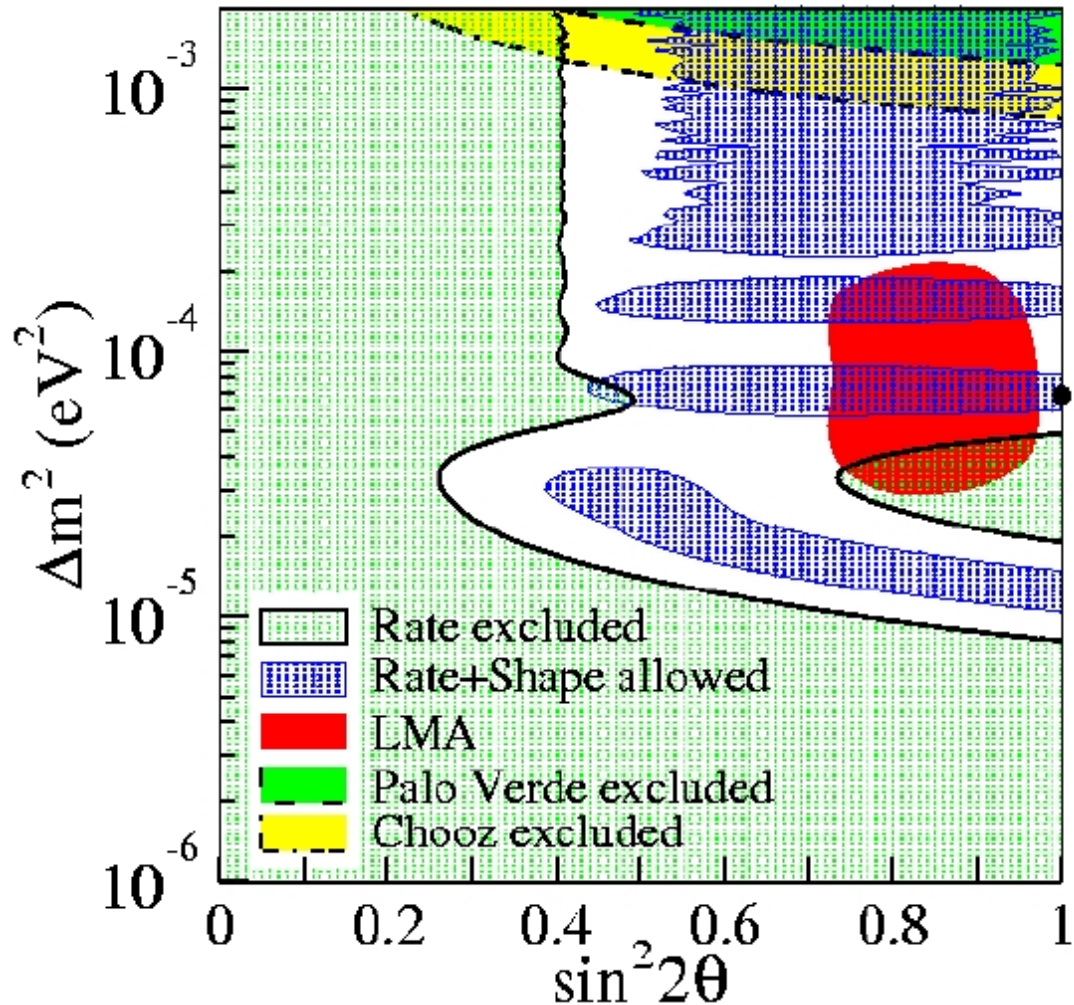

Figure C.7: Results from the KamLAND long-baseline reactor neutrino oscillation experiment. The data show that the LMA solar neutrino allowed region (in red) is now further restricted to two smaller areas centered on larger and smaller mass differences. Future long-baseline experiments will accelerator and reaction neutrinos will measure unknown neutrino parameters such as $\theta_{13}$ and the CP-violating phase $\delta$.

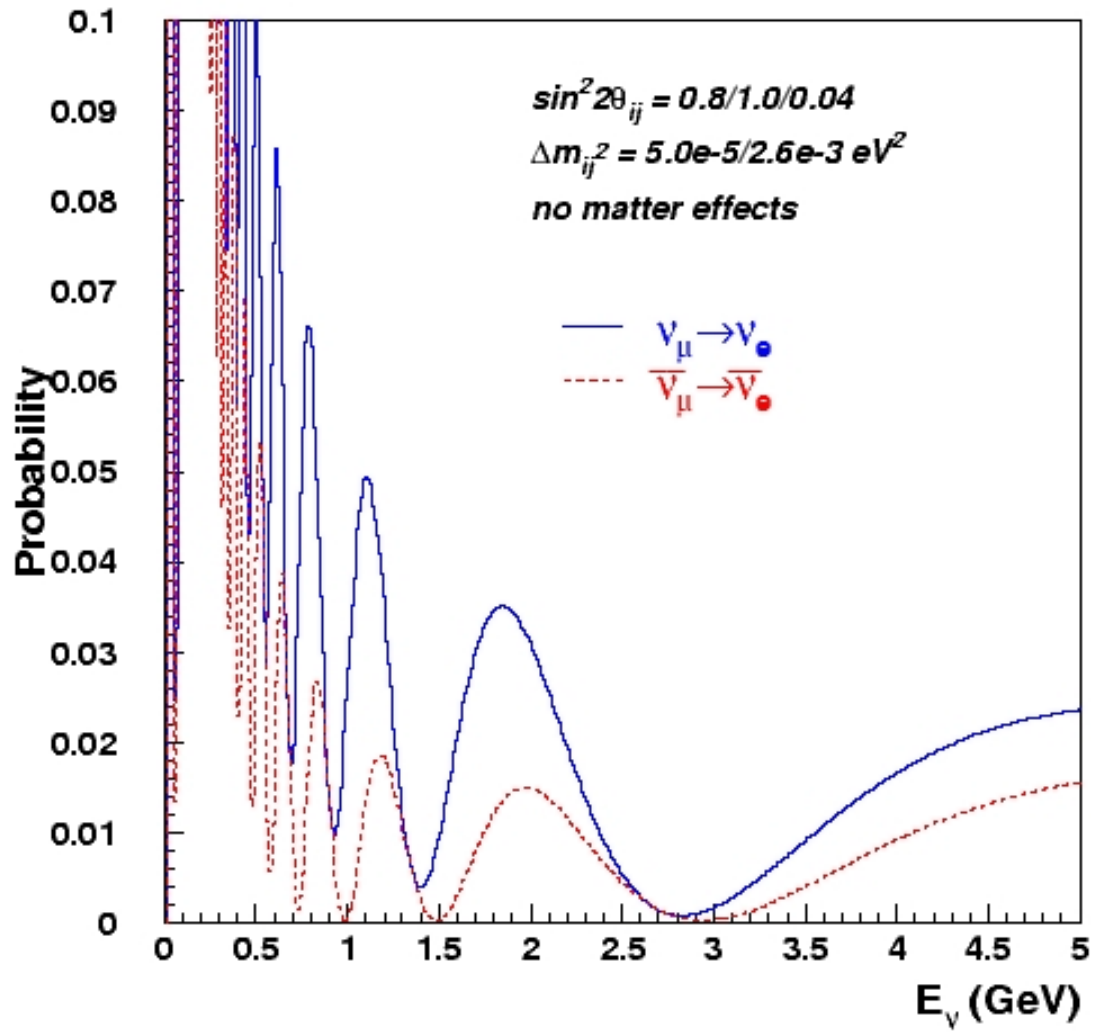

Figure C.8: An example of the possible outcome of oscillation tests of CP violation. This simulation comes from the Brookhaven broad-beam proposal.

***D.1 Atmospheric Neutrino Oscillations: Importance of the Science.*** When primary cosmic-ray protons and nuclei enter the upper atmosphere, the resulting hadronic interactions produce secondary cosmic rays. The spectrum of these secondaries peaks at about 1 GeV, but distribution's tail continues to high energy, falling off approximately as a power-law. One frequent product of hadronic interactions is the pion, which then decays

$$\pi^+ \to \mu^+ + \nu_\mu \to e^+ + \nu_e + \bar{\nu}_\mu + \nu_\mu$$

Thus the atmosphere is a source of cosmic ray neutrinos, which then travel downward and through the earth. An underground detector can observe neutrinos coming from above, as well as those coming from below after traversing the earth. In effect, such a detector, through the zenith angle of the neutrino event, can sample neutrino events with baselines ranging from 10 km (the thickness of the atmosphere) to 13000 km (the diameter of the earth). The latter distance, for GeV neutrinos, implies sensitivity to oscillations with $\delta m^2$ as small as $10^{-4}$ eV$^2$. That is, atmospheric neutrinos provide a marvelous laboratory for studying neutrino oscillations.

Studies of atmospheric neutrinos awaited the advent of the large underground detectors necessary for their detection. In particular, in the 1980s the water Cerenkov detectors Kamiokande and IMB were used to study atmospheric neutrino events producing electrons or muons in the tanks. In 1988 and 1989 these collaborations reported that they were recording too few $\nu_\mu$ events – the first hint of a profound discovery. (The expected ratio of $\nu_\mu/\nu_e$ is not too different from the 2/1 value given by $\pi$ decay above. However, rather sophisticated calculations are done to estimate the incident $\nu_\mu$ and $\nu_e$ fluxes, based on measured laboratory cross sections and including effects like the earth's magnetic field.) By 1990 results were available from a different kind of proton decay detector, the fine-grained iron detectors Frejus and NUSEX. While the results from these smaller detectors had substantial error bars, they did not appear to confirm the atmospheric neutrino anomaly. The program continued with the Soudan 2 fine-grained iron tracking calorimeter, the MACRO detector, and the 50-kton successor to Kamiokande, Super-Kamiokande.

One of the comforting aspects of atmospheric neutrinos is that the measurements are self-normalizing: apart from small geomagnetic effects, the source is isotropic. Thus one can look for new physics by determining whether the muon/electron event ratio depends on zenith angle (or baseline). The data produced by Super-Kamiokande was definitive. Normalizing to the Monte Carlo calculations of the expected ratio of muon to electron events, Super-Kamiokande found for the sub-GeV sample of events, after 1489 days of data, the ratio of ratios

$$\frac{(N_\mu/N_e)_{data}}{(N_\mu/N_e)_{M.C.}} = 0.688 \pm 0.016 \pm 0.050$$

The up-down asymmetry for multi-GeV muon sample was not zero, but

$$\frac{N_{up} - N_{down}}{N_{up} + N_{down}} = -0.303 \pm 0.030 \pm 0.004$$

In contrast, the corresponding asymmetry for electrons is consistent with zero. The up-down muon asymmetry deviates from the expected zero by 10σ, independent of any Monte Carlo input. The Super-Kamiokande binning of these events by zenith angle shows a systematic depletion of the muon events as one sweeps toward downward angles, with the upward-going flux lower than the downward flux by a factor of two. This led to the announcement in 1998 by the Super-Kamiokande collaboration of the discovery of massive neutrinos and neutrino oscillations. Careful analysis shows that the data are consistent with $\nu_\mu \rightarrow \nu_\tau$ oscillations with $\theta_{23} \sim 45°$ and $\delta m_{23}^2 \sim 2 \times 10^{-3}$ eV$^2$.

The properties of atmospheric neutrinos – the large baseline/energy or L/E range accessible (from 1-10$^5$ km/GeV), the self-normalizing properties of up-down comparisons, and the propagation through large amounts of matter – make these neutrinos an attractive source for future studies. In particular, megadetectors built for long-baseline accelerator neutrino studies and for nucleon decay will produce very large data sets. Such detectors will also fully contain higher energy events, which provide better angular resolution. Given an order of magnitude increase in event samples, several goals would be in reach:

- *Detection of the oscillation pattern:* With the improved energy reach and angular resolution, observations of neutrino oscillations should reveal a sinusoidal survival probability. No existing atmospheric neutrino experiment has succeeded in measuring such a pattern. For example, the Super-Kamiokande pattern as a function of L/E, while suggestive of neutrino oscillations, can be fit almost as well by certain neutrino decay models, once one takes into account detector resolution. One would like to see, as a function of increasing L/E, neutrino disappearance and then reappearance.
- *Detection of $\nu_\tau$ appearance:* The favored fit to the Super-Kamiokande results is $\nu_\mu \rightarrow \nu_\tau$ oscillations. Subsequent interaction of the $\nu_\tau$ in the detector will result in $\tau$ appearance at a rate of about 1/kton-y.
- *Detection of matter effects:* In the standard three-flavor scenario matter effects are present if there is a nonvanishing $\theta_{13}$: the $\nu_e \leftrightarrow \nu_\mu$ oscillation can become resonant in matter and significantly modify the oscillation probabilities of electron and muon neutrinos. Matter effects can also arise because of coupling to a sterile neutrino. (Super-Kamiokande uses the absence of matter effects to exclude the pure $\nu_\mu \rightarrow \nu_{sterile}$ oscillation at 99% c.l.) Finally, in detectors sensitive to the sign of the produced lepton, matter effects can lead to different oscillation patterns for neutrinos and antineutrinos. This, as noted in the previous chapter, results in sensitivity to the sign of $\delta m_{23}$, and thus to the mass hierarchy (regular or inverted).
- *Precision measurements of oscillation parameters:* Larger data sets will clearly narrow the allowed ranges for the mixing parameters. At 99% c.l. the Super-Kamiokande results determine $\delta m^2$ only to $\sim$ factor of three, for example. It has been argued that a detector with sensitivity to the sign of the lepton might be able to probe $\sin^2 2\theta_{12}$ to $\sim 0.02$, a result competitive with future long-baseline goals, after 1 Mton-y of data.
- *Study of multi-GeV neutrino interactions in nuclei:* Nuclear response functions and spallation yields are of interest to nuclear structure theorists.

***D.2 Atmospheric Neutrinos: Readiness of Next-Generation Experiments.*** Most of the proposed future detectors that will significantly improve over Super-Kamiokande will be multi-purpose experiments, often primarily focused on proton decay and long-baseline accelerator neutrino physics, but also used as supernova detectors, for studies of atmospheric muons, and for certain solar neutrino physics, such as day-night comparisons.

*Water Cerenkov detectors:* Proposed detectors include UNO, Hyper-Kamiokande, and AQUA-RICH. The first two detectors are qualitatively similar to Super-Kamiokande. For example, UNO would be divided into three cubic compartments, $60 \times 60 \times 60$ m$^3$, with the walls fitted with PMTs sensitive to low-energy neutrinos. One possibility with this detector would be observation of $\tau$ appearance.

The AQUA-RICH detector is based on the ring imaging Cerenkov technique. The proposers envision two spherical detectors equipped with hybrid photodetectors (HPDs) and immersed in a large tank of water. The water in the upper part of the tank is a shield against low energy cosmic rays. The inner surface of the outer sphere is a mirror, focusing the Cerenkov light onto the outer surface of the inner sphere. The inner sphere is densely covered with HPDs to detect the focused Cerenkov rings.

Though some R&D concerns remain, **the technology of water Cerenkov detectors is well developed**.

*Magnetized tracking calorimeters:* While the water detectors are a demonstrated technology for very-large-scale efforts, they also have some drawbacks. One is the lack of sensitivity to the muon charge. Thus an interesting alternative is a large magnetized tracking calorimeter. The muon charge sensitivity allows one to separate neutrino and antineutrino events, and thus to extract the matter effects discussed above. The energy of the hadronic system and the momentum of semi-contained muons can be measured, yielding very good neutrino energy reconstruction. As the angular resolution – the handle on L – is also good, L/E can be well determined. This is helpful in the oscillation pattern extraction.

The long-baseline MINOS detector at Soudan and the NOE part of the ICANOE detector at Gran Sasso will employ this technology. The detector that is planned for atmospheric neutrino oscillation studies at Gran Sasso, MONOLITH, is about an order of magnitude larger (34 ktons, with a fiducial volume of 26 ktons). Thus it will have a fiducial volume similar to Super-Kamiokande, but improved L/E resolution and sensitivity to the muon sign. MONOLITH **employs established technology** and, its proponents argue, could be quickly constructed.

*Liquid argon time projection chambers:* Liquid argon TPCs are sensitive to $\tau$ appearance and to both electron and muon neutrinos down to very low energies. A drawback is the higher cost per kton, compared to iron detectors. Argon TPCs have better angular resolution than water detectors. The 600-ton ICARUS detector is being installed in Gran Sasso, and plans exit to enlarge it to 3.0 ktons. It will be used for $\nu_e$ and $\nu_\tau$ appearance in the CNGS beam from CERN to Gran Sasso. Though limited for atmospheric neutrino studies because of its size, there are proposals to construct super-detectors of mass ~ 30 ktons. Thus this technology is **being demonstrated at a small scale, with plans to extend the technology to Super-Kamiokande masses soon.**

***D.3 Atmospheric Neutrinos: Facilities Requirements.*** We will address facility requirements for very large detectors in the megadetector section.

***D.4 Atmospheric Neutrinos: Summary.*** There are important goals not yet realized in atmospheric neutrino studies, including the observation of the oscillation pattern, increased precision on mixing angles, observation of $\tau$ appearance, and investigation of matter effects in detectors with sensitivity to the sign of the charged lepton. Thus atmospheric neutrino studies will be an important component of the physics program of future underground megadetectors (or Super-Kamiokande-class detectors with new capabilities, such as lepton charge sensitivity).

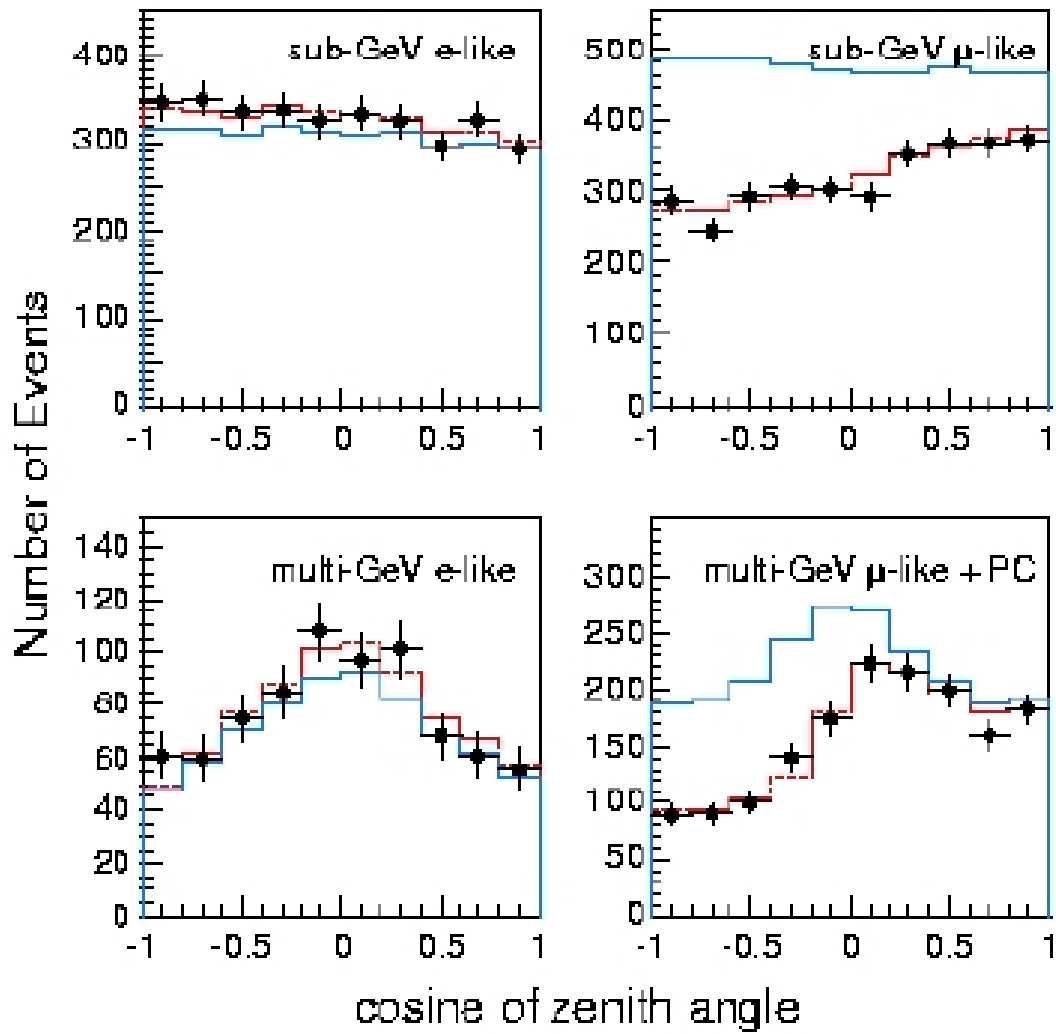

Figure C.9: The Super-Kamiokande atmospheric neutrino results showing excellent agreement between the predicted (blue line) and observed electron-like events, but a sharp depletion in the muon-like events for neutrinos coming from below, through the earth. This depletion at large baselines is seen in both the sub-GeV and multi-GeV data. The results are fit very well by $\nu_\mu \rightarrow \nu_\tau$ oscillations with maximal mixing (red line).

**Cosmology and Astrophysics:** In the past decade cosmology and astrophysics have undergone profound changes. With new ground- and space-based instrumentation these fields have become far more quantitative, and the connections between the underlying nuclear and particle microphysics and astrophysical phenomena far more critical to our understanding. This has produced a fertile intersection of cosmology, astrophysics, and nuclear and particle physics. The recent National Research Council study on "Quarks and the Cosmos" was inspired by the deepening microphysics-macrophysics connections, as well as by the associated sociological changes: many scientists trained in particle and nuclear physics are now applying their skills to cosmological problems.

Underground science has a central role to play in cosmology and astrophysics. Arguably the most important new mystery in physics is the matter/energy budget of our universe. Precision measurements of the cosmic microwave background, large-scale structure, and the light curves of distant supernovae have forced us to accept that the cosmos is filled with dark energy and dark matter, producing a universe very close to critical density. The visible universe open to conventional observations accounts for only ~ 1% of the total matter/energy budget. Deep underground experiments provide our best hope for identifying particle dark matter, and thus for relating dark matter to candidate theories for a more complete "standard model" of subatomic physics.

Core-collapse supernovae are important astrophysical laboratories. They produce prodigious neutrino fluences in all flavors, and these neutrino travel to us over galactic "baselines." The matter effects on oscillations are unique due to the high matter density and the important contributions of neutrino-neutrino scattering to the MSW potential. These neutrinos are a fascinating probe of the deep interior of the supernova. The neutrino light curve provides our best hope for identifying exotic phase (strangeness condensation, quark matter) transitions that might occur during protoneutron star cooling. Such transitions could alter the high-density equation of state, leading to unexpected mass/radius relationships in neutron stars. The neutrino burst as well as correlations between the burst and the gravitational wave signal could help us understand the explosion mechanism. The neutrino light curve controls the neutron-proton chemistry of the hot nucleon gas that is blown off the neutron star, where the r-process likely takes place. Thus neutrino observations would eliminate many current uncertainties in nucleosynthesis calculations. It is very important to have an array of supernova observatories underground to map out the flavor and light curves of the neutrinos from the next galactic supernovae.

Laboratory measurements of nuclear cross sections were vital to the solar neutrino problem, as they allowed astrophysicists to construct a quantitative standard solar model that accurately predicted even the highly temperature-dependent $^8$B neutrino flux. The new stellar laboratories we hope to exploit for fundamental physics – core-collapse supernovae, thermonuclear supernovae like SNIas (the cosmological standard candles crucial to studies of the dark energy), novae, and red giants – similarly need to be understood as well as possible. Energy production in stellar environments occurs through networks of low energy nuclear reactions. Measurements of such reactions in the laboratory are often severely limited by cosmic ray backgrounds. An underground laboratory provides an ideal site for a low-energy accelerator to measure nuclear astrophysics cross sections, free of backgrounds.

***E.1 Dark Matter: Importance of the Science.*** A combination of astrophysical measurements – maps of the cosmic microwave background, large-scale structure surveys, SNIa measurements of the Hubble expansion, gravitational lens maps of galaxy matter distributions – have progressed to the point of providing a reliable inventory of the composition of the universe. Energy densities are usually quoted in terms of the critical density, the density at which a test's particle's kinetic energy due to the Hubble expansion is compensated by the potential energy due to gravitational attraction

$$\rho_{crit} = \frac{3H^2}{8\pi G} \approx 1.88 \times 10^{-29} h^2 g/cm^3$$

Here $h \sim 0.71 \pm 0.04$ is the Hubble constant $H$ in units of 100 km/s/Mpc and $G$ is the gravitational constant. Thus the universe is closed if $\rho < \rho_{crit}$ and open otherwise. Measurements of the first acoustic peak in the angular power spectrum of the cosmic microwave background (CMB) indicate that the universe's total energy density is near the critical value, $\Omega_{tot} = \rho/\rho_{crit} = 1.0 \pm 0.04$, consistent with the prediction of inflation. The various measurements provide a very consistent inventory of the universe, a dark energy density of $\Omega_\Lambda \sim 0.73$, a total matter density of $\Omega_M \sim 0.27$, and a baryonic contribution to the matter density of $\Omega_b \sim 0.044$. The fact that the stars and other measured, luminous matter account for only ~ 1% of the total density means that majorities of both the total matter and the baryonic matter consist of unidentified dark components.

The most direct determination of the cosmological density of baryonic matter comes from the theory of big-bang nucleosynthesis and measurements of primordial light-element abundances. In particular, accurate determinations of the primordial deuterium abundance in low-metalicity gas clouds have been made from the absorption spectra of background quasars. The nucleosynthesis calculations reproduce the observed deuterium abundance only if $\Omega_b h^2 = 0.020 \pm 0.001$. The ratio of amplitudes of odd to even acoustic peaks in the CMB anisotropy spectrum provides an independent determination of $\Omega_b h^2 = 0.0224 \pm 0.0009$, in stunning agreement with the nucleosynthesis value. Thus physics at the time of recombination confirms a result from the nucleosynthesis epoch, three minutes after the big bang. The discrepancy between these determinations and the mass seen in stars and other luminous structure requires most of baryonic matter to be dark. Possibilities include matter hidden in condensed objects – there have been extensive gravitational microlensing searches for massive compact halo objects (MACHOS) near our galaxy, with too few being found to account for mass in the halo – and matter hidden in gas clouds.

Several measurements give consistent results for the total matter density. CMB measurements, particularly the height of the first acoustic peak, require $\Omega_M h^2 = 0.135 \pm 0.009$. Red-shift survey measurements of the shape of the power spectrum for large-scale matter inhomogeneities yield $\Omega_M h^2 = 0.20 \pm 0.03$. These results agree well with those obtained by combining measurements of the baryon density and the baryon-to-total-mass density ratio in clusters, as determined by X-ray measurements or by measurements of the Sunyaev-Zel'dovich distortion of the CMB. The results are consistent with Hubble-expansion observations of distant Ia supernovae, which indicate $\Omega_\Lambda - \Omega_M \sim 0.4$.

The excess of matter over baryonic matter strongly suggests that some type of nonbaryonic particle was produced in the early hot universe, and now exists as background matter, affecting the expansion and large-scale structure of our universe. One possibility is massive neutrinos. If a neutrino has a mass much smaller than 2 MeV, it decouples thermally from other matter while still relativistic. Their current number/energy densities can be related to those of the photons (the cosmic microwave background), which are given by Stefan's law

$$\begin{Bmatrix} n_\gamma \\ \rho_\gamma \end{Bmatrix} = 2\int \frac{d^3q}{(2\pi)^3} \frac{1}{\exp(q/T_\gamma)-1} \begin{Bmatrix} 1 \\ q \end{Bmatrix} = \begin{Bmatrix} 2\zeta(3)T_\gamma^3/\pi^2 \\ \pi^2 T_\gamma^4/15 \end{Bmatrix} \approx \begin{Bmatrix} 408/cm^3 \\ 4.6\times 10^{-34} g/cm^3 \end{Bmatrix}$$

That is, the background photon number and energy densities are fixed by today's CMB temperature, which has been measured to be $T_\gamma \sim 2.72K$. Clearly $\rho_\gamma/\rho_{crit}$ is very small. That would also be the case for neutrinos were it not for neutrino mass. For neutrino masses small compared to the temperature at coupling (so that the neutrinos have a relativistic distribution) but large compared to today's temperature (so that the mass dominates the present energy density) one finds

$$\rho_\nu = 2\sum_i m_\nu(i) \int \frac{d^3q}{(2\pi)^3} \frac{1}{\exp(q/T_\nu)+1} = \frac{3}{4} n_\gamma \left(\frac{T_\nu}{T_\gamma}\right)^3 \sum_i m_\nu(i)$$

The sum extends over the light neutrino masses. $T_\gamma$ is higher than $T_\nu$ due to photon reheating by $e^+ + e^- \rightarrow \gamma + \gamma$ after weak interaction freezeout

$$\frac{T_\nu}{T_\gamma} = \left(\frac{\rho_\gamma}{\rho_\gamma + \rho_{e^-} + \rho_{e^+}}\right)^{1/3} = \left(\frac{4}{11}\right)^{1/3}$$

Thus one finds

$$\rho_\nu = \frac{3}{11} n_\gamma \sum_i m_\nu(i) = 0.0106 \frac{\rho_{crit}}{h^2} \sum_i m_\nu(i)/eV$$

For the three known light neutrinos, the minimum $\rho_\nu$ is obtained for a standard seesaw hierarchy: only one generation is important, with $m_\nu \sim \sqrt{\delta m_{23}^2} \sim 0.055$ eV. The maximum $\rho_\nu$ is obtained for a degenerate hierarchy, three neutrinos each with a mass ~ 2.2 eV (the tritium β decay bound), yielding for $h \sim 0.71$

$$0.0011 \leq \rho_\nu/\rho_{crit} \leq 0.14$$

The lower bound is not too much less than the mass density in the visible stars, while the upper bound (corresponding to three degenerate light neutrinos with $m_\nu \sim 2.2$ eV) is some 100 times larger. In fact, though the exact bound is somewhat in debate, very recent CMB results from WMAP when combined with large-scale structure and supernova Ia data appear to lower the upper bound from 0.14 to 0.026. (We use the analysis of Hannestad, which is somewhat more conservative than the original WMAP analysis.) That is, the bound on $\rho_\nu$ derived from laboratory data is now significantly less restrictive than that imposed by cosmology. (Perhaps a better statement is that tritium β decay mass bounds must be improved in order to avoid the neutrino mass scale becoming a significant uncertainty in cosmological analyses.)

One concludes from the excess of matter over baryonic matter that most of the universe's matter is non-baryonic and (given constraints on light neutrino masses) must involve one or more particles outside the standard model of particle physics.

One important possibility is the axion, a light pseudoscalar predicted in extensions of the standard model in which the strong CP problem is solved by the Peccei-Quinn mechanism. Axions have couplings to **E•B,** and can be produced by fluctuations in this pseudoscalar source term. Axion couplings are severely limited by laboratory searches, red giant and other stellar cooling arguments, and by the duration of the neutrino burst from SN1987A. The primary "open window" corresponds to very weakly coupled axions, with masses of $10^{-2}$ to $10^{-5}$ eV, where the lower bound is established by requiring the cosmic axion contribution to dark matter not exceed $\rho_{crit}$. Thus light axions near this lower limit are most interesting cosmologically.

Axions are produced in the early universe by a nonthermal mechanism in which classical axion field oscillations are excited, yielding a highly degenerate axion Bose condensate that acts as cold dark matter. The production mechanism becomes more effective for lighter axions, so the contribution to $\Omega_M$ increases as the axion coupling and mass drop. A sea of axions of mass $10^{-5}$ eV and typical velocity $\sim 10^{-3}$ is thus a candidate for the bulk of $\Omega_M$. The technique of choice to search for cosmological axions is the conversion $a \rightarrow \gamma$ in a microwave cavity placed in a strong magnetic field. This resonant process requires tuning the cavity to step through the allowed range of axion masses, looking for the microwave photon. This program is partially completed, with new and more sensitive experiments in the construction stage. The microwave cavity experiments do not require underground sites.

Perhaps the leading particle cold-dark-matter candidate comes from supersymmetric extensions of the standard model. Although the energy scale where supersymmetry should appear is not fixed *a priori*, the mass hierarchy problem – the stability of the electroweak scale with respect to radiative corrections – suggests that superpartner masses must be ~ 1 TeV. As explained below, the lightest supersymmetric particle most likely is stable and could account for the nonbaryonic dark matter.

Several arguments favor weak-scale supersymmetry. The gauge coupling strengths measured in accelerator experiments unify at the scale of grand unified theories (GUTs) if the masses of supersymmetric particles are ~ 1 TeV. Precision elecroweak data favor a light Higgs boson, as predicted by the minimal supersymmetric extension of the standard model (MSSM).

In order to prevent baryon- and lepton-number violation in supersymmetric particles, the conservation of R parity is imposed on theories. The R-parity quantum number is +1 for standard-model particles and −1 for the supersymmetric partners (sparticles). Conservation of R parity requires sparticles to be produced in pairs, heavy sparticles to decay into lighter sparticles, and the lightest sparticle to be stable. Thus the lightest supersymmetric particle (LSP) is a leading particle dark matter candidate. The LSP must be a neutral, weakly-interacting massive particle (WIMP), as LSPs with strong or electromagnetic interactions would become bound in anomalously heavy isotopes. Severe bounds exist on the possible abundances of such isotopes.

A natural possibility for the LSP in the neutralino, a linear combination of the wino, bino, and the two higgsinos, the superpartners of the neutral gauge and Higgs bosons. If the neutralino is the stable LSP, it would be present today as a cosmological relic from the early big bang, when temperatures were sufficiently high to produce pairs of supersymmetric particles. Like other weakly interacting particles, WIMPs will fall out of thermal equilibrium with the rest of the universe when the temperature drops to the point that weak rates, which typically vary as $T^5$, can no longer keep up with the Hubble expansion. The result at late times is cold dark matter, massive particles moving nonrelativistically at the time of structure formation. The relic density of any WIMP depends on its annihilation cross section, which by assumption is weak scale. Detailed calculations for the neutralino, in a variety of supersymmetric theories, suggest that the relic density will be sufficient to account for a significant fraction of the dark matter, *i.e.*, ~ 0.1–0.3 of the critical density.

Most direct-detection particle dark matter experiments search for nuclear recoils produced in elastic scattering of neutralinos from nuclei. Searches for inelastic nuclear excitations have also been done in targets where a low-lying excited state can be reached. The subsequent γ-decay provides the signal. The predicted rates depend on the neutralino cross section and the density and velocity distribution of neutralinos in the vicinity of the solar system.

There are many variables affecting neutralino-nucleon cross sections. Candidate supersymmetric models are classified according to the mechanism by which the supersymmetry breaking is communicated from the hidden sector to the visible sector. The most basic mechanism is provided by minimal supergravity (mSUGRA), which arises as a low-energy limit of a supergravity theory. In mSUGRA the broken supersymmetry is communicated to the visible sector via gravity, leading to soft SUSY-breaking masses at the TeV scale. Among the parameters are a universal gaugino mass $m_{1/2}$ and tanβ, a ratio of the vacuum expectation values of Higgs fields. Parameter variations define a range for mSUGRA spin-independent neutralino-nucleon cross sections ~ $10^{-6}$–$10^{-11}$ pb. Frameworks more general than mSUGRA broaden this range.

Experiments limit the WIMP scattering rate, and thus the neutralino-nucleon cross section as a function of the WIMP mass. Some 20 dark matter experiments are currently operating, with all but one of the world's underground laboratories hosting at least one. The current detection limit (90% c.l. sensitivity) is ~ 0.3 events/kg/day, corresponding to a cross section ~ $10^{-6}$ pb. This sensitivity was achieved by the Edelweiss experiment, in which no nuclear recoil events above 20 keV were observed after 7.3 kg-days of exposure of a single 0.33 kg Ge cryogenic detector. Similar sensitivity has been achieved by the ZEPLIN I group, which has 230 kg-days of data using a single-phase liquid Xe detector with a fiducial mass of 3.1 kg. Next-generation improvements to target masses of 10–50 kg should reach cross section sensitivities of ~ few $10^{-8}$ pb within 5 years, while next-to-next generation efforts may reach $10^{-10}$ pb sensitivities (or event rates of 1/100 kg y) in 10 years.

*E.2 Dark Matter: Readiness of Next-Generation Experiments.* Several 100-1000 kg dark-matter experiments are in the planning or proposal stages, some with smaller-scale versions already running or currently under construction.

All of these experiments have very stringent background constraints that require deep underground sites and careful attention to sources of ambient radioactivity. Many of the experiments allow rejection of the otherwise dominant electromagnetic background by determining whether the interacting particle produced an electron recoil or a nuclear recoil. Such discrimination is possible because the energy loss per unit track length is much higher for nuclear recoils than for electron recoils. Experiments may also be able to exploit the expected time dependence of the WIMP signal. If the experiment's energy response, efficiencies, and backgrounds can be kept stable and if a sufficient number of WIMP events is detected, it may be possible to detect a ~ 5% modulation in the WIMP scattering rate and energy spectrum due to the Earth's orbital motion. The presence of this annual modulation could confirm a signal, while the absence can be exploited to identify background, thereby improving an upper bound.

*GENIUS (Ge):* The GENIUS collaboration plans to operate an array of 40-400 high-purity Ge detectors (100kg-1 ton) immersed directly in liquid nitrogen. The liquid nitrogen is stored in a tank 12m in diameter and 12m in height, surrounded by 2m of insulating material (polystyrene foam). The Ge detectors are suspended at the center of the inner tank on a polyethylene support. Several concentric layers containing up to 37 detectors each are envisioned. The detector is designed for both dark matter and double beta decay searches. The signal for WIMP scattering is ionization in the Ge crystals. One of the GENIUS design goals is a reduction of backgrounds by a factor of 1000, relative to best current measurements. This will be achieved by removing almost all materials from the immediate vicinity of the detectors (the crystal mounts and cryostat system are the main background sources in current experiments) and by the immersion in very pure liquid nitrogen. The compact arrangement of Ge crystals provides significant self-shielding.

Preliminary studies suggest high purity Ge detectors work reliably under such conditions: low energy thresholds (2.5 keV) and good energy resolution (1 keV at 300 keV) were achieved for 300-400g crystals operating for up to three weeks in liquid nitrogen. Monte Carlo estimates indicate the background goal is feasible if care is taken to bring the crystals quickly underground, avoiding activation, and if Ge crystal surfaces are kept clean.

The expected sensitivity of GENIUS for spin-independent WIMP scattering is about $10^{-9}$ pb, assuming an exposure of 100 kg y and a background of 1 event/100 kg y/keV.

A small test version, GENIUS-TF, is currently under construction in Gran Sasso. Fourteen natural Ge crystals (40 kg) will be mounted within a 0.064 $m^3$ volume of ultrapure liquid nitrogen. The entire experiment fits within a volume of 2m × 2m × 2m. GENIUS-TF will determine whether "naked" Ge detectors work reliably in liquid nitrogen for extended periods (at least a year) and will be helpful in testing materials radiopurity. **GENIUS is in the research and development stage**.

*Majorana (Ge):* Majorana was discussed previously as a double beta decay detector. The enriched Ge crystals will be arranged in 10 21-crystal modules in order to provide a high amount of self-shielding. Each detector will be axially (2) and azimuthally (6) segmented so that single-scatter events, which are almost always contained in one detector segment, can be distinguished from multiple-scatter events, which usually deposit energy in more than one segment. Two preliminary phases are anticipated, one with a single enriched $^{76}$Ge crystal and a second with 14-18 enriched crystals in one cryostat system.

Estimates of Majorana's performance as a WIMP detector are based in part on experience with currently running Ge diode experiments. The ionization signal from WIMP scattering will be similar to current detectors, but the background rate should be reduced due to the segmentation and to the close packing of the modules. Both of these changes with significantly reduce Compton backgrounds and help with neutron rejection. (Neutrons have a mean free path of several centimeters in Ge and thus can be identified because of scattering in multiple sites.) Pulse shape analysis is expected to provide position information that can be use to reject events on the surfaces of the detectors: surface contamination of the crystals may be a limiting background for this type of dark matter experiment. Another expected advantage of segmented Ge detectors is reduced capacitance and thus a threshold below 1 keV ionization energy.

Majorana plans to run for 10 years with a mass of 500 kg. The long duration will increase sensitivity because of the predicted annual modulation of the event rate and energy spectrum. The necessary detector stability over several years has already been demonstrated in smaller Ge crystal arrays.

Assuming a flat background of 0.005 counts/(keV kg day) between threshold and 20 keV ionization energy, Majorana's cross section sensitivity for spin-independent WIMP-nucleon interactions should be ~ $10^{-9}$ pb. This assumes a 1 keV threshold and an exposure of 5000 kg years. **Majorana is in the latter stages of research and development.**

*CryoArray (Ge, Si):* CryoArray is a large array of germanium and silicon detectors cryogenically cooled to ~ 25mK, the same technology used in CDMS (Cryogenic Dark Matter Search). Simultaneous measurements are made of the ionization in the Ge diodes and the phonon energy produced in an interaction. Together these two measurements allow rejection of electron-recoil background events. Cryogenic detectors share the excellent energy resolution of conventional Ge diodes. In CDMS I the resolution achieved in the phonon and charge channels was less than 1 keV. For events above 10 keV recoil energy the rejection efficiencies for photons and low-energy surface electrons was greater than 99.95% and 95%, respectively. CryoArray is expected to have improved discrimination against surface electrons (>99.5%) because the phonon measurement provides information of the location of the interaction: comparison of phonon-pulse arrival times in a detector's four independent channels localizes interactions in the detector's xy-plane. Events occurring near detector surfaces result in phonon pulses with shorter rise times, allowing their rejection. Si detectors, which are less sensitive to WIMPs than Ge detectors, will be valuable in confirming that a possible signal is due to WIMPs.

CDMS I, with a active fiducial mass of Ge of ~ 0.3 kg, was mounted at a shallow site on the Stanford campus. CDMS II, which will have a total mass of ~ 7 kg and 42 detectors, is currently being deployed at a depth of 2080 mwe in the Soudan mine. This experiment will run through 2006. The detectors for CryoArray will be close packed within a cryostat, with the full experiment likely consisting of several identical cryostats, each with a volume less than 1 $m^3$. Simplifications introduced since CDMS II construction should allow ten times more detector mass to be mounted within a given cold space. This will make materials screening easier. A challenge in constructing a one-ton array at reasonable cost will be developing mass-production techniques for the ~ 1000 detectors. The cold and warm electronics systems developed for CDMS II lend themselves well to mass production.

To reach its goal, sensitivity to 10 WIMP interactions per year or, equivalently, to a WIMP-nucleon cross section of $10^{-10}$ pb, CryoArray requires relatively modest improvements in the technology of CDMS II, provided that the experiment is placed at sufficient depth to render the fast neutron background negligible. The photon background must be improved by approximately a factor of fifty, through a combination of better materials screening and better self-shielding. This goal should be achievable, as it is a factor of three beyond what is currently achievable with Ge diodes. More difficult will be the task of reducing β background by a factor of ~ 300. Better self shielding will be particularly effective for βs, but screening materials for low-energy surface βs will be a challenge. **CryoArray is in the research and development stage**.

**ZEPLIN IV (liquid Xe):** Liquid xenon has some excellent properties as a dark matter detector. It has a high density, 3 $g/cm^3$, so detectors can be compact. It has a high atomic mass (A=131.3), which is favorable for WIMP scalar interactions provided that a low threshold energy can be achieved.

Liquid xenon is an intrinsic scintillator, having high scintillation (λ = 170 nm) and ionization yields because of its low ionization potential (12.13 eV). There are no long-lived radioactive Xe isotopes, and other impurities (such as $^{85}$Kr) can be reduced to very low levels by centrifugation or with a distillation tower and cold trap. Krypton contamination levels as low as 1ppb have been achieved. Liquid xenon is available in large quantities.

The formation of excimer states, bound ion-atom systems, produces scintillation in liquid Xe. If the electrons are drifted in a high electric field ($10^5$-$10^6$ V/cm), a secondary process, called proportional scintillation, can be detected. If the electrons are drifted out of the liquid into Xe gas, the secondary process is electroluminescence, which takes place at much lower fields of a few kV/cm. Electron and nuclear recoils can be distinguished by measuring both the primary scintillation signal and a secondary process.

The recoiling Xe atom loses its energy through both ionization and scintillation. The number of fast scintillation photons is only about 25% of the number of electrons associated with an electron or gamma with the same energy. The number of released free electrons is also small, as the ionization electrons recombine very quickly. The strong recombination leads to the emission of more UV photons. Under a high electric field a nuclear recoil will thus yield a small charge signal and a much larger light signal, while an electron recoil has the opposite behavior. The ratio of charge to light is thus the basis for separating nuclear from electron recoils.

The ZEPLIN IV project, with ZEPLIN II as a prototype, is a proposed one-ton two-phase (liquid and gas) Xe detector. The goal is to measure both scintillation and electroluminescent photons to separate electron from nuclear recoils. Experiments with a 1 kg liquid Xe test chamber have achieved recoil energy thresholds below 10 keV and factors of 1000 in electron discrimination.

Due to the high purity of liquid Xe, the scintillation photons can be collected with a high efficiency while the electrons can be drifted to an anode region for readout. To maintain background discrimination at low energies, it is crucial to detect the few electrons produced in a nuclear-recoil event. The efficiency of this process is better in two-phase Xe than in the single phase. PMTs detect the primary scintillation photons a few nanoseconds after the event, while the electrons are drifted to the gas phase, where electroluminescence occurs. The same PMTs detect the electroluminescent photons after a few tens of microseconds, depending on the electron drift distance.

ZEPLIN II, the prototype for ZEPLIN IV, is under construction at UCLA and will be installed at the Boulby mine, UK. The 35 kg liquid Xe fiducial volume is viewed by seven PMTs placed above the gas phase. The static electric field to drift the free ionization electrons up to the gas phase is shaped by ten copper rings. Two wire frames form the electron extraction field at the liquid-gas phase surface and the electroluminescence field above the surface. The detector is vacuum-insulated with a double-layer chamber made of cast copper vessels.

The one-ton ZEPLIN IV detector will be modeled after ZEPLIN II. Eighty 5-inch PMTs will be placed above the liquid and gas phases, with special care taken to reduce any dead regions in the detector. The insertion of an internal CsI photocathode is being considered for signal amplification. The projected sensitivity of ZEPLIN IV is $\sim 5 \times 10^{-9}$ pb for a background of 2 events/(kg day keV) and 360 days of data.

*XENON (liquid Xe):* XENON is a proposal for a modular liquid Xe experiment with ten time-projection chambers (TPCs), each containing 100 kg of active target Xe. An additional 150 kg of liquid Xe surround the target, providing a veto shield for background events generated in the containment vessel and other materials. XENON will be a two-phase experiment, with detection of both the primary and proportional scintillation signals. The electrons produced in a nuclear or electron recoil are drifted to the gas phase, where they produce proportional scintillation light in a strong electric field.

The primary UV photons are detected by an array of PMTs placed above the liquid-gas interface. To increase the detection efficiency, a CsI photocathode is placed in the liquid and used to convert downward-going photons into photoelectrons. To increase the primary light collection efficiency, the TPC walls will be made of 90% Teflon, which has a 90% diffuse reflectivity at 178 nm. Three signals are thus expected from one event: the prompt scintillation signal, the electroluminescence signal, and the proportional scintillation signal from the CsI photocathode. The difference in arrival time between the first two signals gives the z-coordinate of an event, while the xy position can be determined by reconstructing the spot where the proportional scintillation pulse is produced. The 3D event localization allows background discrimination by fiducial volume cuts. The overall efficiency for background rejection is expected to exceed 99.5%, with visible energy thresholds of a few keV, corresponding to recoil energies below 10 keV.

The liquid Xe TPCs are cylindrical, with a height of 30 cm and an inner diameter of 38 cm. The cylinder, formed by a sandwich of Teflon spacers and thin copper rings for field shaping, contains the active liquid Xe target. The cylinder is closed at the bottom by a thin Cu plate. The inner surface of the plate is coated with CsI. A larger Cu cylinder housing the wire structure for the proportional scintillation field and 37 PMTs in a close-packed hexagonal pattern close the top. The entire cylinder is enclosed in a copper vessel containing the liquid Xe for active shielding. Two rings of 16 PMTs detect the scintillation light from the shield. While the baseline detector design uses low-radioactivity PMTs to read out light, readout schemes based on large-area avalanche photodiodes or gas electron multipliers are also under study.

The extrapolated sensitivity of XENON for one ton of target material and three years of exposure is $4 \times 10^{-10}$ pb. This assumes a background rate of 0.039 events/(ton day keV) (corresponding to a nuclear recoil discrimination of 99.5%) and a visible energy threshold of 4 keV (corresponding to a 10 keV recoil energy in liquid Xe). Depending on the detector readout scheme and the efficiency of the self-shielding, another factor of 4 improvement is expected. A prototype with 7 kg of active material is planned to test all design aspects and to determine backgrounds and the event threshold. **XENON is in the research and development phase.**

***DRIFT-3:*** The DRIFT (Directional Recoil Identification From Tracks) experiment uses a negative-ion TPC. The negative ions are created by the slightly electronegative gas attaching to the electrons produced by a recoil track. Because negative ions diffuse much less than electrons do, no external magnet is needed to inhibit diffusion, and sub-millimeter resolution can be achieved even after drifting the ions for meters. Measurement of the range of the recoil together with the total ionization should allow nearly perfect rejection of background photons and electrons.

A one-cubic-meter prototype with 2-mm resolution and 250 g of $CS_2$ gas is currently running at a depth of 3000 mwe in the Boulby mine. A second-generation version with 0.5-mm resolution and higher pressure (and thus greater mass) has been proposed. DRIFT-3 would be a larger version of the second-generation experiment, with 100 $m^3$ of gas at 160 torr, for a total active mass ~ 100 kg. The small mass will likely limit the experiment's sensitivity to cross sections of ~ $10^{-9}$ pb, corresponding to 10 events/y, depending on the Z of the gas that is used.

DRIFT's primary advantage is in providing the directional axis of the recoil, and quite possibly the direction of the recoil as well. Because WIMPs should have a strong directional asymmetry due to the movement of the sun with respect to the galaxy frame, the directional information can confirm that events are due to WIMPs. The diurnal modulation due to the earth's rotation produces a 10% asymmetry, significantly larger than the annual asymmetry and less subject to possible systematic effects. Were a WIMP signal detected, directional information would provide information about their galactic distribution. **DRIFT-3 is in the conceptual stage.**

*E.3 Dark Matter: Facility Requirements.* This section summarizes the facility requirements of next-generation dark matter experiments. What should an underground laboratory provide in order to optimize these experiments?

*Depth requirements:* The most important cosmic-ray muon background for direct dark matter detection experiments is due to fast neutrons (20-500 MeV) produced outside the detector shielding. These high-energy "punch-through" neutrons are difficult to tag with a conventional local muon veto system, as they can originate several meters within the cavern's rock walls. They scatter in the materials surrounding detectors, generating low-energy neutrons that produce signals in central detectors quite similar to those of WIMPS. Countermeasures currently under study include thick active shields and wide umbrella veto systems deployed in the tunnel rock.

In contrast, current experiments very successfully discriminate against backgrounds that deposit energy through electromagnetic interactions, reducing rates by many orders of magnitude. Consequently fast neutrons have become the limiting factor.

Good benchmarks are provided by existing experiments. The sensitivity of CDMS at the shallow

Stanford Underground Facility (16 mwe) is limited by external neutrons, despite a muon veto system that is more than 99.9% efficient. EDELWEISS is an experiment similar to CDMS, but located at depth in the Laboratoire Souterrain de Modane in the Frejus Tunnel (4800 mwe). Because of its depth, EDELWEISS has detected no WIMP candidates during 7.4 kg-days of exposure. This experiment is currently the most sensitive direct dark matter search. As discussed above, the ten-year goal of several experimental groups is to reach a sensitivity of one event/100 kg/year, more than four orders of magnitude beyond current limits. Thus there is a need to extrapolate backgrounds far beyond current experience.

Estimates of the size of "punch through" neutron backgrounds depend on the muon flux and neutron production cross sections. The former is well measured. At the earth's surface the muon flux is about $170/m^2/s$ and the average energy is ~ 4 GeV. The attenuation at 4500 mwe is nearly a factor of $10^7$: the penetrating flux is $800/m^2/y$ and is much harder, with a mean energy of ~ 350 GeV. (For comparison, the flux at 1700 mwe (the WIPP depth) is 100 times higher, while at 6500 mwe (corresponding to the 7400 ft level at Homestake) it is ~ 25 times lower.)

The neutron cross sections are far more uncertain. Energetic muons produce neutrons in rock through quasielastic scattering, evaporation of neutrons following nuclear excitation, photonuclear reactions associated with the electromagnetic showers generated by muons, muon capture, and secondary neutron production in such processes. The neutron yield as a function of the mean muon energy is approximately a power law, $N \propto \langle E_\mu \rangle^{0.75}$. While various theoretical estimates of the high-energy neutron spectrum at depth have been made, few experiments have been done. An empirical function that is in use, derived from the Monte Carlo muon shower propagation code FLUKA, works reasonably well, with deviations from measurements being a factor ~ 3, particularly at greater depths. This factor is a reasonable estimate of uncertainties in fast neutron predictions.

When such estimates are combined with Monte Carlo simulations, one finds that CDMS-II at the Soudan Mine (2080 mwe) should detect no more than eight single-scatter events due to external high-energy neutrons, during an exposure of 6.8 kg-y. This will not adversely impact the experiment's sensitivity goal of $3 \times 10^{-8}$ pb. But backgrounds will be an issue for subsequent experiments. For example, a one-ton Ge experiment designed to reach a sensitivity of $10^{-10}$ pb would yield about 10 WIMP events/y. Detection at this level would probe a majority of the parameter space of currently favored SUSY models, complementing searches that will be done at the LHC. If performed at Soudan depths, one would expect up to 1170 neutron events/y. Neutrons often can be distinguished by their multiple scattering in experiments with good granularity – detailed Monte Carlo simulations are necessary to estimate the rejection possible because of multiple scattering. If one assumes this technique will provide a factor of ten suppression in the fast neutron background, the resulting signal/noise would still only be 1/10.

Depth reduces the neutron background to an acceptable level. The factor of 25 that would be achieved by going from Soudan to a site at 4500 mwe would reduce the neutron background rate to 50/y. If an additional factor of ten is obtained by vetoing multiple scattering events, the resulting background would be a factor of two below the target WIMP rate (corresponding to a cross section of $10^{-10}$ pb). An additional factor of 25 would be obtained by going to 6500 mwe, the NUSEL-Homestake 7400 ft level, reducing the background to one event per several years. If WIMPS are not observed, this reduction might be important in establishing the tightest possible limits.

While depth is the simplest, most certain, and least expensive solution to the fast neutron background, sites shallower than 4500 mwe could be acceptable with proper attention to background reduction. Sophisticated shielding and vetoing could reduce this background by one to two orders of magnitude. A thick (1-2m) scintillator active veto around the detectors could tag high-energy neutrons as they penetrate inward. In addition the cavern rock, or an outer heavy passive shield, could be instrumented with additional veto detectors in order to catch some part of the shower associated with the initiating muon. Finally, increased granularity in detectors could enhance the multiple scattering rejection rate. While only preliminary studies of the efficacy and cost of these steps have been made, it is likely that the cost would not exceed that of the central dark-matter detectors.

In summary, the goals of experiments envisioned for the next ten years (corresponding to WIMP cross sections no smaller than $10^{-10}$ pb**) could likely be met with shielding of 4500 mwe**. A deeper site (e.g., 6000 mwe) would provide an additional safety net against residual muon-related backgrounds, and may be necessary for next-to-next-generation experiments that plan to reach beyond $10^{-10}$ pb. Shallower sites may be satisfactory if additional investments are made in the detector, including carefully designed shields and high granularity. However, there remains some concern that a possible systematic leakage of muon-related neutron events could render a shield less effective than required, though no specific mechanism for such leakage has been identified, to our knowledge.

*Materials handling issues*: During fabrication and transport of detector components at the earth's surface, high-energy cosmic-ray-induced neutrons, protons, and muons can activate materials through spallation reactions. Knowledge of the cosmic ray spectrum and spallation cross sections is required to calculate activation rates. The flux of cosmic rays and its variation with geomagnetic latitude is well known. The nuclear spallation cross sections are far more uncertain. Neutrons dominate nuclide production at the earth's surface (~95%), with protons (~5%) and muons (~1%) of some importance. There are only a few (n,x) and (p,x) cross sections measured as a function of target mass and energy. Cosmogenic activation programs like SIGMA and COSMO use compiled cosmic ray intensities and semiempirical formulas fitted to available nuclear data to estimate cross sections.

| Isotope | Decay | Half life | Energy deposition in Ge (keV) | Activation (µBq/kg) |
|---|---|---|---|---|
| $^3$H | $\beta^-$ | 12.33 y | $E_{max}$=18.6 | 2 |
| $^{49}$V | EC | 330 d | $E_{K(Ti)}$=5, no γ | 1.6 |
| $^{54}$Mn | EC,$\beta^+$ | 312.3 d | $E_\gamma$=840.8, $E_{K(Cr)}$=5.4 | 0.95 |
| $^{55}$Fe | EC | 2.73 y | $E_{K(Mn)}$=6, no γ | 0.66 |
| $^{57}$Co | EC | 271.8 d | $E_\gamma$=128.4,142.8, $E_{K(Fe)}$=6.4 | 1.3 |
| $^{60}$Co | $\beta^-$ | 5.27 y | $E_{max}$=318, $E_\gamma$=1173.2,1332.5 | 0.2 |
| $^{63}$Ni | $\beta^-$ | 100.1 y | $E_{max}$=66.95, no γ | 0.009 |
| $^{65}$Zn | EC,$\beta^+$ | 244.3 d | $E_\gamma$=1124.4, $E_{K(Cu)}$=9 | 9.2 |
| $^{68}$Ge | EC | 270.8 d | $E_{K(Ga)}$=10.37 | 172 |

Table IE.1 Cosmogenically-produced isotopes in Ge for an exposure of 30 days at sea level, followed by one year of exposure under ground.

In principle calculations or measurements of cosmogenic activation rates in all WIMP detector target materials (and potentially in surrounding materials) must be performed in order to accurately estimate allowed exposure times on the earth's surface and required "cool-down times" below ground. Here we discuss Ge as an example, a proposed target material in several experiments, as extensive calculations have already been done.

The Table above shows cosmogenically produced isotopes in natural Ge after exposure for 30 days on the surface followed by storage below ground for one year. Only isotopes with half-lives longer than 200 days are shown. For $^{68}$Ge a saturation activity is taken, as this isotope cannot be separated during the zone melting process. The calculations were done with a variation of the COSMO program. Limits on $^{68}$Ge and $^{3}$H activation from the CDMS-I 1999 data run are consistent with these calculations. These are the two most troublesome activities. $^{68}$Ge decays by electron capture to $^{68}$Ga, which emits a 10.4 keV X-ray. $^{3}$H $\beta^-$ decay has an endpoint of 18.6 keV.

To estimate expected counting rates due to $^{68}$Ge and $^{3}$H, we take the two extreme cases of GENIUS (which has a background goal of 0.01 events/(kg y keV), with only the charge signal recorded) and CryoArray (which has an electron-recoil rejection efficiency goal of 99.95% and a γ-ray background goal of 4.75 events/(kg y keV)). Two μBq of $^{3}$H would produce ~ 5 events/(kg y keV), which is tolerable for CryoArray but much too high for GENIUS. A $^{68}$Ge activity of 172 μBq produces ~ 2500 events/(kg y keV) in the 10.4 keV line. Although this is a much higher event rate, the excellent energy resolution of Ge detectors will concentrate most of these events in a few keV-width bins centered on 10.4 keV. The activity will also subside after a few years underground. Thus the $^{68}$Ge activity is less problematic.

While limiting surface exposure times to 30 days or less is possible – and while shielded production and transportation schemes are being discussed – it would be preferable to produce the Ge detectors in a shielded facility directly at the NUSEL site. Modest cover of 10-20 mwe is sufficient to eliminate the hadronic component of cosmic-ray-induced showers, which dominates the activation. Thus Ge detector fabrication could be done on a level much shallower than that required for the experiment itself.

Most of the dark matter experiments discussed above use copper. The table below lists the cosmogenic activities in Cu after a surface exposure of 90 days followed by underground storage of one year. The activities, again calculated using a modified version of the COSMO code, must be considered in the context of specific experimental designs. The quantity and location of copper will vary, as well as the degree to which the indicated activities will interfere with specific signals.

| Isotope | Decay | Half life | Q-value (keV) | Activity (μBq/kg) |
|---|---|---|---|---|
| $^{3}$H | $\beta^-$ | 12.33 t | 18.6 | 6.3 |
| $^{22}$Na | EC,$\beta^+$ | 2.6 y | 2842.2, $E_\gamma$=1274.5 | 0.7 |
| $^{49}$V | EC | 330 d | 601.9 | 7.8 |
| $^{54}$Mn | EC,$\beta^+$ | 312.3 d | 1377.1, $E_\gamma$=840.8 | 125 |
| $^{55}$Fe | EC | 2.73 y | 231.4 | 10.1 |
| $^{57}$Co | EC | 271.8 d | 836, $E_\gamma$=128.4,142.8 | 28.5 |
| $^{60}$Co | $\beta^-$ | 5.27 y | 318, $E_\gamma$=1173.2,1332.5 | 8.33 |
| $^{63}$Ni | $\beta^-$ | 100.1 y | 66.95 | 1.02 |
| $^{65}$Zn | EC,$\beta^+$ | 244.3 d | 1351.9, $E_\gamma$=1124.4 | 123.8 |

Table IE2: Cosmogenically produced isotopes in Cu for an exposure of 90 days at sea level followed by one year of below-ground storage.

*Basic facility needs:* The basic facility needs of dark matter experiments are quite similar to those for double beta decay:
- The space required to set up any one of the next-generation detectors is no more than $10 \times 50$ m$^3$. It could be advantageous to mount several dark matter experiments in a single larger hall because of common needs for cranes, water shielding, and a muon veto.
- Experimental rooms must be cleanable, that is, upgradeable to clean-room standards during initial assembly and later operations, with air scrubbed to reduce radon levels. Some experiments require constant radon purging.
- Each experiment requires additional underground laboratory floor space to house data acquisition (~ 50 m$^2$). Modest surface laboratory space (~ 30 m$^2$) and a surface control room with computer links to the laboratory (~ 30 m$^2$) are needed. A surface (or underground) laboratory clean room (class 1000-10000) for staging and assembly of experiments is important.
- Because stability during extended periods of data-taking is critical, air condition is necessary in both the experimental and data acquisition areas.
- Power requirements are modest, no more than 50 kW.
- Access to NUSEL common facilities – machine shops (below and/or above ground), the low-level counting facility, and the radon-free clean storage area – is important.
- Nearly all dark-matter experiments would benefit from an underground facility for electroforming copper. The germanium experiments would be helped by underground facilities for crystal growth and detector manufacturing.
- The typical size of underground experimental crews during installation is 5-10 people. Standard running requires only one or two people accessing experimental areas, with larger numbers during upgrade periods and run commissioning. The principal laboratory requirement is 24/7 access in case of emergency.

*Special facility needs (experiment specific):* Each experiment has specific additional needs not covered in the general list above:
- GENIUS requires room space of about 16 m in diameter and 19 m in height in its conventional configuration, with the liquid nitrogen tank above the ground level of the hall. A two-ton crane, a platform, and a crown wheel would provide access to a class 100 clean room on top on the tank. An alternative configuration would submerge the tank completely below the laboratory floor level, requiring excavation of a hole and coating of the hole with water-proof concrete. This alternative configuration is attractive because the clean room would be at ground level, so that no extra platform or crown wheel would be required. This configuration would be simpler and would be safer in case of a leak or an earthquake. In either configuration, a liquid nitrogen filtering system using active charcoal beds, a nitrogen gas liquefaction system, and a liquid nitrogen supply would be required.
- Majorana's special needs are discussed in Section A.3.
- CryoArray will require an additional room for cryostat pumps and plumbing (25 m$^2$ footprint), a few He liquifiers (~ 20 kW each of power consumption), and storage for a few hundred liters of liquid nitrogen and liquid helium.
- Zeplin IV requires a crane rated for at least two tons and a liquid nitrogen supply.
- XENON electrical power needs are 100 kW, with 3-phase lines. Liquid nitrogen consumption is estimated to be 3000 liters/d, requiring a large on-site tank, with nitrogen lines to the underground laboratory.
- DRIFT III space needs are somewhat larger: configurations of $10 \times 50 \times 6.5$ m$^3$ and $15 \times 15 \times 10$ m$^3$ are under consideration. A 5-ton crane is needed, with enough space above the apparatus for operations. A large gas TPC requires gas-handling equipment, exhaust treatment, and possibly other safety measures.

***E.4 Dark Matter: Summary.*** Direct WIMP searches are motivated by the possibility that the majority of dark matter in the universe consists of long-lived or stable, weakly interacting particles produced in the big bang. The field is relatively young and developing rapidly, with larger next-generation experiments to follow current efforts. The detector techniques under development – conventional Ge diodes, liquid xenon scintillation detectors, Ge and Si phonon-mediated detectors, gaseous TPCS – are quite varied. In addition to measuring the nuclear recoil spectrum of the WIMP, all experiments provide some additional signature more specific to WIMPS, e.g., the spectral form and event rate in multiple materials or signal modulation due to the motion of the sun and earth about the galactic center. Several current-generation experiments are operating in underground laboratories, while others would like to move to some facility like NUSEL soon. Most experiments could operate successfully at 4500 mwe. Generally the space and power requirements are modest. All projects would benefit from shared infrastructure, particularly low-level counting facilities, materials handling facilities, and underground storage and fabrication facilities.

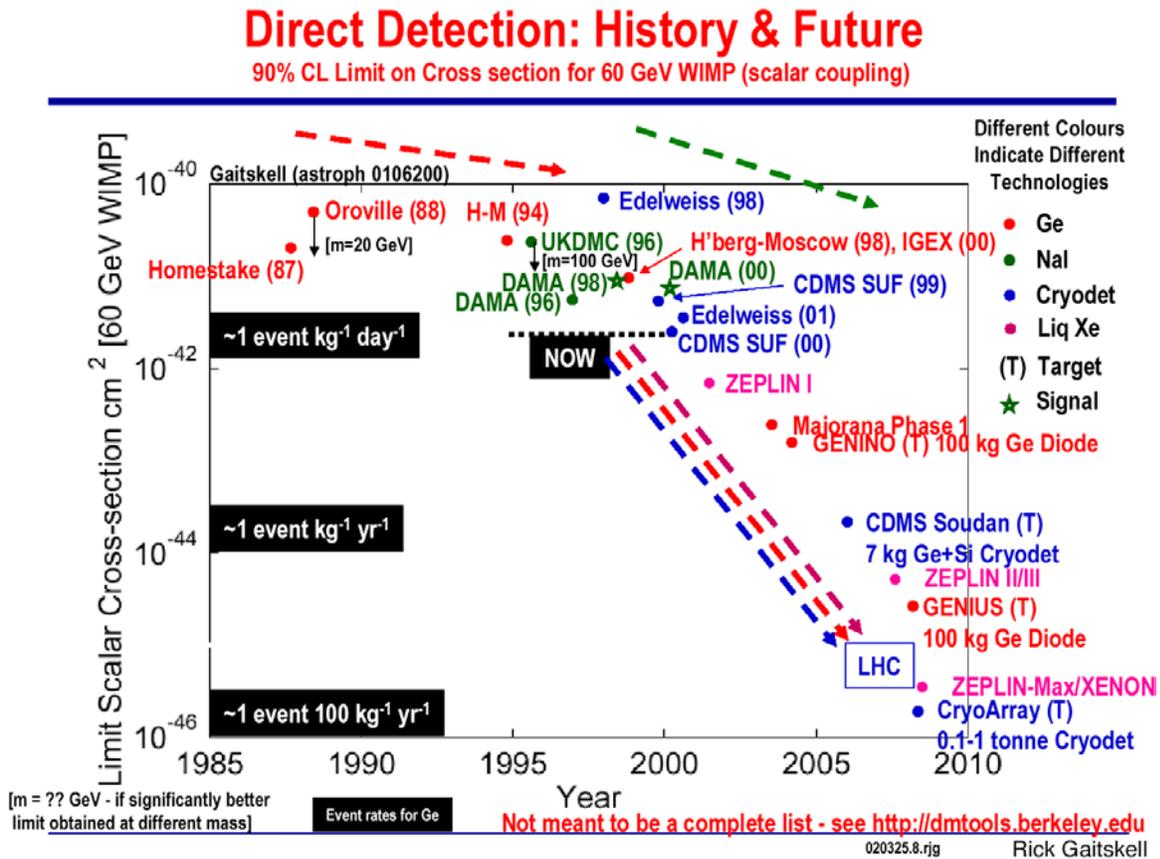

Figure C.10: Projected improvements in dark matter detection limits, given the next-generation efforts now being planned.

***F.1 Supernova Neutrino Detection: Importance of the Science.*** One of the spectacular achievements that resulted from the development of large active solar neutrino and proton decay detectors was the recording of the neutrino burst from Supernova 1987A, which occurred in the Large Magellanic Cloud some 150,000 light years from earth. The coincident detection of events by the Kamioka and IMB collaborations showed that our gross picture of collapse and subsequent protoneutron star cooling – neutrino energy release of ~ $3 \times 10^{53}$ ergs and a cooling time of ~ 3 seconds – was correct. This observation provided many important constraints on particle physics and astrophysics, including bounds on the neutrino mass and neutrino magnetic moment, constraints on anomalous cooling mechanisms such as axions and large Dirac neutrino masses, and even tests of extra-dimension theories.

Core collapse supernovae occur at the end of the life cycle of a massive star, e.g., one of 20 solar masses or so. Such massive stars evolve rapidly through their various hydrostatic burning cycles, until reaching the final stage where explosive Si burning begins to produce an inert iron core. When the iron core reaches the Chandresekar mass it becomes unstable to gravitational collapse. This collapse proceeds rapidly, typically as about 0.6 times the freefall velocity, as the electron gas quickly becomes ineffective in supporting the star: electron capture converts some of the trapped lepton number into escaping neutrinos. The gravitational work done on the collapsing matter also produces nuclear excitations, which further robs energy from the electron gas.

At a density of about $10^{12}$ g/cm$^3$ the neutrinos become trapped: the random-walk diffusion time for an escaping neutrino becomes greater than the collapse time. Thus the remaining lepton number and gravitational energy is locked within the star, trapped until after core bounce.

The collapse of the inner iron core, where the sound speed in nuclear matter exceeds the matter velocity, is homologous. When the central region of the core reaches several times nuclear matter density, the nuclear equation of state abruptly halts the collapse. The resulting trampoline-like rebound produces a pressure wave that moves outward, through the infalling matter and towards the edge of the homologous core (the sonic point). Surrounding infalling shells of matter subsequently reach the same density, rebounding to produce pressure waves that "chase" those already produced. When the edge of the homologous core reaches nuclear density, the compression and subsequent release of the pressure waves result in shock wave formation. The shock wave moves out through the outer iron core and beyond.

The passage of the shock wave through the outer iron core melts the nuclei into a nucleon gas, sharply reducing the neutrino opacity. (The opacity is dominated by coherent neutral current scattering. As the weak charge is approximately the neutron number N, the melting of the nuclei effectively reduces this contribution to the opacity by 1/N.) This and the accompanying enhanced electron capture off free nucleons produces a $\nu_e$ deleptonization pulse, a sharp peak in the neutrino luminosity that lasts for about 3 ms. However, 99% of neutrino emission is associated with the slower cooling of the hot, puffy protoneutron star: neutrinos of all favors leak out on a timescale of about 3 s.

Deep in the interior of the protoneutron star the neutrinos are in flavor equilibrium. As they leak out, however, their decoupling from the matter is flavor-dependent. The heavy-flavor neutrinos, which lack charge current reactions with the nucleons and have reduced cross sections for elastic scattering off electrons, decouple from deeper within the star, where it is hotter. The "neutrinospheres" for electron neutrinos and antineutrinos are further out; the $\nu_e$s are most strongly coupled to the matter because the matter has undergone substantial electron capture, and is thus neutron rich, enhancing $\nu_e$ charged current scattering. Though there is some debate on this matter, the result is a neutrino temperature hierarchy of 8, 4.5, and 3.5 MeV for the heavy-flavor, electron antineutrino, and electron neutrino flavors, respectively. (The corresponding mean neutrino energies are about 3.1 times the temperatures.) This provides an important neutrino oscillation diagnostic, as there is the potential for oscillations to produce a distinctive temperature inversion.

While investigators agree, at least on a qualitative level, in their predictions of the neutrino fluxes, the supernova mechanism is far more controversial. Almost without exception simulations that employ accepted nuclear equations of state and treat the microphysics (such as neutrino diffusion) reasonably (e.g., multigroup flux-limited diffusion or even a full Boltzmann calculation) fail to produce explosions. The initial hydrodynamic shock wave stalls at or near the edge of the iron core, having lost too much energy in melting nuclei and in neutrino emission. The hope has been that strong charge-current neutrino interactions in the hot nucleon soup left in the shock's wake could provide an extra push to force the shock wave outward again. The reasoning is that, since 99% of the released energy resides in neutrinos, this is the place to look for help. But neutrino heating is less effective than one might hope because the distance between the neutrino-emitting protoneutron star surface and the stalled shock wave represents many scale heights. Furthermore, neutrino heating of the matter near the neutron star quickly raises the matter temperature to ~ 2 MeV, where neutrino emission then competes with neutrino heating. That is, no further net energy deposition can occur.

There are hopes that this neutrino-driven "delayed" model will yet prove correct. One can produce an explosion by artificially hardening the neutrino spectrum. Qualitatively it is also appreciated that convection could enhance neutrino energy deposition. By sweeping heated matter to large radii (where the matter expands and cools, before it has a chance to radiate neutrinos), and replacing it with cold matter from larger radii, convection can circumvent the "gain radius" limitation to total neutrino heating. Great effort is currently being invested in 2D and 3D supernova simulations, though the numerical complexity forces shortcuts in the treatment of the neutrino diffusion and other important microphysics. Some of these simulations yield explosions for certain progenitor star choices, though it is unclear whether these results will survive when neutrino diffusion is handled more realistically. Currently there are several terascale supernova simulation efforts underway that have multiD simulations with realistic microphysics as one of their goals. Thus it may soon be known whether convection is a possible solution to the supernova mechanism puzzle.

Understanding supernovae is crucial to many aspects of astrophysics. They are among the most important engines driving the galaxy's long-term chemical evolution. It is believed that about half of the heavy elements are synthesized in the r-process (rapid neutron capture process), with the hot bubble above the protoneutron star being the most likely site. Wonderful new data on r-process nucleosynthesis is coming from observations of metal-poor halo stars, making this field very exciting. Other important nucleosynthesis is associated with the shock wave heating of material in a supernova, a process called explosive nucleosynthesis. Supernova neutrinos are believed to directly synthesize elements like $^{11}B$ and $^{19}F$ through spallation reactions in the carbon and neon zones. This is known as the neutrino process.

Some unusual core-collapse supernovae – suggestive of SNIc explosions, but unusually energetic – have been associated with gamma ray bursts. The nature and origin of gamma ray bursts is one of the key problems in observational cosmology.

As already mentioned, supernovae are marvelous neutrino sources. The detection of these neutrinos in an array of underground detectors could yield important new physics:
- Supernovae produce neutrinos of all flavors, with a temperature hierarchy that may allow us to look for oscillation effects. The matter effects on oscillations – the MSW effect – are especially rich. Because of the higher density, the 1-3 crossing that is not probed by solar neutrinos will cause flavor conversion in supernovae. It is possible that a limit – either an upper bound or a lower bound – on $\theta_{13}$ could be established by a careful measurement of supernova neutrino fluxes. (Flavor conversion requires an adiabatic level crossing, which in turn requires a minimum value for the mixing angle.) In addition, there are important locations in the star where neutrino-neutrino scattering, rather than neutrino-electron scattering, dominates the MSW potential. The effects of this nonlinear coupling are just now being explored theoretically. Supernovae are very likely our only laboratories for studying this interesting aspect of the MSW effect.

- The neutrinos are an important diagnostic of the supernova mechanism. They provide a direct measure of the gravitational binding energy of the neutron star: by following the neutrino light curve observers can monitor the process by which a puffy protoneutron star, perhaps ~ 40-50 km in radius, cools to a compact object of radius ~ 10 km. Similarly, the difference in the total number of electron neutrinos and antineutrinos measures the total lepton number radiated by the core. One of the fascinating questions regarding neutron matter at several times nuclear density is the possibility of new phases, such as kaon condensation or deconfined quark matter. It was recognized several years ago that such a phase transition will normally produce a region of mixed phase, e.g., where droplets of strange matter might coexist with ordinary nuclear matter. (This is a consequence of the two conserved charges, electric and baryon number.) This produces inhomogeneities in the weak charge distribution governing neutral current neutrino scattering, the main reaction trapping the neutrinos. This in turn can alter the neutrino light curve. Thus if a phase change occurs as the protoneutron star cools, it might be detected as a change in the rate of neutrino emission.
- As the protoneutron star cools (and as additional mass falls on to its surface) an interesting possibility is collapse to a black hole. This would appear in the neutrino light curve as a sharp cutoff.
- The neutrinos control conditions crucial to nucleosynthesis. For example, the neutron-proton chemistry of the "hot bubble" above the protoneutron star surface is controlled by the competition between the electron neutrino and antineutrino charge current reactions. Thus measuring the neutrino light curves (and knowing the neutrino parameters governing oscillations) will eliminate uncertainties in models of this nucleosynthesis. Similarly, neutrino process nucleosynthesis depends sensitively on the spectrum of the heavy-flavor neutrinos, which was not measured in SN1987A.
- Important constraints on neutrino masses, including the mass of the $\nu_\tau$, can be derived from supernova neutrino oscillations. Neutrino mass causes a kinematic spreading of the neutrino burst. Several possible sharp time structures in the neutrino light curve – the 3 ms deleptonization burst, the initial sharp rise in the neutrino luminosity, and the abrupt termination of the neutrino flux that would accompany black hole formation – have been discussed as possible "clocks" against which the spreading could be measured. For a galactic supernova the potential mass sensitivity is 1-2 eV. This is no longer so impressive – it is competitive with current tritium mass limits of 2.2 eV, given that we know the mass splittings among the three light flavors, and is less stringent than the WMAP bound. However a supernova constraint has the virtue of being tied directly to the kinematic effects of mass.
- Many supernovae in our galaxy may be obscured optically by the intervening matter. Neutrinos thus provide an early warning that an optical display may soon follow. (It may take hours to a day for the shock wave to reach the outer mantle of the star.) No supernova has ever been observed from such early times, as the light curve turns on. The environment immediately surrounding the progenitor star is probed by the initial stages of a supernova. For example, any effects of a close binary companion upon the blast would occur very soon after shock breakout.
- The simultaneous observation of the gravitational wave, optical, and neutrino signals from a galactic supernova could be very important, placing multiple constraints on the collapse mechanism.
- We will discuss, in the megadetector chapter, exotic possibilities, such as detecting the relic flux of neutrinos from all past supernovae and detecting collapses in our neighboring galaxy Andromeda.

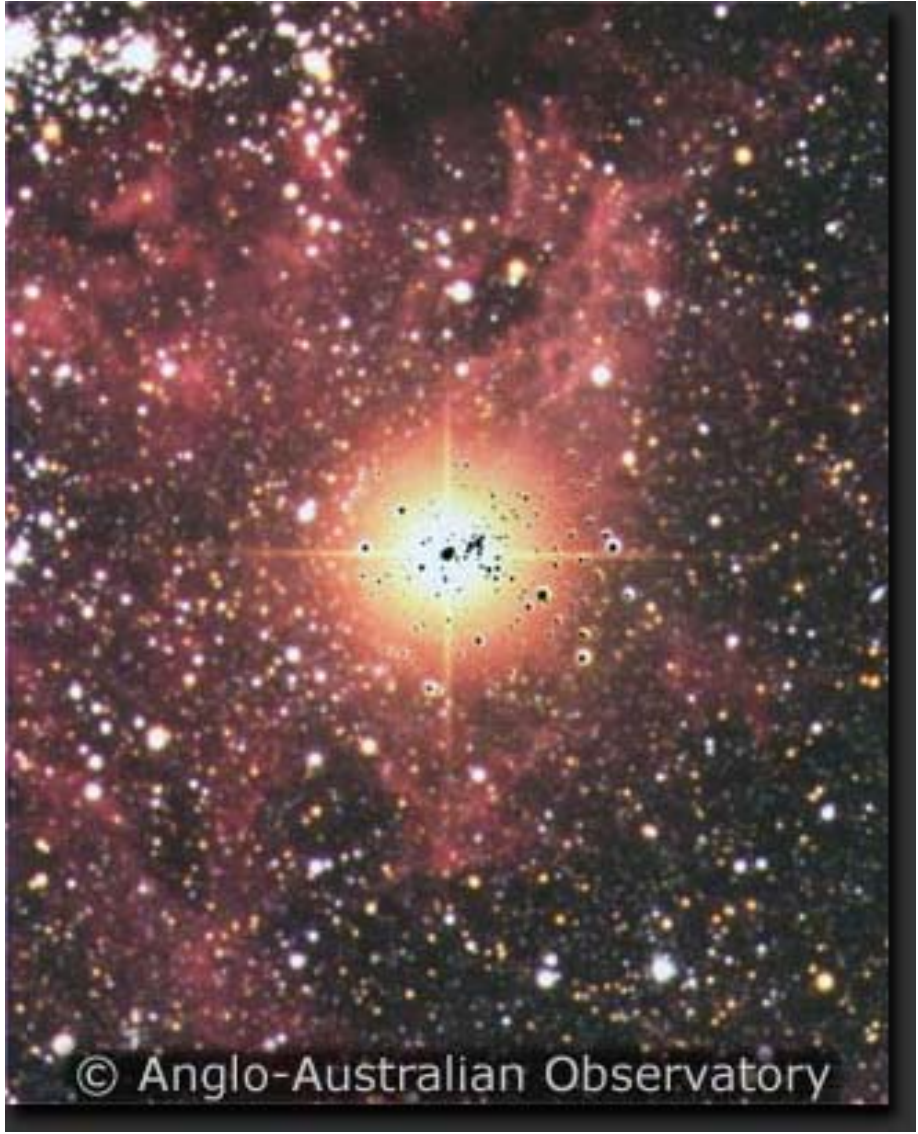

Figure C.11: The optical display of Supernova 1987A, which was observed on earth both optically and via its neutrino burst. The superimposed negative image shows the same star region prior to core collapse. While the optical display and kinetic energy of the explosion are impressive, less than 1% of the energy released in the gravitational collapse is released in this way. The bulk of the energy, ~ 99%, is carried by the neutrino burst.

A-54

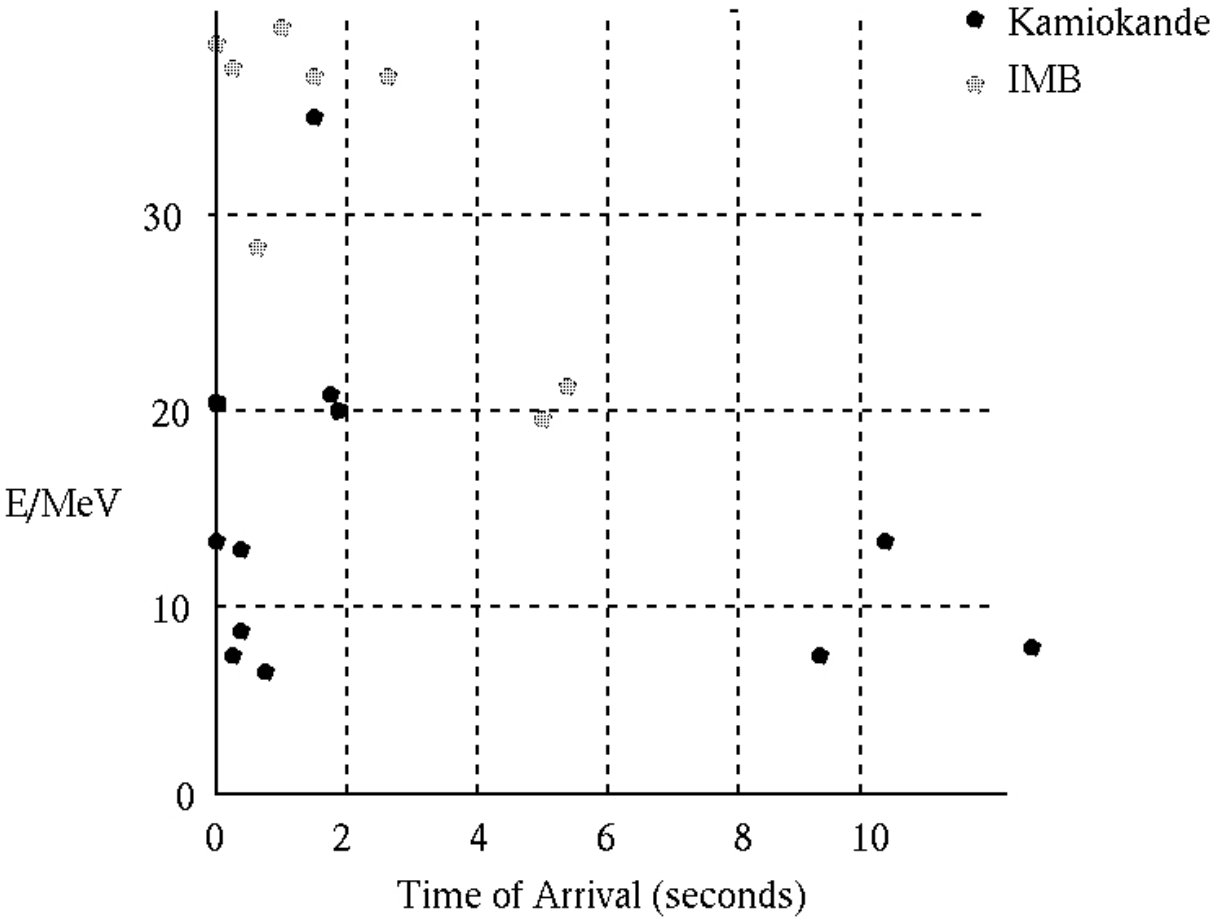

Figure C.12: The IMB and Kamioka neutrino events from SN1987A.

***F.2 Supernova Neutrinos: Current and Future Observatories.*** The most important consideration in preparing for the next galactic supernova is to appreciate that an opportunity presents itself once in a lifetime: the best estimate of the galactic supernova frequency is 1/30y. There are two consequences. First, we need to be prepared at all times with detectors capable of measuring the spectrum, including the flavor and time development, of the next supernova, which could occur at any time. The minimum requirements are detectors operating in both charge-current modes as well as some scheme for measuring neutral current interactions. Second, most (though not all) supernova observatories are likely to be detectors with additional physics goals, such as proton decay or solar neutrinos. Thus a theme of this field is the "supernova watch" – the notion that we should try to coordinate the world's program of underground neutrino detectors to guarantee that the necessary supernova capability is always in place.

Directional sensitivity is also important, as this could help optical observers locate the supernova quickly. Detectors exploiting elastic scattering off electrons provide some directional information. Another possibility, if several detectors are able to measure enough events to determine the turn-on time of the neutrino pulse to millisecond accuracy (or if the deleptonization burst could be used as a clock), is to determine the direction by triangulation.

| Detector type | Target reaction | Spectrum | Timing | Pointing | Flavor |
|---|---|---|---|---|---|
| Scintillator | H (CC) | Y | Y | N | Anti-$\nu_e$ |
|  | C (15.11 MeV $\gamma$) | N | Y | N | All flavors |
| Water Cerenkov | H$_2$O (CC,ES) | Y | Y | Y | Anti-$\nu_e$ |
| Heavy water | D (NC) | N | Y | N | All flavors |
|  | D (CC) | Y | Y | Y | $\nu_e$, anti-$\nu_e$ |
| Long-string water Cerenkov | H (CC) | N | Y | N | Anti-$\nu_e$ |
| Liquid Ar | Ar (CC,ES) | Y | Y | Y | $\nu_e$ |
| High-Z neutron | NaCl,Pb,Fe (CC) | N | Y | N | All flavors |
| Radiochemical | $^{37}$Cl,$^{127}$I,$^{71}$Ga (CC) | N | N | N | $\nu_e$ |

Table IF.1  Supernova neutrino detectors and their principal capabilities.

The table shows the supernova neutrino attributes of the major detector types now in operation. Water detectors are dominated by the charge current (CC) reactions of electron antineutrinos on protons, though the elastic scattering (ES) of neutrinos of any flavor is important because it provides directional sensitivity. Scintillator detectors are also dominated by the CC reaction off protons, though the neutral current (NC) reaction exciting the 15.11 MeV level in $^{12}$C is interesting as a high-threshold process dominantly sensitive to heavy-flavor supernova neutrinos. The signal is the subsequent 15.11 MeV $\gamma$-ray. The CC reactions off deuterium in heavy water detectors produce either an electron or a positron plus two neutrons, the latter a triple-coincidence for antineutrino reactions. The gentle backward peaking of the CC signal provides some directional sensitivity. The deuterium NC signal is a single neutron. Liquid argon detectors can exploit the $^{40}$Ar($\nu_e$,e$^-$)$^{40}$K reaction as well as ES. High-Z targets like Pb have enhanced charged current cross sections for $\nu_e$s because of the Coulomb effects on the outgoing electron and because of the neutron excess. The signal is excitation of states in the continuum that then decay by neutron emission. The NC channel also produces spallation neutrons. Finally, various radiochemical detectors like chlorine, iodine, and gallium are good supernova $\nu_e$ detectors. The specifications for various operating or soon-to-be-operating detectors are given below, along with the events expected from a supernova at the galactic center.

| Detector | Type | Mass (kton) | Location | Events @ 8.5 kpc | Status |
|---|---|---|---|---|---|
| Super-Kamiokande | H2O | 32 | Japan | 8000 | running |
| SNO | D$_2$O/ H$_2$O | 1/1.4 | Canada | 450/300 | running |
| LVD | scintillator | 0.7 | Italy | 170 | online |
| AMANDA | long string | ~ 0.4/pmt | Antarctica |  | running |
| Baksan | scintillator | 0.33 | Russia | 50 | running |
| Borexino | scintillator | 0.3 | Italy | 100 | ~late 2003 |
| KamLAND | scintillator | 1.0 | Japan | 300 | running |
| Icarus | liquid Ar | 2.4 | Italy | 200 | 2005 |

Table IF.2 Sensitivity of operating detectors to supernova neutrinos.

Several new detectors with supernova neutrino sensitivity have been proposed. Most have other primary physics goals, but a few are dedicated supernova neutrino observatories. In most cases, depth is not a major issue for detecting the early burst neutrinos, but detector thresholds and depth can influence how long a detector can follow the neutrino light curve. Detector mass is also crucial.

*Lead-based dedicated observatories:* Most of the dedicated observatories are based on the observation that neutrino-induced spallation neutrons could be detected in relatively inexpensive and easily maintained large detectors. These generally exploit the rapid increase in CC cross sections with Z: a heavy nuclear target often has a CC $\nu_e$ rate that is not too much smaller than the NC rate, despite the three flavors of neutrinos and antineutrinos contributing to the latter. Lead is the choice of detectors like OMNIS because it is a plentiful, inexpensive high-Z material.

Lead-based detectors are sensitive to the higher energy portion of the supernova neutrino flux capable of exciting the nucleus above the neutron-emission threshold. This threshold, in the absence of oscillations, thus favors the hotter heavy-flavor neutrinos. The CC reaction $^{208}$Pb$(\nu_e,e^-)^{208}$Bi$^*$ can produce either single or two-neutron emission, with the ratio sensitive to the temperature of the $\nu_e$ spectrum. The NC reaction $^{208}$Pb$(\nu_x,\nu_x)^{208}$Pb$^*$ is observable via the n, 2n, and γ emission channels.

One interesting form of lead is lead perchlorate. A water-based solution is a good neutron moderator, and the presence of Cl provides an attractive (n,γ) signal, 8.6 MeV in γ-rays. Because lead perchlorate is transparent, it has been considered for Cerenkov detectors, which would provide spectral sensitivity.

The current OMNIS design uses two ktons of lead-slab detectors and 0.5 ktons of lead perclorate, which would produce several thousand events following a supernova at the galactic center. The lead-slab design alternates vertical slabs with planes of scintillator for detecting the neutrons from the lead. OMNIS will have intrinsic timing of better than 100 μs, so that fine structure in the neutrino burst (such as the neutronization burst or black hole formation) can be resolved.

Construction of the lead slab modules for OMNIS could begin immediately, although some research to optimize the detector is still underway. Monte Carlo simulations have been used to determine the optimum slab thickness, balancing neutron detection efficiency against cost, and preliminary engineering designs of these modules have been completed. Considerable R&D must be done to optimize the lead perchlorate detector. The experimenters plan to site the lead-slab modules in the WIPP laboratory and the lead perchlorate modules in NUSEL.

Cosmic-ray-induced neutrons are of some concern, so greater depth is helpful in minimizing this background. However, extreme depths are not needed. The safety of lead perchlorate in an underground laboratory must be studied.

*Radiochemical quasi-real-time detectors:* CC radiochemical detectors may be of interest for supernova neutrino detection because of their low cost and ease of maintenance, and because of their sensitive to a single neutrino flavor, $\nu_e$. Though one would normally favor an active, real-time detector, our best current $\nu_e$ detector SNO is of moderate size and may not run for many more years. An accurate integral measurement of the neutrino fluence in this channel is important because, when combined with the results from water Cerenkov detectors, the net lepton number radiated by the protoneutron star can be determined.

As GNO and SAGE are small detectors and chlorine has been dismantled, the most attractive proposal is one based on the reaction $^{127}$I$(\nu_e,e^-)^{127}$Xe. This detector has been suggested as a modern version of the chlorine detector, but with a considerably larger cross section and with an automated system for flushing the noble gas Xe out of the xenon. Unlike the chlorine experiment, the Xe extraction (99% of the Xe could be removed in a 15 minute purging with helium) would be done twice daily and counted ($\tau_{1/2} \sim 36.3$ d), so that the experiment could also look for solar neutrino day/night effects. This would also mean that the detector would be kept free of Xe, awaiting a galactic supernova, the detector's main purpose. The proposed 3-kiloton detector, with the iodine deployed as a NaI solution, would record approximately 700 $\nu_e$ events from a supernova occurring at the galactic center. The detector could be maintained by a single investigator working part time, and would be active 100% of the time.

A more sophisticated version of this experiment has been suggested where the purging is triggered by an observation of the Cerenkov light produced by the emitted electron. The detector would be arranged as an array of upright cylindrical modules, roughly one meter in diameter, separately flushed, with each viewed by surrounding phototubes. The modular geometry reduces the purging time to ~ 1 minute. Thus a Cerenkov signal could be verified by extraction and counting of a Xe atom – a technique similar to the EXO double beta decay scheme. The timing capability of this detector for supernova neutrino detection would be determined by the electron detection.

*Water Cerenkov megadetectors:* Large water Cerenkov detectors such as UNO will have excellent supernova neutrino detection capabilities, including very high statistics and excellent directional sensitivity. Their mass would allow detection of supernovae in Andromeda (though these are infrequent) and observation of the relic supernova neutrinos produced over the entire history of star formation – capabilities beyond the reach of most other proposed detectors. These issues are discussed in the proton decay/megadetector chapter.

*Liquid argon detectors:* The most important supernova neutrino channel in a liquid argon detector is the CC reaction $^{40}\text{Ar}(\nu_e,e^-)^{40}\text{K}^*$. The cross section is known accurately. (The model-dependent Gamow-Teller contribution has been determined by the same technique used for $^{37}\text{Cl}$, observation of the β-delayed protons following the decay of $^{40}\text{Ti}$, the isospin analog of $^{40}\text{Ar}$.)

A liquid argon drift chamber such as Icarus can detect the primary electron track as well as the secondary tracks from the $^{40}\text{K}^*$ γ decay, with approximately a 5 MeV threshold. Elastic scattering events are also visible. The 2.4 kton Icarus detector is to be sited at Gran Sasso. Icarus would record ~ 200 events for a supernova occurring at the galactic center. The proposed LANNDD detector (Liquid Argon Neutrino and Nucleon Decay Detector), with 70 kilotons of liquid argon in a single chamber, would record thousands of events. The signal would be strongly enhanced by flavor oscillations. The LANNDD detector is a second possibility for detecting the relic supernova neutrinos.

***F.3 Supernova Neutrinos: Facility Requirements.*** As the detectors mentioned above are discussed elsewhere in this Science Book, we focus here on common, general requirements for supernova neutrino physics within NUSEL.

Detector up-time must be virtually 100%. Therefore it is highly desirable for experimenters to have 24/7 access to the detectors, enabling them to respond immediately to any difficulties. Excellent network conductivity is essential for the same reason.

Relative timing of events among different neutrino experiments is crucial for event location by triangulation. Precise timing within each experiment is important in correlating sharp features (like the deleptonization burst or black hole formation) with other detectors or with gravitational wave detectors. During the peak of the burst, the average time between consecutive events in UNO would be less than 100 μs. Therefore one must maintain this kind of timing accuracy within the underground laboratory.

Though discussed elsewhere in the Science Book, we note that almost all massive detectors require specialized excavations. In the case of UNO, the NUSEL-Homestake plan calls for a dedicated hoist and shaft (the Yates) and underground transport of the ~ two megatons of excavated rock into the open cut, using a specially constructed conveyor. The iodine detector is ideal for a vertical crater excavation of a type commonly done at Homestake, with the cavity afterward flooded by water to isolate the array of 1m cylindrical modules from the rock walls. LANNDD and the lead perchlorate detectors involve hazardous materials that will require careful handling; LANNDD should be isolated and separately vented.

*F.4 Supernova Neutrinos: Summary.* Supernova neutrino detection has the potential to provide important new information on neutrino properties (including the effects of neutrino background on the MSW potential), the supernova mechanism, the properties of superdense nuclear matter, the conditions under which r-process, ν-process, and other nucleosynthesis occurs, and black hole formation. Neutrino detection is important to gravity wave and optical observations, verifying that a galactic supernova has occurred (crucial if it is optically obscured) and providing an early warning to optical astronomers that an explosion is imminent. The most massive detectors may detect supernovae in our neighboring galaxy Andromeda and the relic supernova background neutrinos that will constrain the history of star formation.

The general detector requirements include very large masses and flavor sensitivity (requiring multiple channels or multiple detectors). Some detectors must provide good energy resolution, low thresholds, good timing, and pointing capability. It is crucial that detectors be capable of operating for very long times, or that the community arranges for a succession of detectors to provide continuous coverage. As the galactic supernova rate is about 1/30 y, missing the next event would be tragic.

The depth requirements are generally modest, though the most challenging issues (measuring the relic neutrinos, following the neutrino light curve out to very long times) require very quite detectors. Most supernova detectors will serve several purposes, with the ancillary uses (proton decay, solar or atmospheric neutrino detection) often determining the needed depth. Detector stability and reliability, and 24/7 access to guarantee that reliability, are essential.

***G.1 Nuclear Astrophysics.*** Experimental nuclear astrophysics is the study and measurement of nuclear processes in astrophysics, especially those driving both the steady evolution and explosions of stellar systems. The importance of this field is illustrated by the solar neutrino problem: precision nuclear astrophysics measurements were combined with careful modeling, yielding a standard model of main-sequence stellar evolution that accurately predicted solar neutrino fluxes. The credibility of these predictions ultimately led the community to recognize that fundamental new neutrino physics was responsible for the deficits in counting rates.

New neutrino discoveries are part of a broader revolution underway in astrophysics, one driven by marvelous new instrumentation: new-technology earth- and space-based telescopes, underground neutrino observatories, the Compton Gamma Ray Observatory, the Chandra X-Ray Telescope, the Space Infrared Telescope Facility, WMAP, and new large-area cosmic ray detectors. These instruments are now providing information of unprecedented detail on objects within and outside our galaxy. These data often reflect the underlying nuclear and atomic microphysics governing these objects. The nuclear astrophysicist tries to determine that microphysics from laboratory experiments, thereby enabling modelers to deduce the properties of the astrophysical systems being studied.

One major goal of nuclear astrophysics is to understand hydrostatic nuclear burning through the different phases of stellar evolution, determining the lifespans of the stars and the initial conditions at the onset of stellar explosions. Another is the understanding of nuclear processes far from nuclear stability, which characterize nucleosynthesis in novae, X-ray bursts, and supernovae. These also determine the elemental and isotopic abundances observed in stellar atmospheres and in the meteoritic inclusions that have condensed in stellar winds, or detected with gamma ray observatories.

The laboratory measurement of nuclear processes in stellar explosions requires the development of a new generation of radioactive beam facilities to produce exotic short-lived nuclear species and to observe the reactions of these nuclei on the split-second timescales of stellar explosions. This physics is being and will be pursued above ground with large accelerators at facilities like ISAC II, RIA, and Riken. Different techniques are needed for the study of reactions important in the quiescent periods of stellar evolution. A new generation of high intensity, low-energy accelerators for stable beams are needed in order to simulate within human timescales the processes that occur in nature over stellar lifetimes.

While more than thirty years of intense experimental study have helped define the major features of nuclear burning during hydrostatic stellar evolution, so far only two fusion reactions have been studied at the relevant stellar energies. One of these measurements was made with the first underground accelerator experiment, LUNA I conducted at Gran Sasso. Many other rates crucial to stellar modeling have been deduced indirectly from extrapolations of higher energy laboratory data. These extrapolations can be off by orders of magnitude in cases where the underlying nuclear structure is poorly constrained. Furthermore, to obtain empirical information on the effects of stellar plasmas on fusion rates, experimenters must make measurements at very low energies where the screening effects of atomic electrons become most important.

The associated uncertainties complicate the modeling of important phases of stellar evolution, such as the CNO reactions within massive main sequence stars and the reactions on heavier species that govern the later stages of nuclear burning (e.g., the production of the γ-ray source $^{26}$Al in massive stars and novae). Descriptions of the red giant and the asymptotic giant helium- and carbon-burning phases, where the slow neutron capture (s-) process responsible for the origin of more than half of the elements likely occurs, are also limited by nuclear physics uncertainties. Many of these uncertainties can be reduced or eliminated if a suitable low-background underground accelerator facility were established. The problems that could be effectively addressed with such a facility include:

*$^{12}C(\alpha,\gamma)^{16}O$:* The rate of this reaction determines the $^{12}C/^{16}O$ ratio produced by helium burning. This in turn determines the masses within the carbon and oxygen zones of a Type II supernova progenitor, thereby influencing the outcome of the core collapse (whether a neutron star or black hole is formed) and the explosive nucleosynthesis associated with the passage of the shock wave. While calculation of the reaction rate for conditions typical of helium burning requires knowledge of the cross section near $E_{cm} \sim 300$ keV, data exist only above 1 MeV. The extrapolation to lower energies is complicated by two states just below the $^{12}C + \alpha$ threshold. Present data and extrapolations fall far short of the 10% precision at 300 keV necessary to meaningfully constrain astrophysical calculations.

Cosmic-ray background in γ-ray detectors, beam-induced backgrounds (which diminish at the lower energies), and low beam currents have limited past experiments. Several different techniques have been applied to this reaction with comparable levels of success. Two general strategies are under consideration for the future: an intense $^{12}C$ beam in conjunction with a $^{4}He$ gas jet target, γ-ray detectors, and a recoil separator, or a $^{12}C$ target, an intense $^{4}He$ beam, and γ-ray detectors. In either case the γ detection requires a large-solid-angle array of either scintillator or high-purity germanium detectors. A high-current accelerator facility located deep underground would clearly address the factors that limited previous experiments.

*S-process neutron sources, $^{13}C(\alpha,n)^{16}O$ and $^{22}Ne(\alpha,n)^{25}Mg$:* In intermediate-mass AGB stars the $^{13}C(\alpha,n)$ reaction is thought to be the main s-process neutron source, operating at temperatures $\sim 1 \times 10^8$ K. The $^{22}Ne(\alpha,n)$ reaction operates at somewhat higher temperature $\sim (2-3) \times 10^8$ K and is the dominant s-process neutron source in more massive stars. The rates of these reactions determine the neutron fluence during the s-process, and are important to arguments that identify site or sites for the s-process. S-process nuclei contribute significantly to the production of elements between Fe and the transuranics: the origin of the heavy elements was one of the urgent scientific questions recently identified in the NRC Report on "Quarks and the Cosmos."

Although measurements of the $^{13}C(\alpha,n)$ cross section have reached down to $E_{cm} \sim 300$ keV, the extrapolation to the astrophysically important region of 150-200 keV is hampered by poorly constrained subthreshold resonances, yielding a reaction rate uncertain by an order of magnitude. The $^{22}Ne(\alpha,n)$ reaction is thought to be dominated by narrow resonances. The rate is highly uncertain due to the possibility of unobserved weak resonances just above threshold: according to the NACRE compilation the uncertainty at $T \sim 2 \times 10^8$ K exceeds two orders of magnitude. In previous measurements the neutrons produced in these reactions were moderated in polyethylene and then detected, at thermal or epithermal energies, with $^3$He-filled proportional counters. This method yields a high neutron detection efficiency (20–50%) and is insensitive to γ rays or charged particles, but is affected by any background neutron source, including those produced by cosmic rays. While previous experiments exploited active and passive shielding extensively, all were limited in sensitivity by cosmic ray neutrons. A repetition of these experiments in an underground facility thus should yield greatly improved results. The $^{13}C(\alpha,n)$ cross section measurements could be extended to lower energies, while the lower-energy resonances affecting $^{22}Ne(\alpha,n)$ could either be measured or more tightly constrained. In addition, the development of $^3$He-filled proportional counters free of α emitters on the inside walls might be helpful in reaching ultimate sensitivities.

*Light nucleus reactions $^3He(\alpha,\gamma)^7Be$ and $^2H(\alpha,\gamma)^6Li$:* The $^3He(\alpha,\gamma)$ reaction governs the pp chain branching to the ppII and ppIII cycles, leading to the production of $^7Be$ and high energy $^8B$ solar neutrinos in the sun. The Q-value for the $^3He(\alpha,\gamma)$ reaction is 1586 keV and the γ-ray spectrum consists of lines at ~ 1.6, 1.2, and 0.43 MeV. The energy of the lowest data point so far measured is $E_{cm} \sim 107$ keV, while the most effective energy (the Gamow peak) for this reaction in the sun is ~ 22.9 ± 12.8 keV. The $^2H(\alpha,\gamma)^6Li$ reaction is responsible for $^6Li$ production in the big bang. Laboratory measurement extend only to 600 keV, while the energy range relevant for big bang nucleosynthesis is ~ 50–200 keV. An accurate cross section at big-bang energies will be important if $^6$Li is observed in future studies of abundances in metal-poor halo stars. (The primary process by which galactic $^6$Li is thought to be produced is cosmic-ray spallation, a secondary process that becomes much more effective at late times, when the interstellar medium is richer in C and O.) Non-standard big-bang models, such as those postulating inhomogeneities, tend to produce more $^6$Li, and thus can be constrained by observational limits on big-bang $^6$Li nucleosynthesis, if the production cross section is known accurately.

*Hydrogen-burning scenarios:* There are significant resonant and nonresonant cross section uncertainties at low energies for several hydrogen-burning reactions occurring in the CNO, NeNa, and MgAl cycles (or chains). These reactions are important in both main sequence and more evolved stars. The required studies include (p,γ) and some (p,α) reactions on isotopes of N, O, Ne, Na, Mg, and Al. The CNO cycle operates in the sun, producing 1.5% of solar energy as well as neutrino fluxes that, if measured, would directly determine the metallicity of the solar core. A comprehensive understanding of the CNO, NeNa, and MgAl cycles is important in interpreting isotopic abundance anomalies ($^{17}$O, $^{22}$Ne, and $^{26}$Al) reflecting nucleosynthesis prior to solar system formation.

***G.2 Nuclear Astrophysics: Readiness of a Next-Generation Facility.*** The advancements in precision astrophysics put continual pressure on the nuclear astrophysics community to improve cross section measurements, a few of which were discussed above. The one underground accelerator now operating, LUNA II at Gran Sasso, is limited in a number of ways: most of the recommendations for improving LUNA I were not implemented because of space and other constraints at Gran Sasso. This has restricted LUNA's impact to pp-chain reactions. The study of stellar He burning processes requires substantially higher beam energies than those available at LUNA. At NUSEL-Homestake the US nuclear astrophysics community would combine the passive shielding provided by depth with the latest detector technology for active event identification and background rejection. New, commercially available accelerator technology will provide a one to two orders of magnitude improvement in the beam intensity. The Homestake facility would be developed in three stages:

- The first-stage development will concentrate on γ ray and neutron detector development (moderated neutron detector, differential large scintillation detector) to identify optimal strategies for lowering the intrinsic background in facility detectors. This would include selecting materials for chamber, target, and shields.
- In second-stage development a 0.6-1.0 MeV 1-10 mA light-ion (p and α) accelerator will be set up for a variety of experiments and for target development (focused on reducing beam-induced background). As every reaction produces unique background problems, it is likely that each reaction will require 6 months to a year of effort.
- In the third stage, and in parallel with the efforts described above, a heavy ion linear accelerator with recoil mass separator and gas jet targets will be developed for reaction measurements requiring reverse kinematics. Note that the study of stellar He-burning processes requires substantially higher beam energies than those available at LUNA II. This stage will include the development of a high-intensity ECR source. Measurements with inverse kinematics generally have lower beam-induced backgrounds. Stage three developments are crucial to the measurement of the $^{12}$C(α,γ)$^{16}$O reaction as well as for selected proton-induced reactions.

A final proposal for the NUSEL underground accelerator facility will require about three years of effort. A US collaboration has been formed to complete this work. The study of underground detection techniques will be performed in collaboration with the LUNA group, and the development of active shielding and event-tracking techniques will be carried out with at several existing US low-energy accelerator facilities. While a final proposal will require two to three additional years of effort, NUSEL plans call for finished space to be available on the 4850 ft level no earlier than FY07. Thus it is anticipated that **an underground accelerator will be one** **of the earliest facilities installed at NUSEL.**

***G.3 Nuclear Astrophysics: Facility Requirements.*** The following are the basic requirements for completing the three-stage program:
- Stage I: Space requirement 5m ×10m × 3m high; 50 kW of electrical power; ventilation and chilled water.
- Stage II: Space requirement 15m × 10m × 5m high; 200 kW of electrical power; ventilation and chilled water; an overhead crane.
- Stage III: Space requirement 15m × 30m × 5m height; 900 kW of electrical power; ventilation and chilled water; an overhead crane.

The desired location within NUSEL-Homestake is 4850 ft, as this provides ample shielding. The technical aspects of the proposal are currently being developed, with a core group of investigators meeting regularly. All three of the stages will need strong technical and infrastructure support from the participating accelerator laboratories.

***G.4 Nuclear Astrophysics: Summary.*** There remain significant uncertainties in the reaction rates for the pp chain, affecting the precision of standard solar model neutrino flux predictions. The uncertainties in many charge-particle reactions that drive late stellar evolution, including the helium-burning reaction $^{12}C(\alpha,\gamma)^{16}O$ crucial to the evolution of Type II supernova progenitors, are much larger. A better understanding of these reactions would be of great help to astrophysics, with a more quantitative model of the SNIa "standard candle" being one possible outcome. Charged particle reactions are thought to provide the neutron fluences necessary to s-process nucleosynthesis, which is responsible for about half the heavy elements between Fe and Pb.

The installation of an accelerator laboratory deep underground, utilizing recent advances in accelerator, detector, and data-handling technologies, could lead to major improvements in our understanding of the nuclear reactions that power the stars and drive explosive astrophysical environments. Such a facility would have no counterpart elsewhere in the world.

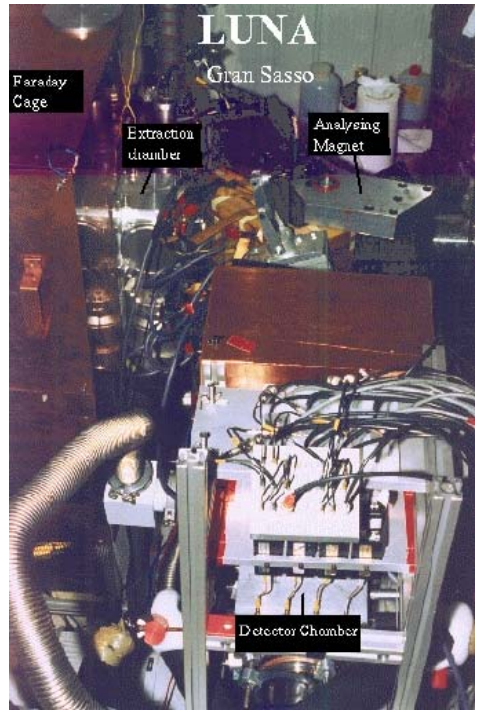
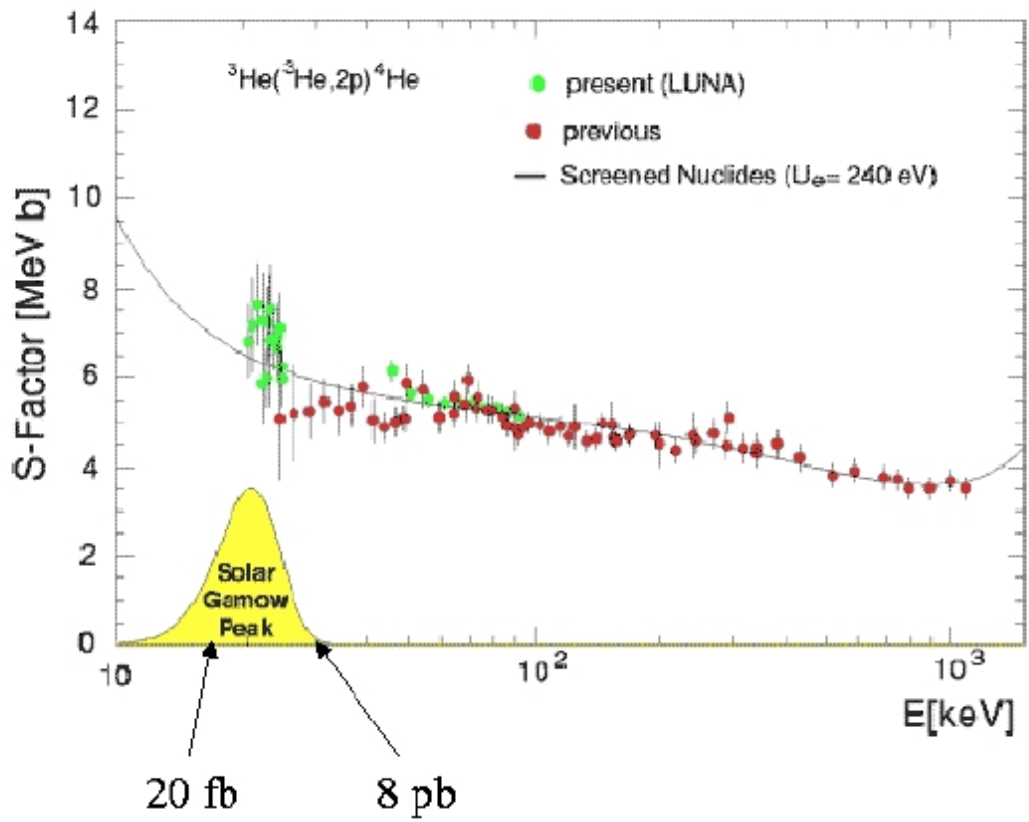

Figure C.13: The LUNA II facility at Gran Sasso and a pp chain Gamow peak measurement.

**Megadetector Physics:** The capacity to excavate a stable, deep megacavity for a next-generation proton decay experiment is one of the strongest arguments for the NUSEL-Homestake site. Homestake provides access to the extremely stable Yates rock formation through the Yates shaft, which could be dedicated to the excavation and construction of a detector like UNO. As noted in the Overview, the integrity of Homestake rock for such an excavation has been established: there is a long history of studies of large-span excavations at Homestake. The proposed site has already been studied thoroughly and is directly accessible, available for additional coring and rock mechanics studies. As the Yates mining capacity is one megaton/year, a UNO-sized excavation would require less than two years. The rock would be disposed on site, through a conveyor built on the 600 ft level that would transport the rock from the Yates shaft to the Open Cut. The excess pumping capacity (through the Ross pump column) could drain UNO in about 90 days. It is unclear to us that whether any other site could meet the excavation, stability, rock disposal, and pumping requirements of UNO.

The driver for a megadetector is the physics, particularly the opportunity to do a next-generation nucleon decay experiment while at the same time meeting the needs of the very-long-baseline neutrino oscillation program aimed at detecting CP violation. Furthermore there is a long list of other important physics – atmospheric, solar, and supernova neutrino measurements – that would be advanced by the construction of the right detector at the right depth.

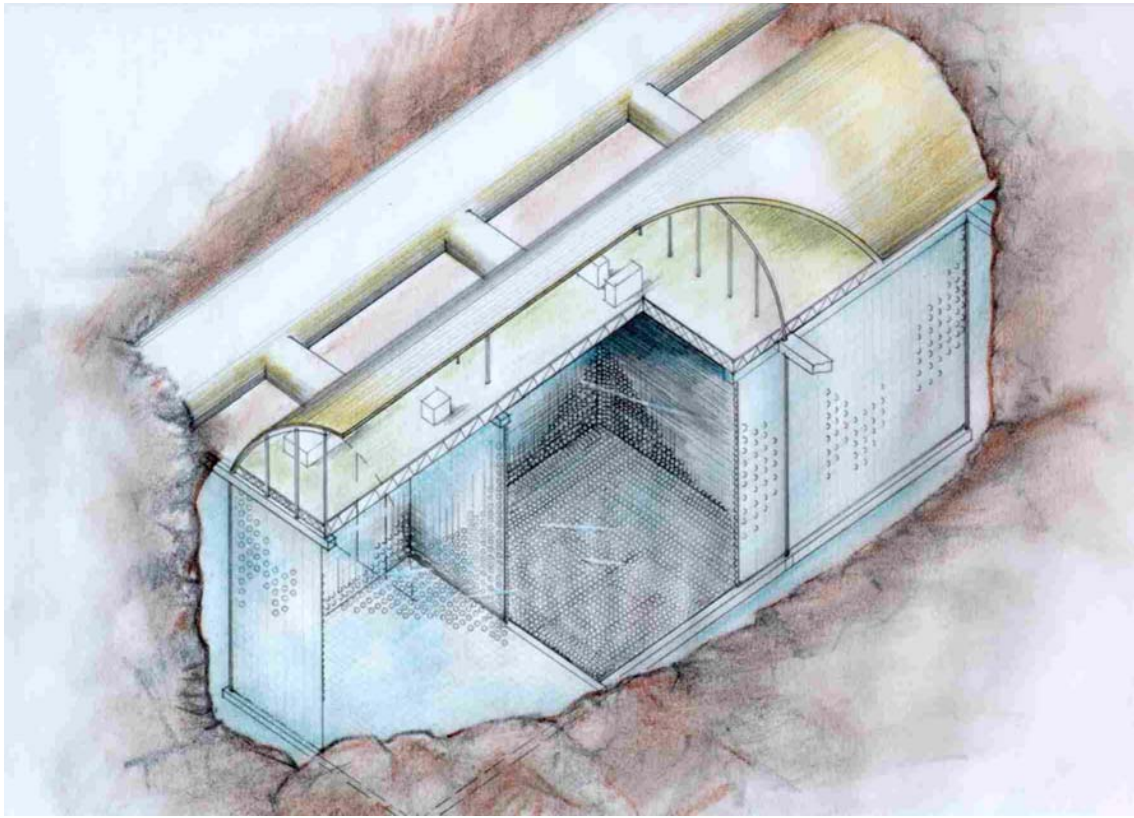

Figure C.14: Artist's rendering of UNO, a planned 0.5 megaton water Cerenkov detector for proton decay, long-baseline neutrino physics, and atmospheric and supernova physics.

***H.1 Nucleon Decay: Importance of the Science.*** While current experiments show that the proton lifetime exceeds $\sim 10^{33}$ years, its ultimate stability has been questioned since the early 1970s. At that time theorists began the development of Grand Unified Theories (GUTs) to extend the standard model, an effort to unify three fundamental forces, the strong, electromagnetic, and weak, operating between the quarks and leptons. The dramatic meeting of the strengths of the three forces at $\sim 10^{16}$ GeV in supersymmetric GUTs greatly encouraged this effort. It is also consistent with the pattern of masses that is emerging from recent neutrino discoveries, which implies a right-handed Majorana seesaw mass $\sim 0.3 \times 10^{15}$ GeV, close to the GUT scale. But perhaps the most generic and crucial prediction of GUTs is that nucleons will decay into leptonic matter such as a positron and a meson, revealing quark-lepton unity. That is, the matter comprising our universe is not absolutely stable.

Certain early versions of GUTs based on the gauge group SU(5), including minimal supersymmetric SU(5) models, predict relatively short lifetimes for the proton, such as branches to the $e^+\pi^0$ and $\nu K^+$ modes with partial lifetimes of $10^{28}$ to $10^{32}$ years. These predictions have been excluded by the IMB, Kamiokande, and Super-Kamiokande experiments. A class of well-motivated GUT theories based on the gauge group SO(10) and supersymmetry are now the focus of interest. These successfully describe the masses and mixings of all the quarks and leptons, including neutrinos, and also account for the origin of the excess of matter over antimatter through the process of leptogenesis, as discussed in the neutrino section of the Science Book. These theories place a conservative upper bound on the proton lifetime that is within a factor of ten of the current lower limit. This makes the discovery potential for proton decay in a next-generation experiment – one that exceeds Super-Kamiokande by at least an order of magnitude – rather high.

From a broader perspective, the discovery of proton decay would provide physicists with a unique window on physics at truly short distances, less than $10^{-30}$ cm. Such distances cannot be probed by any other means: current high-energy accelerator experiments probe to $10^{-17}$ cm. The discovery of proton decay would provide crucial support to the notion of unification and to ideas like leptogenesis. Because the predictions of a well-motivated class of theories are not far above current limits, the need for next-generation proton decay experiments is compelling. Those theories include supersymmetric SO(10), flipped SU(5), and string-derived SU(2)×SU(2)×SU(4) models. The modes most favored are $\nu K^+$ and $e^+\pi^0$ with lifetimes less than $\sim 2 \times 10^{34}$ and $10^{35}$, respectively. Should proton decay be discovered in these channels, searches for related modes, including $\mu^+\pi^0$ and $\mu^+ K^0$, could provide additional constraints on GUTs.

The "classical" proton decay mode $p \rightarrow e^+\pi^0$ can be detected efficiently with low background. The present best limit on this mode, $\tau/\beta > 5.7 \times 10^{33}$ y (90% c.l.), comes from the 92 kton-y exposure provided by Super-Kamiokande. The detection efficiency of 44% is dominated by final-state $\pi^0$ absorption and by charge-exchange for protons decaying in $^{16}$O. The expected background is 2.2 events/Mton-y.

The mode $p \rightarrow \nu K^+$ is more difficult to detect in water Cerenkov detectors due to the unobserved neutrino. The present limit in Super-Kamiokande is the result of combining several channels, the most sensitive of which involves observation of the $K^+$ through $K^+ \rightarrow \mu^+\nu$ in coincidence with a $\gamma$ ray from the deexcitation of $^{15}$N (produced because a proton in $^{16}$O decayed). Monte Carlo studies suggest that this mode should remain background free for the foreseeable future. The present limit on this mode is $\tau/\beta > 2.0 \times 10^{33}$ y (90% c.l.).

***H.2 Nucleon Decay: Readiness of Next-Generation Experiments.*** Both the lifetime and decay modes of the nucleon are unknown *a priori*. Thus the lifetime could range from just above present limits to values many times larger. It follows that it is appropriate to measure progress logarithmically, motivating order-of-magnitude improvements in detector size. The efficiency for detecting the $e^+\pi^0$ mode is dominated by pion absorption effects in the nucleus, and cannot be improved significantly. An order-of-magnitude improvement in this mode can be achieved by running Super-Kamiokande for an additional 30-40 years, or by constructing an order-of-magnitude larger experiment.

The favored decay modes of the nucleon are also unknown *a priori*, and the signatures of possible decay modes are generally quite distinct. Thus detectors have to be sufficiently versatile to achieve good sensitive to most or all of the kinematically allowed channels. The need for long running times requires a detector technology that is reliable, so that repairs are infrequent. The long running time also makes it important for the proposed detector to be able to address other physics questions, while the proton decay search is ongoing.

A variety of proton-decay detector technologies have been discussed:

*Water Cerenkov detectors:* Water Cerenkov detectors enjoy several advantages over alternatives. The medium is inexpensive, the technology is mature and well tested, and the ability to deploy the technology on very large scales is demonstrated, due to the long experience with Super-Kamiokande. Currently there is an international collaboration exploring a next-generation version of Super-Kamiokande. The Japanese project is Hyper-Kamiokande, while the US version is UNO (Underground Nucleon Decay and Neutrino Observatory). There have been discussions in Europe of mounting such a detector in the Frejus tunnel. The total mass of the proposed UNO detector is 650 kilotons. A veto shield makes up the outer 2.5m of the detector; the fiducial volume cut is 2m inside of that boundary. The active volume is 440 kilotons, 20 times larger than the Super-Kamiokande detector. The structure is "rural mailbox" style, a rectangle 60m × 60m ×180m, divided into three cubic sections. The inner section has higher phototube coverage, 40%, while the coverage over the remainder of the detector is 10%. The inner section is important for solar neutrino physics and also for detecting the 6 MeV γ accompanying p→$K^+$ν for a proton bound in $^{16}$O.

Detailed Monte Carlo studies, including full reconstruction of simulated events, indicate that water megadetectors could reach the goal of an order-of-magnitude improvement on anticipated nucleon decay limits from Super-Kamiokande. With sufficient exposure, clear discovery of nucleon decay into $e^+\pi^0$ would be possible even at lifetimes ~ few × $10^{35}$ y, if selection criteria are tightened to reduce background counts. For example, with a detection efficiency of 18%, the expected background is only 0.15 events/Mton-y, ensuring a signal/noise of 4/1 even for a proton lifetime of $10^{35}$ y. Any of the proposed megadetectors would provide a decisive test of supersymmetric SO(10) GUTs by reaching a sensitivity of ~ few × $10^{34}$ y for the ν$K^+$ mode.

HEPAP recently classified the science potential of such a next-generation detector as **absolutely central**. The report noted that the UNO technology is well tested and **may not require significant R&D**. However, it noted that a rigorous professional civil and mechanical engineering design of UNO depends on the choice of final site. The report concluded that the detector could be completed within 10 years of ground breaking.

*Modular water Cerenkov detectors:* An alternative design that has been suggested for a megaton water Cerenkov experiment is an array of ten 100 kton cylindrical detectors, 50m in diameter and 50m in height. Studies indicate that these could be placed as deep as 6950 ft at Homestake, on the circumference of a 250m circle centered on the No. 6 shaft. This location accesses another section of the Yates formation, rock known for its high integrity. (This is the same area of the formation where the 7400 ft main laboratory development would be done.) The construction of such chambers has been rather thoroughly investigated: the rock is removed through a tunnel near the cylinder base. When construction is completed, the connection between the chamber and this tunnel is closed, and the chamber is finished with a concrete liner and with an inner water-tight plastic liner.

The proposal excavation plan utilizes the No. 6 Winze and Ross for excavation. In current NUSEL planning this would not be available. However excavation through the No. 4 Winze and Yates is still quite practical. While there is additional transport of rock, the cost is not large. Nor does the use of the smaller No. 4 Winze limit the rate at which the excavation could proceed. The argument for the deeper location is that the background needs of new detectors have been underestimated historically. In the case of Homestake, the placement of the detectors at 6950 ft is feasible, in the rock mechanics/engineering sense, and the fractional increase in the cost is small. The gain is more than an order of magnitude additional suppression in the background from cosmic rays.

*Liquid argon detectors:* It has been noted elsewhere in the Science Book (see, for example, the supernova neutrino discussion) that the feasibility of very large mass liquid argon detectors is being explored. Such detectors have advantages over water detectors. The sensitivity to the supersymmetry-favored p→$K^+\nu$ mode is enhanced by about an order of magnitude due to the extraordinary bubble-chamber-like pattern recognition capabilities. Thus the three-kton ICARUS detector is projected to be as sensitive to this mode as Super-Kamiokande. Due to the pattern recognition quality, a single observed event could be convincing evidence of proton decay. The proposed massive successor to ICARUS, the 70-kton LANNDD detector, has been discussed previously. The drawback of liquid argon detectors is the smaller mass, which limits the sensitivity to modes like $e^+\pi^0$ that are readily observed in water.

Very-large-scale liquid argon detectors are not yet a demonstrated technology. Thus there is **significant R&D remaining to be done**. A serious engineering issue is ensuring safety if detectors like LANNDD are placed in a multipurpose underground laboratory.

*Scintillation detectors:* The feasibility of a very-large-mass scintillation detector for proton decay has been discussed. The KamLAND experiment is considered a prototype for such applications. It will have enhanced sensitivity to the p→$K^+\nu$ mode by directly observing the $K^+$ by dE/dx and observing the subsequent $K^+\rightarrow\mu^+$ decay. KamLAND will also have sensitivity to very difficult modes like n→$\nu\nu\nu$ by observing associated nuclear γ rays, though their limit will be sharply limited by background.

*H.3 Nucleon Decay: Facility Requirements.* Among the facility requirements for next-generation nucleon decay experiments are:
- Civil engineering/rock mechanics: The civil engineering facility requirements for a detector like UNO are formidable. They include the identification of a site capable of sustaining a massive cavern and of mining, transporting, and disposing of at least 2 Mtons of rock. We believe Homestake is the only US site that has demonstrated that such construction is clearly feasible.
- A recent HEPAP report on future large DOE projects concluded that the optimal depth for the full proposed scientific program of UNO is at least 4000 mwe. Many arguments leading to this conclusion are connected with the use of UNO for atmospheric, solar, and supernova neutrino physics. However depth is important for some proton decay modes. The use of nuclear γ rays as a tag for p→$K^+\nu$ requires sufficient depth so that accidental spatial and temporal coincidence of a low-energy event due to cosmic ray muon spallation products is avoided. A study of this background as a function of depth for the $K^+\nu$ mode has not yet been done, though it is clear that the muon background is reduced by about an order of magnitude for every 1600 mwe below Super-Kamiokande depths (2700 mwe). (The difference between the 4850 ft level at Homestake and Super-Kamiokande's depth is approximately 1700 mwe.)

    There are arguments for siting low-threshold detectors even deeper. The signature for the mode n→$\nu\nu\nu$ is a very low energy coincidence (a 2 MeV γ ray from a bound-state decay in $^{11}$C followed by the β decay of the ground state (20.4m); or neutron emission from $^{11}$C

followed by the β decay of $^{10}$C (Q = 3.65 MeV, 19.3 s)). The planned KamLAND search for this decay is inhibited by the cosmic-ray-induced fast neutron rate of 5000/d, as fast neutron knockout reactions on $^{12}$C mimic the decay. If a large scintillation detector were located at 7000 mwe, the resulting $10^3$ suppression (relative to KamLAND) of the fast neutron background would result in a proportional improvement in the ννν mode limit.
- There must be a reasonable method of filling the detector. A possibility at Homestake would be to use the water seeping into the mine, which is very clean. A water purification system could be installed. The fill could proceed at 500 gpm, corresponding to 210 days for a 0.6 Mton detector.
- The laboratory must be able to handle a large-scale construction project. One would expect several hundred physicists to be involved. A Homestake advantage is the possibility of dedicating the Yates hoist to the experiment, and of optimizing that hoist for the construction.

*H.4 Nucleon Decay: Summary.* The stability of matter is clearly one of the deepest question in nature. Recent neutrino mass discoveries hint of grand unification, and GUTs generically predict nucleon decay. The favored theories predict lifetimes within about an order of magnitude of current bounds. We have the capacity to do a next-generation experiment that extends limits by an order of magnitude. One possibility uses an established technology, large water Cerenkov detectors. Other ideas include massive liquid argon or scintillator detectors, which offer advantages for certain favored decay channels.

A megadetector for nucleon decay, long baseline neutrino experiments, and for astrophysical neutrino studies is likely to be the signature project for NUSEL. Homestake provides an ideal site, and with the creation of a modern laboratory, designed to facilitate construction and operations, would be the ideal venue.

*I. Other Megadetector Uses.* This section summarizes and extends previous discussions about ancillary uses of a megadetector – other than proton decay or long-baseline physics. Most of these uses benefit from a relatively deep site.

One natural application of a large proton decay detector is to study the zenith-angle dependence of the solar neutrino rate. Such day-night studies are possible even with the energy calibration difficulties of very large detectors. Dividing the events into only day and night bins, Super-Kamiokande measures a D/N asymmetry of $0.33 \pm 0.22$ (statistical error only). Thus ~ 20 Super-Kamiokande-years will be required to measure a $3\sigma$ effect if current indications persist. A detailed determination of the zenith-angle dependence, especially the crucial effect of Earth's core, is probably out of reach. A megadetector would provide the needed event rate and might be designed to keep potential systematic effects small. As muon-induced delayed $\beta$ activities tend to be the most serious background in solar neutrino experiments, significant depth is important.

As discussed in the atmospheric neutrino section of the Science Book, several of the outstanding challenges in this subfield require very large detectors. This includes the measurement of the oscillation pattern (disappearance and then reappearance as a function of L/E), the observation of $\nu_\tau$ appearance, and precision measurements of the parameters governing the oscillations.

Because nucleon decay detectors are designed to operate for long periods, an ideal second application is to supernova neutrino physics. (Recall that the rate of galactic supernovae is about one in 35 years.) There is keen interest in measuring the neutrino "light curve" out to very long times, due to processes like kaon condensation that may occur only after the protoneutron star has radiated most of its lepton number. Such a phase change can alter the neutrino opacity of the star, changing the neutrino emission rate. Another example is a delayed collapse into a black hole, which would lead to a sharp termination of neutrino emission. The ability of a detector to follow the neutrino light curve depends on the detector's mass (the event rate) and the detector threshold (whether late, low-energy events can be detected). Ideally one would like a "smart" detector that could respond to an initial burst by preparing to trigger on very low energy events that, otherwise, might be discarded.

A second possibility is the detection of the constant flux of relic supernova neutrinos, a quantity that would be of great significance cosmologically in that it integrates over the earliest generation of massive stars to now. Recently it has been argued that the Super-Kamiokande limit is already significant, approaching within a factor of three the range of fluxes that come from plausible models of star formation. This is an example of a problem where any detection is important, but where a great deal of additional information resides in the spectrum and its redshifts. Background rates, and thus depth, are crucial because this is an isotropic low-energy signal.

The detection of extragalactic supernova neutrinos is also within reach of a megaton detector: event counts for a collapse in Andromeda would be a few tens. (But Andromeda collapses occur less than once per century.) Scintillation detectors would be capable of measuring terrestrial antineutrinos of interest to geophysicists.